\documentclass[fleqn,usenatbib]{mnras}
\usepackage{lineno}

\usepackage{newtxtext,newtxmath}

\usepackage[T1]{fontenc}
\usepackage{orcidlink}

\DeclareRobustCommand{\VAN}[3]{#2}
\let\VANthebibliography\thebibliography
\def\thebibliography{\DeclareRobustCommand{\VAN}[3]{##3}\VANthebibliography}

\usepackage{graphicx}	
\usepackage{amsmath}	
\usepackage{bm}
\usepackage[mathscr]{euscript}
\DeclareMathOperator*{\argmax}{arg\,max}

\title[FlowSN2]{2Flow2SN: Dusty Supernova Cosmology with Photometric Redshifts and Contamination using Neural Simulation-Based Inference}

\author[B. M. Boyd et al.]{Benjamin M. Boyd$^{1}$\thanks{E-mail: \href{mailto:bmb41@cam.ac.uk}{bmb41@cam.ac.uk}}\orcidlink{0000-0002-0622-1117
}, Maximilian Autenrieth$^{1,2}$\orcidlink{0009-0006-2068-5950}, Kaisey S. Mandel$^{1,2}$\orcidlink{0000-0001-9846-4417},
Matthew Grayling$^{1}$\orcidlink{0000-0002-6741-983X},\newauthor Madeleine Ginolin$^{1}$\orcidlink{0009-0004-5311-9301}, Lisa Kelsey$^{1}$\orcidlink{0000-0003-0313-0487}, and Stephen Thorp$^{1}$\orcidlink{0009-0005-6323-0457} \\ 
$^{1}$Institute of Astronomy and Kavli Institute for Cosmology, University of Cambridge, Madingley Road, Cambridge, CB3 0HA, UK\\
$^{2}$Statistical Laboratory, DPMMS, University of Cambridge, Wilberforce Road, Cambridge, CB3 0WB, UK}

\date{Accepted XXX. Received YYY; in original form ZZZ}

\pubyear{\the\year{}}

\begin{document}
\label{firstpage}
\pagerange{\pageref{firstpage}--\pageref{lastpage}}
\maketitle


\begin{abstract}
Photometric Type~Ia supernova (SN~Ia) cosmology in the era of the Vera C. Rubin Observatory's Legacy Survey of Space and Time (LSST) demands statistical frameworks that simultaneously account for selection effects, photometric redshift uncertainties and non-Ia contamination at scale. We present \textit{FlowSN2}, a modular statistical framework combining hierarchical Bayesian modelling and simulation-based inference (SBI), that reframes photometric SN cosmology from a regression task to a joint inference problem over light curve summaries and photometric redshifts. By employing neural likelihood estimation via normalising flows, \textit{FlowSN2} approximates the data distributions of selected objects in a form that can be reused across different cosmological models, achieving amortisation over the Hubble diagram. This amortisation enables inference across alternative cosmological parameterisations, such as $w_0w_a\mathrm{CDM}$, away from the fiducial training cosmology. The framework integrates a standardisation model that disentangles intrinsic SN variations from extrinsic host-galaxy dust effects and adopts a principled Bayesian resampling approach to propagate
SN classification uncertainty,
without requiring an explicit contaminant likelihood. 
Across simulated tests spanning five cosmological parameter sets, our framework recovers true cosmological and population parameters, with posteriors in close agreement with both tractable analytical solutions and pure-Ia baselines. Conveniently, the framework enables Bayesian model comparison at negligible additional computational cost, providing a valuable tool for addressing debates surrounding SN standardisation and evolving dark energy. Capable of scaling to over a hundred thousand SNe without retraining, \textit{FlowSN2} offers an interpretable and statistically principled solution for cosmological inference with upcoming wide-field photometric surveys.
\end{abstract}

\begin{keywords}
methods: statistical -- cosmological parameters -- surveys -- supernovae: general -- cosmology: observations
\end{keywords} 


\section{Introduction}

Type Ia supernovae (SNe~Ia) are standardisable candles that are useful in helping us infer properties of our Universe. It is thought that the thermonuclear explosions come from similar binary progenitor systems involving at least one white dwarf \citep{ruiter2025}. The similar initial conditions result in homogeneity in how their luminosities correlate with apparent light curve properties. Particularly, SNe that appear bluer in colour are often brighter, as are those whose light curves evolve more slowly \citep{phillips}. These correlations are conventionally expressed in a linear standardisation model of the SN's absolute magnitude \citep{tripp1998}, which may be combined with the apparent magnitude to calculate its distance modulus using the inverse square law. 

To constrain cosmological properties, inferred SNe distances, along with redshifts, are plotted on the Hubble diagram. This led to the discovery of the accelerating expansion of the Universe, providing evidence for dark energy \citep{riess1998,perlmutter1999}. More recently, SNe distances \citep{brout2022,des5yr,rubin2023}, combined with constraints from baryon acoustic oscillations \citep{desi2025} and the cosmic microwave background \citep{planck2020}, have provided evidence for an evolving dark energy equation of state. Current debates centre on whether the preference towards the alternative cosmological parameterisation is a signal of new fundamental physics or merely a residual artefact from survey systematics \citep{efstathiou2024,vincenzi2025,dhawan2025}. Recent recalibration of the DES 5YR SN~Ia sample \citep{dovekie, popovic2026} using DA white dwarfs \citep{damodel} reduced the significance of the evolving dark energy results from $\sim 4 \sigma$ to $\sim 3 \sigma$, emphasising the importance of scrutinising the full analysis pipeline. 

An important systematic to consider in SN~Ia cosmology is the influence of selection effects. Malmquist bias \citep{malmquist1922} is a common selection effect whereby our flux-limited surveys preferentially detect brighter objects. Selection effects can also arise from colour preferences, cadence-driven sensitivity to light curve duration and the spatio-temporal observing plan \citep{lochner2018,lochner2022}. Further selection cuts are often applied in post-processing depending on data quality and light curve fits. If left unaccounted for, these selection effects can strongly bias constraints on both cosmological parameters and inferred SN~Ia population parameters. 

The inferences of population parameters not only inform our astrophysical understanding of SNe~Ia, but also feed back into informing their standardisation. SNe~Ia in higher mass host-galaxies have been observed to be apparently brighter than those in lower mass galaxies \citep{kelly2010,sullivan2010,lampeitl2010,childress2013,kelsey2021,kelsey2023}, in a phenomenon known as the ``mass-step''. This has motivated investigations as to whether the mass-step originates from intrinsic differences in SN progenitors or differing dust properties in the galactic environment \citep{thorp2022,brout2021,murakami2025,ginolin2026}. Investigations into the mass-step have benefited from fitters that explicitly model dust parameters separately from intrinsic SN parameters \citep{mandel2009,mandel2014,mandel2017,mandel2022,thorp2021,popovic2021,grayling2024}. Due to the biases that selection effects can impose on population parameter inferences, studies on extrinsic and intrinsic population parameters, as well as the mass-step, are often conducted on volume-limited samples \citep{ginolin2025b,murakami2025,ginolin2026,giunta2026}.

Systematic uncertainties in the modern era will become even more challenging as we move to larger wide-field photometric surveys, such as the Vera C. Rubin Observatory’s Legacy Survey of Space and Time \citetext{LSST, \citealt{lsst2019}} and the Nancy Grace Roman Space Telescope \citetext{Roman, \citealt{roman2018}}, that will have orders of magnitude more SNe than current samples. Previously, SN~Ia cosmological studies have benefited from a spectroscopic redshift with negligible observational uncertainties. The future of SN cosmology will instead have photometric redshift estimates that are substantially more uncertain and often multimodal \citep{kind2013,jones2019,thorp2024pop,autenrieth2024improved,chen2025}. Furthermore, SN~Ia classification has historically required a spectrum to identify distinct silicon absorption features. Whilst surveys such as TiDES \citep{tides} are rising to the challenge to spectroscopically classify as many transients as possible, many of today's surveys will classify most SNe~Ia using photometry alone, utilising neural network approaches \citetext{e.g. \citealt{muthukrishna2019rapid,boone2019,moller2020,qu2021}}. Photometric classifications will be imperfect, so cosmological analyses must account for non-Ia contamination.

The increasing complexities of SN~Ia cosmology have necessitated bespoke inference pipelines that correctly handle uncertainties and systematics. Hierarchical Bayesian modelling provides an interpretable statistical framework for modelling SN populations, including physical priors and latent parameter relationships whilst consistently propagating uncertainty \citep{mandel2009,mandel2011,mandel2014,mandel2017,mandel2022,march2011,scalzo2014,shariff2016,thorp2021,grayling2024,sarin2026}. These models may be extended to account for selection effects in SN~Ia cosmology by assuming the probability a SN is selected is a function of its observed properties \citep{rubin2015,march2018,rubin2023,rubin2026,hoyt2026}. The probability of selection is assumed to take a simplified analytical form, such as a smooth sigmoid, allowing the likelihood to be analytically tractable. In practice, the true probability of selection may not be well represented by a simple parametric form after several layers of selection and quality cuts inherent in modern survey pipelines. 

An alternative approach to modelling selection effects in SNe~Ia samples is by using survey simulators \citep{kessler2009,skysurvey,dai2026,jaxsnana}. Modern simulators are able to emulate surveys to high fidelity, implementing true observing catalogues, instrumental systematics and data processing pipelines. They also are able to forward model different SN population models, physical dust effects and host-galaxy association. A popular method that uses these simulations to account for selection effects is BEAMS with Bias Corrections \citetext{BBC, \citealt{kessler2017}}. One-dimensional BBC involves simulating SNe under an assumed cosmology and fitting them to estimate their distance. The simulated SNe are then split into redshift bins, where the average difference between the true simulated distance and the fitted distance in each bin can be calculated.  The average differences in distances, known as bias corrections, are applied to the real SNe distance estimates before Hubble diagram regression to infer cosmological parameters. It is expected that SNe in the highest redshift bins will have the largest bias correction due to Malmquist bias. BBC can also be performed in higher dimensional bins, for example in colour and light curve timescale. The method also employs Bayesian Estimation Applied to Multiple Species \citetext{BEAMS, \citealt{beams}}, using an explicit mixture model approach for handling non-SN~Ia contamination. This methodology is able to account for non-trivial selection processes, although there is still a known dependence on the fiducial cosmology assumed in the bias correction simulations. Investigations have found the influence of the assumed fiducial cosmology is insignificant for present day analyses \citep{camilleri2024}, however, with scaling sample sizes it may become a dominant systematic. Furthermore, what the correct binning procedure to use in the uncertain photometric redshift regime remains unclear, with early forecasts using the analysis pipeline showing significant biases \citep{mitra2025}.

Simulation-based inference (SBI) provides a natural workflow for learning probabilistic relationships between global parameters and observed data \citep{cranmer2020}. The probabilistic relationships are commonly approximated with flexible neural networks that are trained on simulations and applied to real data at inference time. Since neural networks are universal function approximators, we are able to approximate distributions that are analytically intractable. SBI has been used in the context of SN~Ia cosmology \citep{alsing2019,karchev2023,karchev2025}, capturing complexities such as selection effects and non-Gaussian photometric redshift uncertainties implicitly through the survey simulations. One can infer a variety of SN dust and host galaxy population parameters by including these effects in the simulator's forward model \citep{karchev2024,karchev2025cigars,popovic2026sbi}. Other SBI approaches can also help choose which model, such as the parameterisation for the host galaxy mass-step, the data supports using the Bayesian evidence \citep{jeffrey2023,karchev2023b,popovic2026sbi}. 

A drawback of using neural networks to learn the full probabilistic relationship between global parameters of interest and data is the loss of interpretability. This also requires rerunning simulations and retraining for each cosmological and SN~Ia model we wish to constrain. Instead, a best-of-both-worlds approach is to use analytical hierarchical modelling for the upstream parameter relationships and then use SBI only for the downstream relationships that are intractable to derive. This modular SBI approach to SN~Ia cosmology defines the \textit{FlowSN} methodology \citep{boyd2024,flowsn}. \textit{FlowSN} combines SBI with hierarchical Bayesian modelling, using flexible normalising flows \citep{tabak2010density,tabak2013family} to approximate the intractable observed data likelihood conditioned on the latent distance modulus and redshift. Conditioning on latent parameters allowed for upstream cosmological parameters to be changed with ease, enabling the same neural network to reliably constrain a variety of cosmological models beyond the fiducial training cosmology. \textit{FlowSN} was tested on state-of-the-art SNANA survey simulations \citep{kessler2009}, where it was found to be less biased and had improved uncertainty calibration over the traditional BBC methodology \citep{kessler2017}. The original formulation of the framework used the simplest linear SN~Ia standardisation model \citep{tripp1998} and was only applicable to spectroscopic surveys with negligible redshift uncertainties. 

In this \textit{FlowSN2} work we expand our methodology for the future of photometric SN~Ia cosmology. The problem is reframed from a regression on the Hubble diagram of distance as a function of a well-determined spectroscopic redshift, to a joint problem where we consider photometric redshifts as data with uncertainties. In practice, this means including an additional observable data dimension in our first normalising flow and then training a second network to learn the joint distance modulus-redshift distribution under the influence of selection. This combination allows for amortisation on the Hubble diagram, enabling the model to constrain a range of cosmological parameterisations. We also get the added benefit of being able to infer SN rates under the influence of selection. The SN standardisation is upgraded to the Simple-BayeSN model \citep{mandel2017}, which allows for explicit separation of extrinsic dust and intrinsic SN parameters, as well as host-galaxy mass-steps. We tackle the possibility of photometric contamination from non-Ia objects, which may have misspecified likelihoods, using the modular Bayesian resampling approach proposed by \cite{autenrieth2026}. This treatment is convenient as it requires no retraining of our likelihood approximations and no explicit likelihood model for the contaminant population. Finally, we show that with little additional cost, we can perform model comparison in these complex settings, allowing us to choose which model, be it the mass-step parameterisation or evolving dark energy, the data supports. 

The general \textit{FlowSN2} statistical model is defined in Section~\ref{sec:method}, where we outline the assumed forward model, posterior construction, model comparison procedure, contamination treatment and the training of the normalising flows. In Section~\ref{sec:toy}, we show results on a simplified forward model that has an analytical likelihood solution, useful for verifying that \textit{FlowSN2} accurately reproduces the corresponding analytical-likelihood posteriors. We also include demonstrations of inference under contamination, scalability and model selection. In Section~\ref{sec:conc}, we conclude with future avenues for the work and its implications as a powerful framework for SN~Ia cosmology.

In Appendix~\ref{app:analytical}, we derive the analytical likelihood solution for our survey simulations, used to validate our method. We also include further experimental details, such as training ranges and priors, in Appendix~\ref{app:cov_samp} to Appendix~\ref{app:cmb_bao_prior}. Further investigation into posterior biases and calibration is detailed in Appendix~\ref{app:further}.

\section{Method}
\label{sec:method}

In this section, we define the general methodology used in the \textit{FlowSN2} statistical framework that can be applied to realistic photometric SN cosmology analyses. In Section~\ref{sec:forward}, we describe the general generative forward model assumed in our methodology. Section~\ref{sec:hbm} describes the assumptions and procedures needed to construct the SN likelihood and sample the posterior on cosmology and population parameters. We discuss how to combine likelihoods from multiple surveys in Section~\ref{sec:surveys}. Section~\ref{sec:evidence} shows us how to re-use our likelihoods to compute Bayes factors and perform model selection. In Section~\ref{sec:contam_theory}, we describe how to perform inference under the influence of contamination from non-Ia objects. The procedure used to train our two normalising flows is described in Section~\ref{sec:training}. Finally, we give details on our implementation of the framework in Section~\ref{sec:implementation}.

\subsection{Forward Model}
\label{sec:forward}

We first establish the general  probabilistic generative model we have designed to account for the stochastic and uncertain effects that influence the observed light curve summary statistics and redshifts following selection. Within this framework, variables associated to an individual supernova are marked with the subscript $s$, thereby distinguishing them from the global parameters that characterise the population across all SNe.
\subsubsection{Photometric Redshifts}
\label{sec:photz}

The true CMB-frame redshift $z_s$ for an individual SN with index $s$ is sampled from a distribution composed of a volumetric rate $R(z)$ and a differential co-moving volumetric component $\text{d}V_\text{c}/\text{d}z$, according to
\begin{equation}
\label{eq:samp_z}
z_s \sim P(z|\, \xi_{\text{rate}},\bm{C}) \propto \frac{R(z;\,\xi_{\text{rate}})}{1+z}\ \bigg|\frac{\text{d}V_\text{c}(z;\,\bm{C})}{\text{d}z}\bigg|.
\end{equation}
In this work, the volumetric rate is assumed to follow a power-law $R(z;\,\xi_{\text{rate}}) \propto (z+1)^{\xi_{\text{rate}}}$ bounded by $z_{\text{low}}< z_s < z_{\text{max}}$, where $\xi_{\text{rate}}$ represents a rate parameter describing its steepness. Time dilation is accounted for by the $(1+z_s)$ factor in the denominator \citep{dilday2008}, and $V_\text{c}$ denotes the co-moving volume, which is itself a function of redshift and the cosmological parameter values $\bm{C}$ \citep{hogg1999}.

The observer-frame or heliocentric redshift $z_s^{\text{hel}}$ results from the perturbation of the CMB-redshift by peculiar velocities $v^{s}_\text{pec}$ and the observer's velocity relative to the CMB rest-frame projected along the line of sight to each SN $v^{s}_{\odot}$, expressed as
\begin{equation}\label{eq:zhel}
z_s^{\text{hel}} = \bigg(z_s+1\bigg)\bigg(\frac{v^s_{\odot}}{c}+1\bigg)\bigg(\frac{v^s_{\text{pec}}}{c}+1\bigg)-1.
\end{equation}
A SN's peculiar velocity $v^s_{\text{pec}}$ is typically unknown. The peculiar velocity sampling distribution $v^s_{\text{pec}} \sim P(v_{\text{pec}})$ is usually assumed to be Gaussian ($\sigma_{\text{pec}}\approx 300~\text{km s}^{-1}$), though may be further constrained with observations \citep{do2026} or simulations \citep{okumura2014}. This updated version of the \textit{FlowSN} methodology works in both regimes where either the normalising flow learns the general influence of peculiar velocities from simulations or it may be conditioned on $z_s^{\text{hel}}$ given known peculiar velocities. We assume $v^{s}_{\odot}$ can be calculated deterministically using the coordinates of the SN and our Sun's speed relative to the CMB ($\approx 369~\text{km s}^{-1}$). The speed of light in $\text{km s}^{-1}$ is denoted by $c$.

We then measure heliocentric redshifts according to
\begin{equation}\label{eq:zhel_hat}
\hat{z}_s^{\text{hel}}\sim P(\hat{z}^{\text{hel}}|\ z_s^{\text{hel}},\sigma_{z,s} ), 
\end{equation}
\cite{flowsn} only considered spectroscopic redshifts, where uncertainty on the measurement was assumed to be very small $\sigma_{z,s} \approx 10^{-5}$. This work considers photometric redshift uncertainties where the sampling distribution may be broader $\sigma_{z,s} \approx 0.03(1+z_s^{\text{hel}})$, non-Gaussian or even multimodal.

We deterministically convert the measured heliocentric redshift into the measured CMB-frame redshift $\hat{z}_s$ using 
\begin{equation}
     \hat{z}_s = \frac{(1+\hat{z}^{\text{hel}}_s)}{\Big(1+\frac{v^s_{\odot}}{c}\Big)}-1.
\end{equation}
Further corrections can also be made to account for peculiar velocities using velocity field maps \citep{carrick2015,stiskalek2026vel}.

\subsubsection{Luminosity Distance}
In the Cosmic Microwave Background (CMB) rest frame, the luminosity distance $d^s_\text{L}$ is defined as\begin{equation}\label{eq:dl_cmb}d^s_\text{L} = \big(z_s+1\big)d_\text{c}(z_s;\ \bm{C}),
\end{equation}
where the transverse co-moving distance $d_\text{c}$ (expressed in parsecs) is determined by the true CMB-frame redshift $z_s$ and the cosmological parameter vector $\bm{C}$. To transform this quantity into the observer frame, the luminosity distance $d^s_\text{L}$ is modified by kinematic corrections such that
\begin{equation}
\label{eq:dl_obs}
\bar{d}^s_\text{L} = \bigg(\frac{v^s_{\odot}}{c}+1\bigg)\bigg(\frac{v^s_{\text{pec}}}{c}+1\bigg)^2 d^s_\text{L}.
\end{equation}
Both velocity terms account for Doppler shifting, with the peculiar velocity entering quadratically to additionally compensate for relativistic beaming effects \citep{davis2011}.

For clarity, we define the baseline CMB-frame distance modulus as\begin{equation}
\label{eq:mu_cmb}
\mu(z;\, \bm{C}) = 5 \log_{10}\bigg(\frac{d_{\text{L}}}{10 \ \text{pc}}\bigg) = 5 \log_{10}\Bigg(\frac{1}{10\text{ pc}}\big(z+1\big) d_\text{c}(z;\ \bm{C})\Bigg),\end{equation}whereas the full observer-frame distance modulus $\bar{\mu}$ is given by
\begin{multline}
\label{eq:mu_obs}
\bar{\mu}(z, v_{\text{pec}}, v_{\odot};\, \bm{C}) = 5 \log_{10}\bigg(\frac{\bar{d}_{\text{L}}}{10 \ \text{pc}}\bigg) \\
= 5 \log_{10}\Bigg(\frac{1}{10\text{ pc}}\big(z+1\big)\bigg(\frac{v_{\odot}}{c}+1\bigg)\bigg(\frac{v_{\text{pec}}}{c}+1\bigg)^2 d_\text{c}(z;\ \bm{C})\Bigg).
\end{multline}

\subsubsection{Dusty Simple-BayeSN Forward Model with Mass-Step}
\label{sec:simple-bayesn}
The original \textit{FlowSN} model \citep{flowsn} used the simple \cite{tripp1998} formula to standardise SNe. In this work, to distinguish between intrinsic SN colour and extrinsic dust, we use a generalised version of the Simple-BayeSN probabilistic generative model \citep{mandel2017}. 

The intrinsic light curve shape $x_s$, which is alternatively referred to as the stretch depending on the SN~Ia spectral energy distribution (SED) model, is assumed to be sampled from a population distribution
\begin{equation}\label{eq:xpop}
x_s \sim P(x \mid \, z_s, \bm{\theta}_x).
\end{equation}
This population distribution is parametrised by $\bm{\theta}_x$, that typically defines properties such as the population's central value and width. The specific nature of these population distributions has been evaluated extensively in the literature \citep{guy2007, betoule2014, dimitriadis2025, ginolin2025a}. Additionally, we explicitly include a conditioning on $z_s$ because previous work demonstrates that these demographic distributions evolve significantly across cosmic time \citep{nicolas2021}. In \cite{mandel2017}, the light curve shape population distribution is assumed to be Gaussian, whereas in this \textit{FlowSN} formalism we have freedom to simulate more complex distributions and learn the influence on observed data distributions with the flexible normalising flow.

Similarly, the intrinsic SN colour is drawn from the general population distribution 
\begin{equation}
\label{eq:cpop}
c^{\text{int}}_s \sim P(c^{\text{int}}|\, x_s, z_s,\bm{\theta}^{\text{int}}_c).
\end{equation}
We parametrise the distribution with $\bm{\theta}^{\text{int}}_c$, which can influence the shape of the colour distribution \citep{grayling2025,ginolin2026,giunta2026}. We condition on $x_s$ to allow for correlations with light curve shape and also on $z_s$ to consider the population distribution evolving with time \citep{popovic2025}. As with the light curve shape population distribution, \textit{FlowSN} is able to model any arbitrary intrinsic colour population distribution that may be non-Gaussian or asymmetric. The extrinsic colour excess or dust reddening is denoted by $E_s \equiv E(B-V)$ for SN $s$. The colour excess per SN is drawn from
\begin{equation}
\label{eq:epop}
E_s \sim P(E| \,\tau),
\end{equation}
where the variable $\tau$ represents the prior mean average. The prior represents the population distribution of $E_s$ values along the line of sight, from the SN through the host galaxy. The $E_s$ variable is physically constrained to be $E_s \geq 0 $. \cite{mandel2017}, following \cite{jha2007}, chooses an exponential distribution for $E_s$, while other 
studies have called for the use of alternative priors such as the Weibull distribution \citep{duarte2026}. Some forms for the population prior have analytical convenience when integrating with Gaussian distributions to marginalise over latent parameters. \textit{FlowSN} is able to implicitly marginalise over colour excess $E_s$ regardless of the form of the dust prior. The apparent colour is modelled as the sum of the intrinsic colour and the colour excess such that
\begin{equation}
    c^{\text{app}}_s = c^{\text{int}}_s + E_s.
\end{equation}

In \cite{mandel2017}, the extinguished apparent magnitude is modelled as
\begin{equation}
\label{eq:simple_bayesn}
    m_s = M^{\text{int}}_0 + \bar{\mu}_s +
    \alpha x_s + \beta_{\text{int}} c_s^{\text{int}} + R_B E_s + \gamma \delta^{\mathcal{M}}_s + \epsilon_s^{\text{int}},
\end{equation}
where $\bar{\mu}_s = \bar{\mu}(z_s, v^s_{\text{pec}}, v^s_{\odot};\, \bm{C})  $ and the $\alpha$ coefficient captures the usual ``broader-brighter'' width-luminosity relation \citep{phillips}. The $\alpha$ coefficient is modelled as a single constant, although it has been observed to have a non-linear light curve shape dependence \citep{larison2024,ginolin2025a}, which would be a trivial extension to this work. The Simple-BayeSN \citep{mandel2017} parametrisation deviates from the traditional \cite{tripp1998} formula due to its explicit separation of the intrinsic ``redder-dimmer'' colour coefficient $\beta_{\text{int}}$ and extrinsic dust relation $R_B \equiv A_B/E(B-V)$ parameters, where $A_B$ is $B$-band dust extinction. As a result of the explicit dust modelling, the residual $\epsilon^{\text{int}}_s$ scatter is attributed to intrinsic effects $\epsilon^{\text{int}}_s \sim N(0,\sigma^2_{\text{int}})$. The intrinsic absolute magnitude constant for a SN with light curve shape $x_s=0$ and intrinsic colour $c_s^{\text{int}}=0$ is denoted by $M^{\text{int}}_0$.

We model a simple mass-step \citep{kelly2010,sullivan2010,lampeitl2010,childress2013,ginolin2026} using a $\gamma$ coefficient such that 
\begin{equation}
    \delta^{\mathcal{M}}_s = \begin{cases}
0 & \text{ if }  \mathcal{M}_s \leq \mathcal{M}_{\text{step}} \\
     1 & \text{ if }  \mathcal{M}_s > \mathcal{M}_{\text{step}} \end{cases},
\end{equation}
where $\mathcal{M}_s$ is the mass of the host galaxy in which SN $s$ resides, drawn from the galaxy population distribution $\mathcal{M}_s \sim P(\mathcal{M}_s|\, \bm{\theta}_{\text{gal}})$ with population parameters $\bm{\theta}_{\text{gal}}$ that may be arbitrarily complex. The threshold $\mathcal{M}_{\text{step}} \approx 10^{10} M_{\odot}$. Separating intrinsic SN and extrinsic dust parameters using a similar threshold \citep{mandel2017,rubin2026,wojtak2026} is straightforward in the \textit{FlowSN2} methodology, as we demonstrate in our model selection experiments in Section~\ref{sec:model_comparison}.

For notational clarity, the population-level parameters for the SN model are represented by the vector $\bm{\Theta}_{\text{SN}} = (M^{\text{int}}_0,
    \alpha, \beta_{\text{int}} ,R_B, \tau, \gamma, \sigma_{\text{int}}, \bm{\theta}_x, \bm{\theta}^{\text{int}}_c)$.

\subsubsection{Light Curve Data and Summary Statistic Fits}
\label{sec:lc}
We assume the light curve simulation and summary statistic fitting to be the same as what was modelled in \cite{flowsn}. We reproduce the assumptions here for convenience.

The simulated latent variables $\bm{d}_s=(m_s,c^{\text{app}}_s,x_s)^T$ dictate the underlying SED of the supernova \citep[e.g.,][]{mandel2022, kenworthy2021}. We characterise the SED $F(T, \lambda_o ;\ \bm{d}, z^{\text{hel}})$ as a continuous function mapped across observer-frame time $T$ and wavelength $\lambda_o$, as well as the latent parameters and the heliocentric redshift. To evaluate the latent flux $\mathrm{f}_{s,i}$ for a given observation $i$ of SN $s$ recorded at observer-frame time $T_{s,i}$, the SED is integrated against the transmission efficiency function $\mathcal{T}_X(\lambda_o)$ of passband $X$ according to
\begin{equation}
\label{eq:flux_int}
    \mathrm{f}_{s,i} = 10^{0.4\times Z_{s,i}} \times \Bigg (\frac{\int^{\infty}_0 \lambda_o \cdot F(T_{s,i}, \lambda_o;\bm{d}_s, z^{\text{hel}}_s) \cdot \mathcal{T}_X(\lambda_o) \cdot \text{d} \lambda_o}{\int^{\infty}_0 \lambda_o\cdot F_{\text{std}}(\lambda_o)\cdot \mathcal{T}_X(\lambda_o)\cdot  \text{d} \lambda_o}\Bigg),
\end{equation}
where $F_{\text{std}}(\lambda_o)$ denotes the spectral flux density of a spectrophotometric reference standard, measured in $\text{ergs cm}^{-2}$ $\text{s}^{-1}$ $\text{\AA}^{-1}$ \citep[e.g.,][]{ narayan2019,damodel}. The instrumental zeropoint is defined via a reference magnitude within the given passband as $Z_{s,i}= 27.5 + m^{\text{std}}_{s,i}$. Generally, SED frameworks transform the temporal and spectral axes into the rest frame \citep{mandel2022, kenworthy2021}. Specifically, the rest-frame phase $t_{s,i}$ relative to the epoch of maximum $B$-band brightness is evaluated as $t_{s,i} \equiv (T_{s,i} - T^{\text{max}}_s)/(1+z_s^{\text{hel}})$, where $T^{\text{max}}_{s}$ represents the estimated observer-frame date of peak luminosity. The rest-frame wavelength is also calculated via $\lambda_r \equiv \lambda_o/(1+z^{\text{hel}}_s)$.

The empirical light curve data point $\hat{\mathrm{f}}_{s,i}$ is subject to diverse stochastic instrumental perturbations. For mathematical convenience, these random variations are modelled under the assumption of Gaussian measurement uncertainties
\begin{equation}\label{eq:lc_eq}
\hat{\mathrm{f}}_{s,i} \sim N(\mathrm{f}_{s,i}, \sigma_{\mathrm{f},s,i}^2),
\end{equation}
where $\sigma_{\mathrm{f},s,i}$ denotes the localised flux measurement uncertainty. The complete photometric dataset for a given supernova $s$, comprising $N_s^{\text{obs}}$ data points, is gathered into the vectors $\bm{\hat{\mathrm{f}}}_s = \{\hat{\mathrm{f}}_{s,i}\}_{i=1}^{N_s^{\text{obs}}}$ for the fluxes and $\bm{T}_s = \{T_{s,i}\}_{i=1}^{N_s^{\text{obs}}} $ for the corresponding observer-frame epochs.

Standard cosmological pipelines generally extract light curve summary statistics by fitting SN~Ia SED or light curve models to data using specialised optimisation or sampling algorithms, e.g. SALT \citep{guy2007, guy2010, kenworthy2021}, SNooPy \citep{burns2011}, or BayeSN \citep{thorp2021,mandel2022,grayling2024,grayling2026a,grayling2026b}. By fitting the SN~Ia model to the photometric data and observed heliocentric redshift $\hat{z}^{\text{hel}}_s$, these algorithms estimate the observed summary parameters $\bm{\hat{d}}_s$. This optimisation frequently employs a maximum likelihood framework, expressed as
\begin{equation}\label{eq:salt_fit}
\argmax_{\bm{d}_s} P(\bm{\hat{\mathrm{f}}}_s | \, \bm{T}_s, \bm{d}_s, \hat{z}^{\text{hel}}_s) \rightarrow \bm{\hat{d}}_s, \bm{\Sigma}_s,
\end{equation}
where $\bm{\Sigma}_s$ represents the $3\times3$ observational covariance matrix capturing the uncertainties and correlations between the estimated parameters. This covariance is typically calculated from the inverse Hessian of the negative log-likelihood function evaluated at the maximum likelihood estimate $\hat{\bm{d}}_s$. Beyond point estimation, alternative strategies for parameter extraction include MCMC \citep[e.g.,][]{mandel2009} and variational inference \citep{uzsoy2024}.

\begin{figure}
\centering
\includegraphics[width=1.\linewidth]{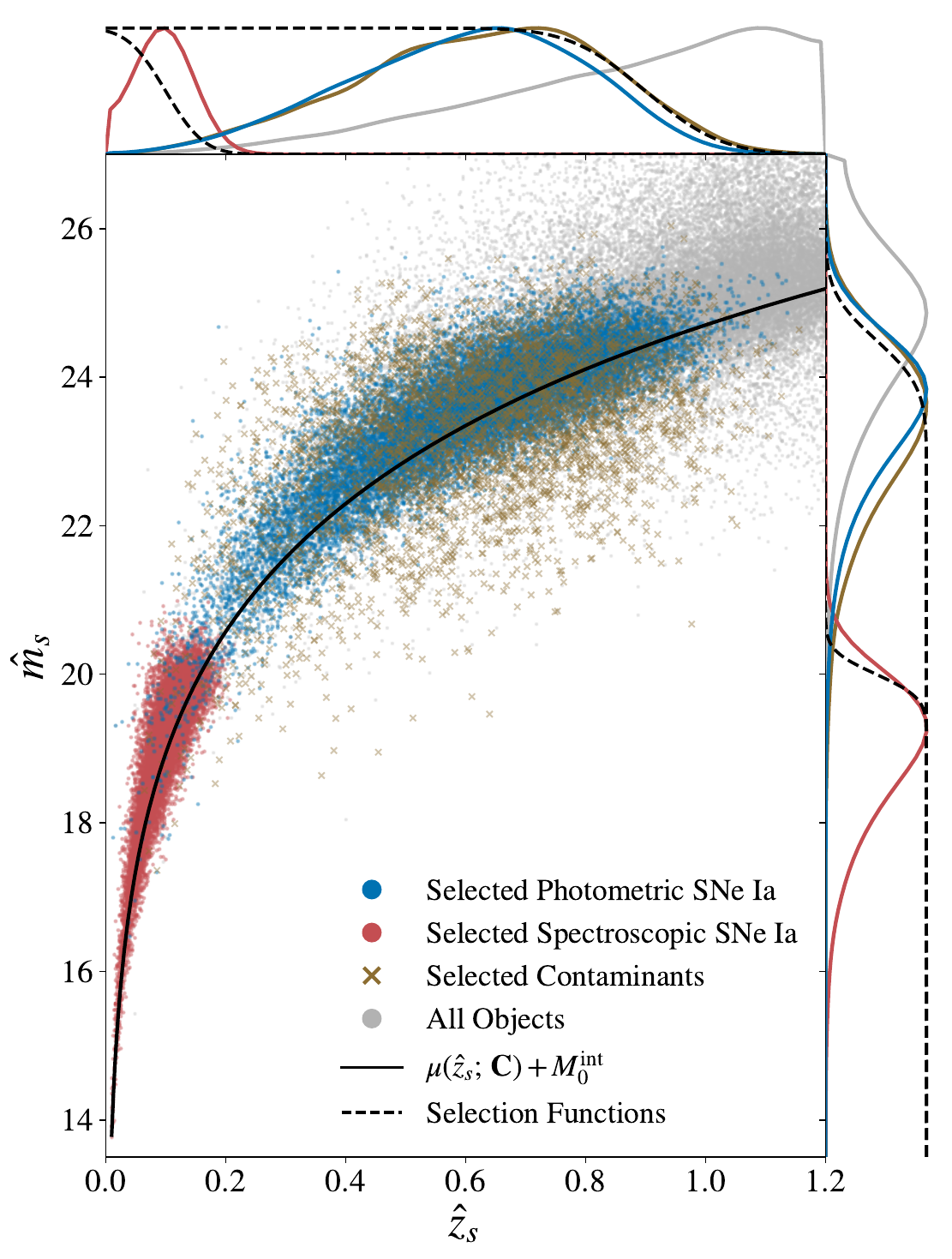}
    \vspace{-0.6cm}    \caption{Illustration of our simulated surveys used in experiments. Two independent surveys are simulated. We first include a low redshift spectroscopic survey and then a high redshift photometric survey, with larger redshift uncertainties and non-Ia contaminants. The top and right panels are the normalised marginal distributions. Both surveys have their own selection function in observed redshift and apparent magnitude space, shown in the marginals with dashed lines. }
    \label{fig:surveys} 
\end{figure}

\subsubsection{Selection}
\label{sec:sel}
Each survey has a distinct selection function that could depend on the light curve data, summary statistics, heliocentric redshift or other factors. We define an indicator variable
\begin{equation}\label{eq:detection}
I_s = \begin{cases}
0 & \text{not selected}\\
1 & \text{selected}\end{cases},
\end{equation}
where $I_s$ specifies whether SN $s$ has been selected based on the selection criteria. A SN can fail to be selected either due to falling outside a given survey's observational footprint or because the recorded data fails to satisfy the quality cuts. Hereafter, for simplicity, symbols featuring an ``$\text{obs}$'' subscript or superscript explicitly denote supernovae that are both successfully observed and satisfy these selection cuts. We include an illustration of how survey selection effects influence low redshift and high redshift distributions in Figure~\ref{fig:surveys}.

\subsubsection{Classification and Contamination}
\label{sec:class}

A photometric analysis that has no spectroscopic confirmation of a SN being of type Ia must consider contamination from non-SN~Ia objects such as fast-evolving transients, cataclysmic variables and core-collapse SNe \citep{vincenzi2021}.

To model the influence of contamination, we consider a true marginal SN~Ia fraction over the modelled redshift range before selection $\pi_{\text{Ia}}$. We then use this fraction to parametrise a Bernoulli trial to determine whether an object will be a SN~Ia or a contaminant
\begin{equation}
    \mathcal{Q}^{\text{Ia}}_s \sim  \text{Bernoulli}(\pi_{\text{Ia}})
\end{equation}
where $\mathcal{Q}^{\text{Ia}}_s$ acts as an indicator to determine which population parameters are used in the data-generating processes described from Section~\ref{sec:photz} to Section~\ref{sec:lc}, such that
\begin{equation}
    (\bm{\hat{d}}_s, \hat{z}_s) \sim \begin{cases}
P(\bm{\hat{d}}, \hat{z} |\,\bm{\Theta}_{\text{SN}}=\bm{\Theta}^{\text{Ia}}_{\text{SN}},\xi_{\text{rate}} = \xi_{\text{rate}}^{\text{Ia}} ,\bm{C}) & \text{if } \mathcal{Q}^{\text{Ia}}_s=1 \\
      P(\bm{\hat{d}},\hat{z} | \,\bm{\Theta}_{\text{SN}}=\bm{\Theta}^{\text{non-Ia}}_{\text{SN}},\xi_{\text{rate}} = \xi_{\text{rate}}^{\text{non-Ia}} ,\bm{C})& \text{if }  \mathcal{Q}^{\text{Ia}}_s=0\end{cases}.
\end{equation}
For the purpose of generating contaminants in our simulation experiments, we assume that contaminants follow the same general data-generating process as the SNe~Ia, but with different population parameters and poor standardisation properties e.g. $\bm{\Theta}^{\text{non-Ia}}_{\text{SN}}=(\alpha\approx0, \beta_{\text{int}}\approx0,\gamma \approx 0 ,\sigma_{\text{int}}\approx 1,...)$ and differing rates. This assumption is not required for inference, since the Bayesian resampling approach in Section~\ref{sec:contam_theory} does not construct an explicit contaminant likelihood.

The SNe~Ia and contaminants both pass through the type-agnostic selection function defined in Section~\ref{sec:sel}, where each object is either selected $I_s=1$ or unselected $I_s=0$. After selection the objects are all given SN~Ia classification scores $\mathcal{C}^{\text{Ia}}_s=f^{\text{class}}(\bm{X}^{\text{class}}_s)\in [0,1]$, where the observed classifier inputs $\bm{X}^{\text{class}}_s$ may include summaries or other data sources such as the light curve \citep{muthukrishna2019rapid,boone2019,moller2020,qu2021}. We assume the classification score is an estimate of the probability of being an SN~Ia given selection and the classification inputs $\mathcal{C}_s^{\text{Ia}} \approx P(\mathcal{Q}_s^{\text{Ia}}=1 \mid I_s=1, \bm{X}^{\text{class}}_s)$. In Section~\ref{sec:contam_theory}, we describe how we use these classification scores in our \textit{FlowSN2} cosmological inference under the influence of contamination.

\begin{figure}
\centering
\includegraphics[width=1.0\columnwidth]{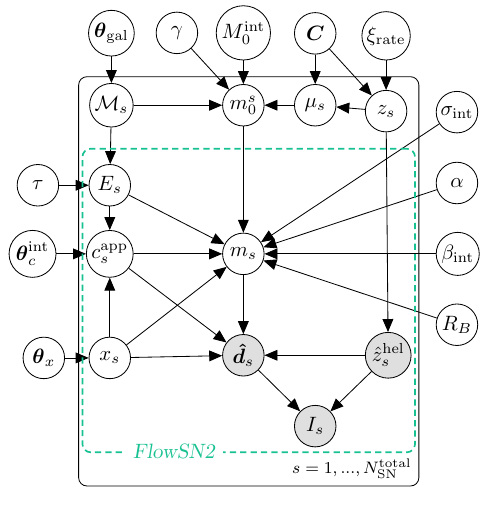}
    \vspace{-0.4cm}
    \caption{The \textit{FlowSN2} graphical model outlining the dependencies between
global parameters $\bm{\Theta}=(\bm{C},\xi_{\text{rate}},M^{\text{int}}_0,\alpha,\beta_{\text{int}},\tau,R_B,\gamma,\sigma_{\text{int}},\bm{\theta}^{\text{int}}_c,\bm{\theta}_x)$ to be inferred, supernova-level latent variables $(\mu_s,z_s,E_s,m_0^s,m_s,c^{\text{app}}_s,x_s)$ and observed data shaded
in grey $(\hat{\bm{d}}_s,\hat{z}^{\text{hel}}_s,I_s)$. We assume host galaxy masses $\mathcal{M}_s$, that are determined by galaxy population parameters $\bm{\theta}_{\text{gal}}$, are known. Any parameter may be split or conditioned on the host-mass, but here we show it influencing the latent SN magnitude and colour excess for illustration. Similarly, it is trivial to introduce redshift evolution to upstream parameters. The green dashed area highlights the dependencies learnt with the $q_1^{\bm{\phi}_1}$ normalising flow. This modular SBI framework enables top-level parameters to be defined explicitly, while lower-level latents within the green bounds are marginalised implicitly. Deciding which latent parameters are captured by the flow is a design choice guided by practical convenience. Each parameter is clearly defined in Section~\ref{sec:method}. Note, we only have observed data for selected SNe with $I_s=1$ and train the normalising flows using only these examples.}
    \label{fig:graphical_model}
\end{figure}

\subsection{Posterior Construction}
\label{sec:hbm}

\begin{figure*}
\centering
\includegraphics[width=1.0\textwidth]{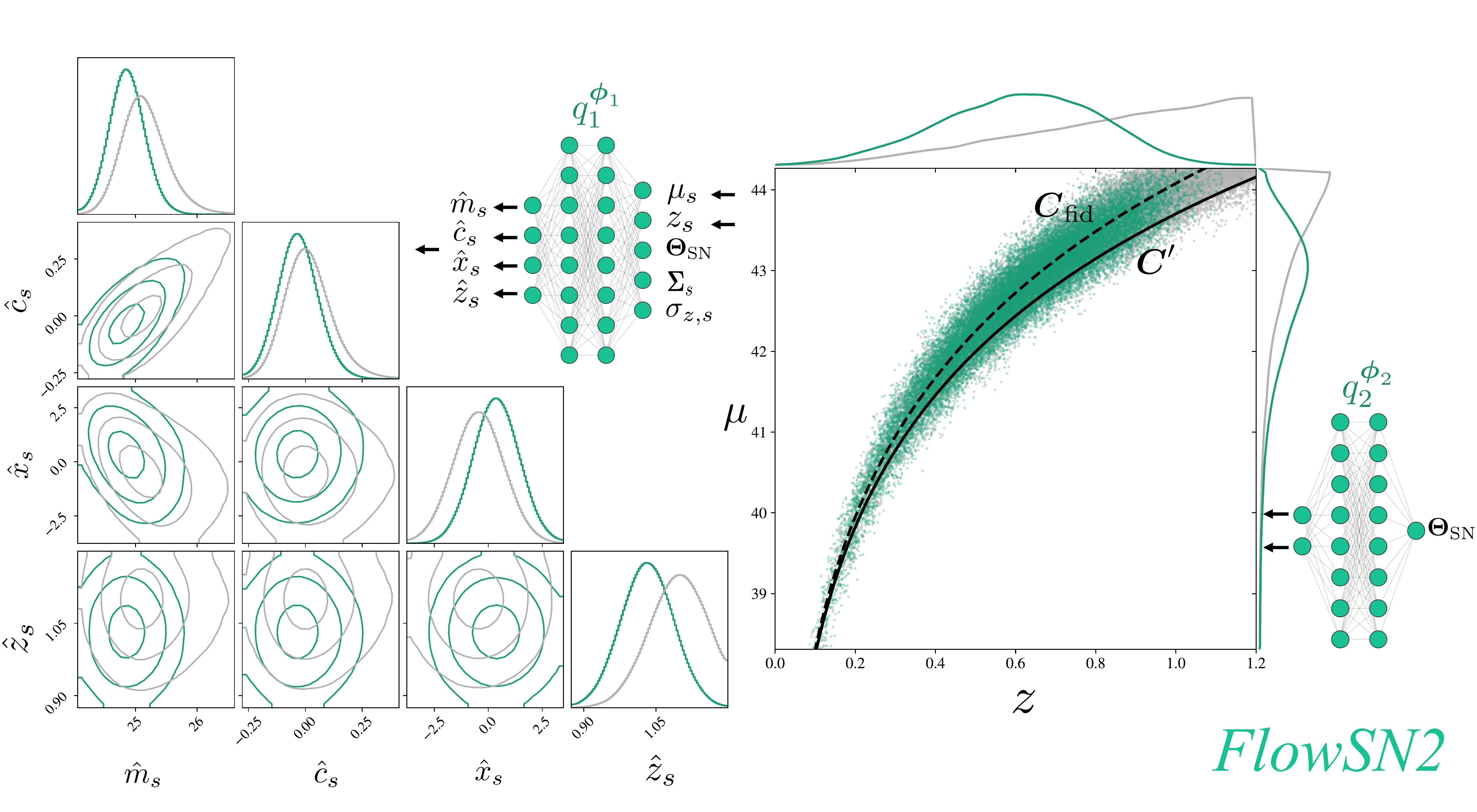}
    \vspace{-0.8cm}    \caption{Illustration of the \textit{FlowSN2} pipeline. The first normalising flow $q_1^{\bm{\phi}_1}$ learns the observed light curve summary statistics $(\hat{m}_s,\hat{c}_s,\hat{x}_s)$ and observed redshifts $\hat{z}_s$ distributions under the influence of selection effects, conditioned on the latent redshift $z_s$ and distance modulus $\mu_s$. The second normalising flow $q_2^{\bm{\phi}_2}$ learns the latent redshift and distance modulus distribution under the influence of selection. The latent redshift distribution under the influence of selection decreases at higher redshifts as Malmquist bias takes effect. The learnt  $q_2^{\bm{\phi}_2}$ represents a noisy Hubble diagram centred around a fiducial cosmology $\bm{C}_{\text{fid}}$ (dashed line), where the introduction of the noise allows us to have support over other sets of cosmological parameters $\bm{C}'$ (solid line) beyond the fiducial standard $\Lambda$CDM model. To emphasise the influence of selection, we also show the distributions of the unselected SNe in grey. Both normalising flows are trained using survey simulations, then are subsequently used during inference in Hamiltonian Monte Carlo sampling of the posterior on cosmological parameters and other SN population parameters of interest.
    }
    \label{fig:graphic} 
\end{figure*}

The fundamental change in the new \textit{FlowSN2} framework, compared to the \cite{flowsn} approach, is the shift from treating SN~Ia cosmology as a regression of distance as a function of known (spectroscopic) redshifts, to jointly modelling light curve summary statistics and photometric redshift data within a unified likelihood. In this subsection we will derive this new framework assuming a pure-Ia sample and address class contamination in Section~\ref{sec:contam_theory}. A probabilistic graphical model illustrating parameter dependencies is in Figure~\ref{fig:graphical_model}.

The posterior of the global parameters of interest $\bm{\Theta} = (\bm{\Theta}_{\text{SN}}, \xi_{\text{rate}}, \bm{C})$ conditioned on all the data, framed as a joint inference problem, is 
\begin{equation}
\label{eq:joint_post}
P( \bm{\Theta} |\ \bm{\hat{D}}_\text{obs}, \bm{\hat{Z}}_\text{obs}, \bm{I}_\text{obs} ) \propto P( \bm{\hat{D}}_\text{obs},\bm{\hat{Z}}_\text{obs}|\ \bm{I}_\text{obs}, \bm{\Theta}) P(\bm{\Theta} |\ \bm{I}_\text{obs}),
\end{equation}
where $\bm{\hat{D}}_\text{obs}~=~\{\bm{\hat{d}}_s\}_{s=1}^{N_{\text{SN}}^{\text{obs}}}$ denotes the observed summary statistics for the selected sample and $\bm{\hat{Z}}_\text{obs}~=~\{\hat{z}_s\}_{s=1}^{N_{\text{SN}}^{\text{obs}}}$ denotes the redshift measurements associated with the selected SNe. The selected sample is represented by the indicator vector of ones $\bm{I}_\text{obs}~=~\{I_s = 1\}_{s=1}^{N_{\text{SN}}^{\text{obs}}}$. The combined data likelihood across all the observed SNe is denoted by $P( \bm{\hat{D}}_\text{obs},\bm{\hat{Z}}_\text{obs}|\ \bm{I}_\text{obs}, \bm{\Theta})$. Our posterior and likelihood are conditioned on light curve summary fit covariances $\bm{\Sigma}_s$ and photometric redshift uncertainties $\sigma_{z,s}$, though we suppress them in our notation here and throughout.  The prior on our global parameters $P(\bm{\Theta} |\ \bm{I}_\text{obs})$ includes $\bm{C}$, conditional on only the selected SNe\footnote{This is equivalent to conditioning on the number of observed SNe $N_{\text{SN}}^{\text{obs}}$, although we do not use this directly to give additional constraining power as other methods do \citep{rubin2015,karchev2025}.} $\bm{I}_\text{obs}$. We also make the approximation that $P(\bm{\Theta} \mid \bm{I}_{\text{obs}}) \approx P(\bm{\Theta})$, implying that the number of selected SNe is non-informative for the global parameters, including cosmological parameters. This neglects the weak constraining power one can gain from selected SN counts which have been incorporated in other models \citep{karchev2025}.

The combined data likelihood for the $N_{\text{SN}}^{\text{obs}}$ selected SNe is written as the product of the individual observed SN data likelihoods, such that
\begin{equation}
\label{eq:big_lik}
P( \bm{\hat{D}}_\text{obs},\bm{\hat{Z}}_\text{obs}|\ \bm{I}_\text{obs}, \bm{\Theta})  =  \prod_{s=1}^{N_{\text{SN}}^{\text{obs}}} P\big(\bm{\hat{d}}_s, \hat{z}_s \big|\,I_s=1,\bm{\Theta}\big).
\end{equation}
Each SN’s observed data are assumed to be conditionally independent of those of other SNe. We also assume that the selection probability for a given SN is conditionally independent of the selection of the others. There may be some covariance systematics between SNe, however, these relationships can be incorporated as conditional likelihood information to maintain the hierarchical structure of the model.

We split the observed data likelihood into cosmology conditionally independent and cosmology-dependent components, whilst marginalising over the true latent redshift and CMB-frame distance modulus such that  
\begin{multline}
\label{eq:joint_lik_2d}
P(\bm{\hat{d}}_s, \hat{z}_s |\, I_s = 1,\bm{\Theta}) =
\\
\int^{\infty}_{-\infty}\int^{\infty}_{0} P(\bm{\hat{d}}_s,\hat{z}_s|\, I_s = 1,\mu_s, z_s,\bm{\Theta}_{\text{SN}}) \\ \times P(\mu_s,z_s|\, I_s=1,\bm{\Theta}_{\text{SN}},\xi_{\text{rate}},\bm{C}) \,\text{d}z_s \text{d}\mu_s ,
\end{multline}
where we have used the fact that, given $\mu_s$ and $z_s$, the observed data are conditionally independent of the cosmological and rate parameters. Since the CMB-frame distance modulus is a deterministic function of the true redshift, $\mu_s=\mu(z_s;\, \bm{C})$, the second factor in the integrand is $P(\mu_s,z_s|\, I_s=1,\bm{\Theta}) =  \delta_{\rm D}\big(\mu_s-\mu(z_s;\, \bm{C})\big)\,P(z_s|\, I_s=1,\bm{\Theta})$ and the marginalisation collapses into a one-dimensional integral
\begin{multline}
\label{eq:joint_lik_1d}
P(\bm{\hat{d}}_s, \hat{z}_s |\, I_s = 1,\bm{\Theta}) =
\\
\int^{\infty}_{0} P(\bm{\hat{d}}_s,\hat{z}_s|\, I_s = 1,\mu(z_s;\, \bm{C}), z_s,\bm{\Theta}_{\text{SN}}) \\ \times P(z_s|\, I_s=1,\bm{\Theta}_{\text{SN}},\xi_{\text{rate}},\bm{C}) \,\text{d}z_s .
\end{multline}

Following a similar approach to \cite{flowsn}, we define the cosmological model conditionally independent data likelihood using an auxiliary density $g_1$
\begin{multline}
\label{eq:g1}
    P(\bm{\hat{d}}_s,\hat{z}_s|\, I_s = 1,\mu_s, z_s,\bm{\Theta}_{\text{SN}})\\=g_1\big(\bm{\hat{d}}_s, \hat{z}_s \big|\,m^s_0=M^{\text{int}}_0+\mu_s+ \gamma \delta^{\mathcal{M}}_s,z_s,\bm{\Sigma}_s,\sigma_{z,s},\bm{\Theta}_{\text{SN}}\big).
\end{multline}
We then use a first normalising flow $q_1^{\bm{\phi}_1}$ with learnable parameters $\bm{\phi}_1$ as a flexible density estimator \citep{tabak2010density,tabak2013family} to approximate $g_1$
\begin{equation}
    g_1\big(\bm{\hat{d}}, \hat{z} \big|\,m_0,z,\bm{\Sigma},\sigma_{z},\bm{\Theta}_{\text{SN}}\big) \approx q_1^{\bm{\phi}_1}( \bm{\hat{d}}, \hat{z} |\, \bm{\psi}_1)
\end{equation}
where $\bm{\psi}_1 = (m_0,z,\bm{\Sigma},\sigma_{z},\bm{\Theta}_{\text{SN}})$. Details of $q_1^{\bm{\phi}_1}$ training using simulations are provided in Section~\ref{sec:q1_train}.

The cosmology-dependent latent redshift distribution of the selected SNe follows from Bayes' theorem,
\begin{multline}
\label{eq:zsel}
    P(z_s |\, I_s =1,\bm{\Theta}_{\text{SN}},\xi_{\text{rate}},\bm{C})\\ = \frac{P(I_s=1|\,\mu(z_s;\, \bm{C}),z_s, \bm{\Theta}_{\text{SN}} )\,P(z_s |\, \xi_{\text{rate}},\bm{C})}{\int^{\infty}_{0} P(I_s=1|\,\mu(z;\, \bm{C}),z,\bm{\Theta}_{\text{SN}} )\,P(z |\, \xi_{\text{rate}},\bm{C}) \,\text{d}z }  \\
    =\frac{\bar{w}(z_s)\, P(z_s |\, \xi_{\text{rate}},\bm{C})}{\int^{\infty}_{0} \bar{w}(z)\, P(z |\, \xi_{\text{rate}},\bm{C}) \,\text{d}z } \\ \text{where }  \bar{w}(z_s)  = \frac{P(I_s=1|\,\mu(z_s;\,\bm{C}),z_s, \bm{\Theta}_{\text{SN}})}{P(I_s=1|\, \bm{\Theta}_{\text{SN}}, \xi_{\text{rate}}=1.5, \bm{C}=\bm{C}_1)} \\ =
    \frac{P(\mu(z_s;\,\bm{C}),z_s|\,I_s=1, \bm{\Theta}_{\text{SN}}, \xi_{\text{rate}}=1.5, \bm{C}=\bm{C}_1)}{P(\mu(z_s;\,\bm{C}),z_s|\,
    \xi_{\text{rate}}=1.5, \bm{C}=\bm{C}_1)}
\end{multline}
where we have used the observation that the selection probability $P(I_s=1|\, \mu_s, z_s,\bm{\Theta}_{\text{SN}},\xi_{\text{rate}}, \bm{C}) \equiv P(I_s=1|\, \mu_s, z_s,\bm{\Theta}_{\text{SN}}) $ is conditionally independent of cosmological and rate parameters given $\mu_s$ and $z_s$ are known. The selection probability is therefore a property of the survey alone, shared by every cosmology, and the second equality follows by applying Bayes' theorem under a fiducial cosmology $\bm{C}_1$ and fiducial rate $\xi_{\text{rate}}=1.5$. Note that the constant $P(I_s=1|\, \bm{\Theta}_{\text{SN}}, \xi_{\text{rate}}=1.5, \bm{C}=\bm{C}_1)$ cancels between the numerator and denominator of Eq.~\eqref{eq:zsel} and never needs to be evaluated. Rather than implement a classifier to approximate the selection probability, we instead approximate $P(\mu_s,z_s|\,I_s=1, \bm{\Theta}_{\text{SN}}, \xi_{\text{rate}}=1.5, \bm{C}=\bm{C}_1)$ directly, and evaluate $P(\mu_s,z_s|\,\xi_{\text{rate}}=1.5, \bm{C}=\bm{C}_1)$ analytically using Eq.~\eqref{eq:samp_z}.

We define the latent redshift--distance modulus distribution for selected SNe using a
second auxiliary density $g_2$
\begin{multline}
\label{eq:g2}
P(\mu_s,z_s|\,I_s=1, \bm{\Theta}_{\text{SN}}, \xi_{\text{rate}}=1.5, \bm{C}=\bm{C}_1)\\= g_2(\Delta\mu_s, z_s|\, \bm{\Theta}_{\text{SN}})
\end{multline}
where $\Delta\mu_s = \mu_s + M^{\text{int}}_{0} + \gamma \delta^{\mathcal{M}}_s - M^{\text{int}}_{0,\text{fid}} - \mu(z_s;\,\bm{C}=\bm{C}_1)$ is the magnitude zero-point of the SN relative to its fiducial value. We then use a second normalising flow $q_2^{\bm{\phi}_2}$ with learnable parameters $\bm{\phi}_2$ as a flexible density estimator to approximate $g_2$
\begin{equation}
    g_2(\Delta\mu, z|\, \bm{\Theta}_{\text{SN}}) \approx q_2^{\bm{\phi}_2}( \Delta\mu, z |\,\bm{\Theta}_{\text{SN}})
\end{equation}
When simulating the training set we scatter the distance modulus about its fiducial
value, $\mu_t = \mu(z_t;\,\bm{C}_1) + \varepsilon^{\mu}_t$ with $\varepsilon^{\mu}_t
\sim N(0,\sigma^2_{\mu})$ and $\sigma_{\mu} \approx 0.25$, \emph{before} applying the survey
selection. Without this scatter the simulated
pairs $(\mu_t, z_t)$ would all lie on the single curve $\mu_t = \mu(z_t;\,\bm{C}_1)$,
and $q_2^{\bm{\phi}_2}$ could not be evaluated away from it. The scatter instead gives
support over $(\mu_s, z_s)$ pairs away from the fiducial cosmology at inference time, as the
trial values of $\bm{C}$, $M^{\text{int}}_0$ and $\gamma$ vary. All alternative cosmologies that lie in the broadened training distribution have support. The scatter must also be
accounted for in the normalisation of $\bar{w}(z_s)$, such that
\begin{multline}
P(\mu_s, z_{s} | \, \xi_{\text{rate}} =1.5,\bm{C}=\bm{C}_1) \\ \approx N(\Delta\mu_s|\,0,\sigma^2_{\mu})\,P(z_s|\,\xi_{\text{rate}} = 1.5,\bm{C}=\bm{C}_1) .
\end{multline}
Further details on the training of $q_2^{\bm{\phi}_2}$ are in Section~\ref{ssec:q2_train}. 

The joint SN data likelihood used in \textit{FlowSN2} in practice is therefore
\begin{multline}
\label{eq:final_lik}
    P(\bm{\hat{d}}_s, \hat{z}_s |\, I_s = 1,\bm{\Theta})  \\ \approx \frac{\int^{\infty}_{0} q_1^{\bm{\phi}_1}(\bm{\hat{d}}_s, \hat{z}_s | \, \bm{\psi}^s_1)\,\bar{w}(z_s)\, P( z_{s} |\, \xi_{\text{rate}},\bm{C})\,\text{d}z_s}{\int^{\infty}_{0} \bar{w}(z_s)\, P(z_s|\, \xi_{\text{rate}},\bm{C}) \,\text{d}z_s} \\
 \text{where }  \bar{w}(z_s)  \approx \\ \frac{ q_2^{\bm{\phi}_2}(\Delta\mu_s,z_s|\,\bm{\Theta}_{\text{SN}})  }{N(\Delta\mu_s|\,0,\sigma^2_{\mu})\,P(z_s|\,\xi_{\text{rate}} = 1.5,\bm{C}=\bm{C}_1)} ,
\end{multline}
where both one-dimensional integrals are numerically computed using the trapezium rule. Figure~\ref{fig:graphic} illustrates the distributions our normalising flows learn and how they interact in the framework.

\subsection{Combining Surveys}
\label{sec:surveys}

In practice, we have SNe from several distinct surveys, each with its own
selection function and its own measurement uncertainties. In the simulated experiments that follow we use two surveys, a low-redshift spectroscopic sample and a high-redshift
photometric one, but the treatment below applies to any number. We index the surveys by
$j = 1,\dots,N_{\text{surv}}$, and write $\bm{\hat{d}}^j_s$, $\hat{z}^j_s$ and $I^j_s$
for the summary statistics, redshift measurement and selection indicator of the $s$th SN
of survey $j$, of which $N^{\text{obs}}_j$ are selected.

The global parameters $\bm{\Theta} = (\bm{\Theta}_{\text{SN}}, \xi_{\text{rate}}, \bm{C})$
are shared by every survey. The cosmology, the SN population and the volumetric rate are
properties of the Universe rather than of any instrument, and it is precisely this
sharing that allows the surveys to be combined informatively. What differs between
surveys is the selection function and the observational uncertainties.

Conditioned on $\bm{\Theta}$, the observed data of a SN in one survey are independent of
those of a SN in any other, and the selection of a given SN is independent of the
selection of the others both within and across surveys. The combined data likelihood is
therefore the product over surveys of the per-survey likelihoods, each of which is
itself the product over that survey's selected SNe,
\begin{equation}
\label{eq:combine}
P( \bm{\hat{D}}_\text{obs},\bm{\hat{Z}}_\text{obs}|\ \bm{I}_\text{obs}, \bm{\Theta}) = \prod_{j=1}^{N_{\text{surv}}} \prod_{s=1}^{N^{\text{obs}}_j} P\big(\bm{\hat{d}}^j_s, \hat{z}^j_s \big|\,I^j_s=1,\bm{\Theta}\big),
\end{equation}
which is Eq.~\eqref{eq:big_lik} with the product over the full sample split into a
product over surveys. Each factor of Eq.~\eqref{eq:combine} takes the form derived in Section~\ref{sec:hbm}.
Since both auxiliary densities are conditional on selection, we train two normalising
flows per survey, $q_1^{\bm{\phi}^j_1}$ and $q_2^{\bm{\phi}^j_2}$, with learnable
parameters $\bm{\phi}^j_1$ and $\bm{\phi}^j_2$, and the importance weight becomes
survey-dependent, $\bar{w}^j(z_s)$. 

\subsection{Bayesian Evidence for Model Selection}
\label{sec:evidence}

We re-use our \textit{FlowSN2} formulation (without retraining our normalising flows) to calculate the Bayesian evidence for model selection. The evidence is defined as
\begin{multline}
P( \bm{\hat{D}}_\text{obs},\bm{\hat{Z}}_\text{obs}|\ \bm{I}_\text{obs},\mathscr{M}_1) \\= \int P( \bm{\hat{D}}_\text{obs},\bm{\hat{Z}}_\text{obs}|\ \bm{I}_\text{obs}, \bm{\Theta},\mathscr{M}_1) P(\bm{\Theta} | \,\bm{I}_\text{obs},\mathscr{M}_1) \text{ d} \bm{\Theta}
\end{multline}
where we include the implicit conditioning on our assumed model $\mathscr{M}_1$. Here we assume $P(\bm{\Theta} |\, \bm{I}_{\text{obs}},\mathscr{M}_1) = P(\bm{\Theta}|\,\mathscr{M}_1)$, implying that the number of selected SNe is non-informative for the global parameters, including cosmological parameters, as done in Eq.~\eqref{eq:joint_post}. Our priors must also be proper so we can integrate over the full prior volume. Since we marginalise over latent parameters, leaving $\approx 15$ global parameters, we can practically calculate the evidence of our model with modern nested sampling codes \citep{handley2015a,handley2015b,buchner2021,ashton2022}.

It is then helpful to compare evidence ratios of differing models using Bayes factors
\begin{equation}
    \text{BF}_{1,0} = \frac{P( \bm{\hat{D}}_\text{obs},\bm{\hat{Z}}_\text{obs}|\ \bm{I}_\text{obs},\mathscr{M}_1)}{P( \bm{\hat{D}}_\text{obs},\bm{\hat{Z}}_\text{obs}|\ \bm{I}_\text{obs},\mathscr{M}_0)},
\end{equation}
where the assumed generative models, $\mathscr{M}_0$ and $\mathscr{M}_1$, may differ, for example choosing the Simple-BayeSN \citep{mandel2017} parameterisation to explicitly model dust over the standard \cite{tripp1998} formula. One can interpret the ratios to determine whether the data favours one model over the other using the Jeffreys' scale \citep{jeffreys1961theory}. We note that Bayes factors can be sensitive to the choice and width of parameters priors, so in practice these should be varied before model comparison results are declared.  

In the special case of comparing a null hypothesis model $\mathscr{M}_0$, that is nested within a more complex model $\mathscr{M}_1$ with identical priors on mutual parameters, we may use the Savage-Dickey density ratio \citep{savagedickey} such that
\begin{equation}    
\label{eq:savdickey}
\text{BF}_{1,0} = \frac{P(\theta = \theta_0 |\, \bm{I}_\text{obs}, \mathscr{M}_1)}{P(\theta = \theta_0 |\,\bm{\hat{D}}_\text{obs}, \bm{\hat{Z}}_\text{obs}, \bm{I}_\text{obs}, \mathscr{M}_1)} ,
\end{equation}
where $\theta$ is a parameter that is freely inferred in the complex model $\mathscr{M}_1$, but held constant at $\theta_0$ in the nested null model $\mathscr{M}_0$. The term $P(\theta = \theta_0 |\, \bm{\hat{D}}_\text{obs}, \bm{\hat{Z}}_\text{obs}, \bm{I}_\text{obs}, \mathscr{M}_1)$ is the marginal posterior density evaluated at $\theta_0$, while $P(\theta = \theta_0 |\,\bm{I}_\text{obs}, \mathscr{M}_1)$ represents the marginal prior density evaluated at $\theta_0$. This property is particularly convenient in the case of our SN analysis as it makes use of our precomputed $\mathscr{M}_1$ posteriors. Furthermore, there are several nested model comparisons of interest. For example, we can evaluate whether dark energy is evolving ($w_a \neq 0$) versus constant ($w_a = 0$), as well as testing for the presence of an intrinsic host-galaxy mass-step ($\gamma \neq 0$ versus $\gamma = 0$) against extrinsic differences in dust. In Section~\ref{sec:model_comparison}, we demonstrate that \textit{FlowSN2} is able to identify the correct data-generating model using the Savage-Dickey density ratio, whilst incurring negligible additional computational cost.

\subsection{Modular Bayesian Inference under Contamination}
\label{sec:contam_theory}

In this subsection, we outline how we make inferences under the influence of non-Ia contamination in our sample. We utilise the modular\footnote{
This approach is considered modular as it prevents feedback from the downstream scientific model into classification by fixing class probabilities, and excludes potentially misspecified contaminant likelihood contributions from inference on cosmological and SN~Ia population parameters. This should be distinguished from when we describe \textit{FlowSN2} as a modular SBI framework, that uses normalising flows to amortise part of the SN~Ia likelihood, whilst keeping other upstream parameter relationships analytical.} Bayesian approach proposed by \cite{autenrieth2026}. This approach is particularly convenient as it does not require an explicit likelihood for the suspect component, here the non-Ia population, protecting downstream inference from potentially misspecified contaminant models while propagating classification uncertainty. We note that the \textit{FlowSN2} framework is also compatible with a BEAMS-like mixture model approach \citep{beams}, although this would require an explicit contaminant likelihood and could bias inference if that likelihood is misspecified. Below, we outline the modular Bayesian resampling procedure adopted in \textit{FlowSN2} and refer the reader to \citet{autenrieth2026} for its theoretical development and empirical validation.

We adopt the two modular interventions proposed by \citet{autenrieth2026} to protect inference on parameters of scientific interest from potentially misspecified mixture components. First, classification scores obtained from the classifier $f^{\text{class}}(\bm{X}^{\text{class}}_s)\rightarrow \mathcal{C}^{\mathrm{Ia}}_s$, defined in Section~\ref{sec:class}, are held fixed during downstream inference, preventing feedback from the scientific model into classification. Here, $\bm{X}^{\text{class}}_s$ denotes the observed information used for classification, which includes summary statistics $\hat{\bm{d}}_s$ and redshifts $\hat{z}_s$, and may also include additional data such as light curve data. Classification scores may be obtained from classifiers trained on representative labelled observations or simulations, provided their calibration is appropriate for the target survey, which can be assessed using representative labelled observations where available. The classification scores need not recover the exact conditional class probabilities. A sufficient condition is that, among selected SNe with similar observed light curve summaries and redshifts, the average matches the true fraction of SNe~Ia. Under this condition, the method can recover the same cosmological and SN~Ia population parameter values as an analysis with known SN types in sufficiently large samples \citep{autenrieth2026}. Second, we treat the non-Ia population as a suspect component, since a misspecified contaminant likelihood could bias inference. We therefore exclude its likelihood contributions from the downstream analysis, restricting inference to the trusted SN~Ia model, at the potential cost of information that a correctly specified contaminant model could provide. Together, these interventions substantially simplify the computational treatment of contamination by allowing us to reuse the existing SN~Ia inference framework across sampled class allocations, as described below.

We resample $M$ independent class allocations, each assigning a binary label to every selected ($I_s=1$) object according to its classification score $\mathcal{C}^{\text{Ia}}_s$
\begin{equation}
    L^m_s \sim \text{Bernoulli}\big(\mathcal{C}^{\text{Ia}}_s\big)  \text{ for } s=1,\dots,N^{\text{obs}}_{\text{SN}} \text{ and } m=1,\dots,M.
\end{equation}
The indicator $L^m_s=1$ signals that object $s$ is retained in the SN~Ia analysis, while $L^m_s=0$ indicates that object $s$ is excluded for allocation 
$m$. The data used in allocation $m$ are therefore
\begin{equation*}
\bm{\hat{D}}^m = \{\bm{\hat{d}}_s : L^m_s = 1\}, \quad \bm{\hat{Z}}^m = \{\hat{z}_s : L^m_s = 1\}, \quad \bm{I}^m = \{I_s : L^m_s = 1\}.
\end{equation*}
Following \cite{autenrieth2026}, the modular Bayesian posterior $P'(\bm{\Theta} |\, \bm{\hat{D}}_\text{obs}, \bm{\hat{Z}}_\text{obs}, \bm{I}_\text{obs})$ is approximated by averaging the normalised posteriors of 
Eq.~\eqref{eq:joint_post} over the $M$ resampled label allocations
\begin{equation}
\label{eq:post_avg}
P'(\bm{\Theta} |\, \bm{\hat{D}}_\text{obs}, \bm{\hat{Z}}_\text{obs}, \bm{I}_\text{obs})
\approx
\frac{1}{M}\sum_{m=1}^{M}
P\big(\bm{\Theta} |\,\bm{\hat{D}}^m, \bm{\hat{Z}}^m,\bm{I}^m\big),
\end{equation}
where $\bm{\Theta} = (\bm{\Theta}^{\text{Ia}}_{\text{SN}}, \xi_{\text{rate}}^{\text{Ia}}, \bm{C})$.
A non-Ia contaminant likelihood \citep{beams} is not included, so we make no inferences about the non-Ia contaminant population parameters $\bm{\Theta}^{\text{non-Ia}}_{\text{SN}}$.

For model comparison, we proceed in a similar way. 
For each resample, we evaluate the Bayes factor on the retained sample and combine the results as
\begin{equation}
\label{eq:bf_avg}
\text{BF}'_{1,0} = \big[\text{BF}'_{0,1}\big]^{-1}
\approx
\left[\frac{1}{M}\sum_{m=1}^{M}
      \frac{P\big(\bm{\hat{D}}^m, \bm{\hat{Z}}^m|\,\bm{I}^m, \mathscr{M}_0\big)}
           {P\big(\bm{\hat{D}}^m, \bm{\hat{Z}}^m|\,\bm{I}^m, \mathscr{M}_1\big)}\right]^{-1},
\end{equation}
where the numerator and denominator are evaluated on the same resampled dataset $(\bm{\hat{D}}^m,\bm{\hat{Z}}^m)$, 
so each term is a well-defined Bayes factor for that resampled data and no contaminant likelihood is required. For nested models, assuming a common prior across class allocations, the Savage-Dickey density ratio evaluated on the modular posterior is
\begin{equation}
\label{eq:bf_sd_avg}
\text{BF}'_{1,0}
\approx
\frac{\sum_{m=1}^M P(\theta = \theta_0 |\, \bm{I}^m,\mathscr{M}_1)}
     {\sum_{m=1}^{M} P\big(\theta = \theta_0 |\,\bm{\hat{D}}^m, \bm{\hat{Z}}^m,\bm{I}^m, \mathscr{M}_1\big)},
\end{equation}
reusing the per-resample marginal posterior densities used to construct the modular posterior 
in Eq.~\eqref{eq:post_avg}.

\subsection{Normalising Flow Training}
\label{sec:training}
 In this subsection, we describe the general simulations and procedures needed to train our normalising flows $q_1^{\bm{\phi}_1}$ and $q_2^{\bm{\phi}_2}$.

\subsubsection{Training Simulations}
In our training, we assume the general forward model described in Section~\ref{sec:forward}, starting with population-level parameters, leading to intermediate latent parameters and resulting in observed summaries and redshifts for the selected SNe $\bm{\Theta} \rightarrow (\mu_t,z_t) \rightarrow (\bm{\hat{d}}_t,\hat{z}_t)$. Both normalising flows can be trained using the same training simulations under a single fiducial combination of cosmology parameters $\bm{C}_1$, fiducial SN~Ia rate $\xi^{\text{Ia}}_{\text{rate,fid}}$ and fiducial intrinsic absolute magnitude $M^{\text{int}}_{0,\text{fid}}$. We also assume no intrinsic mass-step $\gamma=0$ in our training. Since we follow the modular Bayesian resampling prescription for handling contamination \citep{autenrieth2026}, as described in Section~\ref{sec:contam_theory}, we only need to simulate a pure SN~Ia sample to train our SN~Ia model, avoiding the need to train on contaminants.

Our simulations vary each of the other SN population parameters $\bm{\Theta}^t_{\text{SN}} = (
    \alpha_t, \beta^t_{\text{int}} ,R^t_B, \tau_t, \sigma^t_{\text{int}}, \bm{\theta}^t_x, \bm{\theta}^{\text{int},t}_c)$ for each simulated object $t$ such that $\bm{\Theta}^t_{\text{SN}} \sim P_{\text{train}}(\bm{\Theta}_{\text{SN}} )$. The training distribution need not be equal to our inferential priors over the population parameters $P_{\text{train}}(\bm{\Theta}_{\text{SN}} )\neq P(\bm{\Theta}_{\text{SN}} )$, though they must be supported in the same range. If it is not possible to vary the population parameters per-event in a particular simulator, the alternative options are either assuming the population parameters to be constant or conditioning flow $q_1^{\bm{\phi}_{1}}$ on latent $\bm{d}_t$, although this will incur three latent parameters per SN at inference time. 

    Our second requirement for the training simulator is that we can apply a per-event smearing to the latent SN distance such that 
    \begin{equation}
        \mu_t = \mu(z_t; \, \bm{C}_1) + \epsilon^{\mu}_t
    \end{equation}
    where $\epsilon^{\mu}_t \sim N(0,\sigma_{\mu}^2)$. This smearing is applied to the latent variable during simulation and \textit{before} selection. The purpose of the smearing is to allow support over pairs of $(z_s,\mu_t)$ beyond $\bm{C}_1$, including those from a different cosmological model e.g., $w$CDM and $w_0w_a$CDM. We find a smearing of $\sigma_{\mu} \approx 0.25$ to provide sufficient support over typical cosmological parameter priors. 

    The relevant parameters that should be logged to train the flows include the fiducial parameters $(\bm{C}_1, \xi^{\text{Ia}}_{\text{rate,fid}},M^{\text{int}}_{0,\text{fid}})$, per object varying population parameters $\bm{\Theta}^t_{\text{SN}}$, post-smearing latent distance $\mu_t$, latent redshift $z_t$, observed summaries $\bm{\hat{d}}_t$ along with covariances $\Sigma_t$ and observed redshift $\hat{z}_t$ with measurement uncertainty $\sigma_{z,t}$. If the photometric redshift estimate is provided using probability quantiles, we recommend randomly sampling from this to generate $\hat{z}_t$ and $\sigma_{z,t}$. We find simulating $N_{\text{train}}\approx25$ million selected SNe to be sufficient to train both flows.

\subsubsection{Training Flow $q^{\bm{\phi}_1}_1$}
\label{sec:q1_train}

A normalising flow \citep{tabak2010density,tabak2013family} is a flexible density estimator, helpful for approximating intractable distributions. We opt for using a normalising flow over a mixture density network \citep{bishop1994} because it provides the non-Gaussian flexibility needed to model the non-analytical selection effects inherent in realistic surveys. Our first normalising flow is similar to the normalising flow used in the \textit{FlowSN} methodology \citep{flowsn}, but extends it to model redshifts as data. The target likelihood to be learnt is the four-dimensional auxiliary density defined in Eq.~\eqref{eq:g1}. We can describe $g_1$ as the combined SN summary statistic and redshift distribution given the true underlying parameters. The normalising flow $q_1^{\bm{\phi}_1}$, used to approximate $g_1$, is defined as 
\begin{equation}
\label{eq:nf}
q_1^{\bm{\phi}_1}\big( \bm{\hat{d}}, \hat{z} |\, \bm{\psi}_1=(m_0,z,\bm{\Sigma},\sigma_{z},\bm{\Theta}_{\text{SN}})\big) = N(\bm{u}_0 |\,\bm{0},\bm{I}) \prod^L_{l=1} \  \biggr\rvert \frac{\partial f_l^{\bm{\phi}}}{\partial \bm{u}_{l-1}}\biggr\rvert^{-1},
\end{equation}
where $\bm{u}_0 \sim N(\bm{0},\bm{I})$, $\bm{u}_l  = f_l^{\bm{\phi}}\big(\bm{u}_{l-1}|\, \bm{\psi}_1=(m_0,z,\bm{\Sigma},\sigma_{z},\bm{\Theta}_{\text{SN}})\big)$  and $(\bm{\hat{d}},\hat{z}) = f^{\bm{\phi}}_L(\bm{u}_{L-1}|\, \bm{\psi}_1)$. We stack $L$ transforms $f$ with learnable parameters $\bm{\phi}_1$. Evaluating normalising flows using the chain rule is straightforward as their architectures are explicitly designed with lower triangular Jacobian matrices, allowing the determinant to be computed simply as the product of the diagonal elements. In this study, we use the Masked Autoregressive Flow (MAF) architecture \citep{papamakarios2017masked}, which is constructed by stacking Masked Auto-encoder for Distribution Estimation (MADE) transform blocks \citep{germain2015made}.

The objective is to minimise the negative log-density of the normalising flow with respect to $\bm{\phi}_1$ at each step over the joint training distribution of the observed data $(\hat{\bm{d}},\hat{z})$ and conditional context $\bm{\psi}_1$, such that
\begin{equation}
   \min_{\bm{\phi}_1} \; \mathbb{E}_{(\bm{\hat{d}}_t,\hat{z}_t,\bm{\psi}_t)\sim P_{\text{train}}(\bm{\hat{d}},\hat{z},\bm{\psi})}\Big[ -\log q^{\bm{\phi}_1}_1\big(\bm{\hat{d}}_t,\hat{z}_t \big|\ \bm{\psi}^t_1\big)\Big],
\end{equation}
where we sample the training data from the joint distribution $P_{\text{train}}(\bm{\hat{d}},\hat{z},\bm{\psi}_1)$. The objective is optimised at each training iteration using stochastic gradient descent in normalising flow parameter-space $\bm{\phi}$, where the expectation is approximated with empirical averages over mini-batches $\{ (\hat{\bm{d}}_t,\hat{z}_t, \bm{\psi}^t_1) \}_{t=1}^B$ of size $B$

\begin{equation}
    \min_{\bm{\phi}_1}\; \frac{1}{B} \sum_{t=1}^B -\log q^{\bm{\phi}_1}_1 \big(\bm{\hat{d}}_t,\hat{z}_t\big|\ \bm{\psi}_1^t\big).
\end{equation}
This optimisation is equivalent to minimising a Monte Carlo approximation of the forward Kullback-Leibler (KL) divergence between $g_1$ and $q_1^{\bm{\phi}_1}$ \citep{papamakarios2021}. 

We note that by design, we explicitly condition the normalising flow on latent $z_t$ and $\mu_t$ (through $m^t_0=M^{\text{int}}_{0,\text{fid}}+\mu_t$), before peculiar velocities are introduced in the simulations. This means the normalising flow implicitly learns the influence of peculiar velocities, as well as the influence of selection effects and measurement uncertainties, on the observed data.

\subsubsection{Training Flow $q^{\bm{\phi}_2}_2$}
\label{ssec:q2_train}
In our new \textit{FlowSN2} methodology, we introduce a second normalising flow. The purpose of this flow is to learn the two-dimensional latent $(z,\mu)$ distribution under the influence of selection. This problem can be re-parametrised in the $\mu$ dimension as the difference between simulated training distance (after smearing) $\mu_t$ and the distance under the assumed fiducial cosmology such that $\Delta\mu_t = \mu_t - \mu(z_t;\,\bm{C}=\bm{C}_1)$. We define the auxiliary density $g_2(\Delta \mu,z |\,\bm{\Theta}_{\text{SN}})$ in Eq.~\eqref{eq:g2}. To train the normalising flow $q^{\bm{\phi}_2}_2$, with trainable parameters $\bm{\phi}_2$, to approximate $g_2$, we define the objective
\begin{equation}
   \min_{\bm{\phi}_2} \; \mathbb{E}_{(\Delta \mu_t,z_t,\bm{\Theta}^t_{\text{SN}})\sim P_{\text{train}}(\Delta \mu,z,\bm{\Theta}_{\text{SN}})}\Big[ -\log q^{\bm{\phi}_2}_2\big(\Delta \mu_t,z_t \big|\ \bm{\Theta}^t_{\text{SN}}\big)\Big].
\end{equation}
We use the same MAF architecture and mini-batch optimisation procedure as used for $q^{\bm{\phi}_1}_1$, described in Section~\ref{sec:q1_train}.

\subsection{Implementation}
\label{sec:implementation}
In this subsection, we describe the technical implementation of the \textit{FlowSN2} framework. To improve training stability when training $q_1^{\bm{\phi}_1}$, we firstly recentre observable dimensions so that the flow effectively learns residual distributions. Specifically, we subtract true redshifts from observed redshifts ($\hat{z}_t \rightarrow \hat{z}_t - z_t$) and subtract population means $\bm{d}^t_0 = (m_0^t, c^{\text{int}}_0, x_0)^\top$ from light curve summaries ($\bm{\hat{d}}_t \rightarrow \bm{\hat{d}}_t - \bm{d}^t_0$). Following this centring step, all dimensions (including contextual variables) are standardised to zero mean and unit variance via $x_t \rightarrow (x_t - \bar{x})/s_x$, using the sample mean $\bar{x}$ and standard deviation $s_x$ of the training set. 

For $q_1^{\bm{\phi}_1}$, the architecture consists of $L=4$ stacked Masked Auto-encoder for Distribution Estimation (MADE) blocks, permuting the conditional feature ordering between successive blocks. Each block comprises two hidden layers with 64 units each, totalling 24,872 learnable parameters. For $q_2^{\bm{\phi}_2}$, we adopt a lighter architecture comprising $L=2$ MADE blocks with two hidden layers of 16 units each, amounting to 1,236 learnable parameters. It takes $\sim$200 kB per survey to store our flow weights. To ensure network predictions satisfy the requirement of at least second-order continuous differentiability for downstream sampling, we employ GELU activation functions throughout. 

Training was performed on 2745 mini-batches of size $B = 8{,}192$ SNe, reserving $10\%$ of the training data as a validation set. We save the model checkpoint yielding the lowest validation loss over 20 epochs for $q_1^{\bm{\phi}_1}$ and 10 epochs for $q_2^{\bm{\phi}_2}$. We implement the normalising flows using \textit{FlowJAX} \citep{flowjax}, built upon \textit{JAX} \citep{jax2018github} to facilitate seamless integration with the \textit{numpyro} \citep{numpyro} hierarchical Bayesian modelling framework. On a single NVIDIA A100 GPU, training both normalising flows requires approximately 10 minutes in total.

To evaluate the numerator of Eq.~\eqref{eq:final_lik}, we perform numerical integration via the trapezium rule across a grid centred on the measured redshift, with a total grid width dictated by its observational uncertainty. The denominator is integrated across the full redshift range. Owing to its independence from individual SN observations, the denominator requires evaluation only once (or twice if population parameters are split by host-galaxy mass).

To sample the posterior, we use Hamiltonian Monte Carlo (HMC; \citealt{duane1987}) via the No-U-Turn Sampler (NUTS; \citealt{betancourt2014, betancourt2017}) implementation in \textit{numpyro} \citep{pyro, numpyro}. Unlike traditional random-walk Markov Chain Monte Carlo algorithms, HMC uses gradient information of the target posterior density to simulate physical particle dynamics, allowing it to explore high-dimensional, highly correlated parameter spaces with high efficiency and faster convergence. 

To supply HMC with the required gradients across both our neural normalising flows and cosmological functions, we rely on automatic differentiation and just-in-time (JIT) compilation provided by \textit{JAX} \citep{jax2018github}. This allows exact gradients to be computed via backpropagation without resorting to computationally expensive finite-difference approximations. Cosmological distance calculations leverage \textit{wcosmo}\footnote{\url{https://github.com/ColmTalbot/wcosmo}} for $w$CDM models and a custom JAX implementation for $w_0 w_a$CDM models.

For a sample of $\approx 2000$ SNe, sampling a single posterior chain takes under 10 minutes on an NVIDIA A100 GPU. Incorporating Bayesian resampling to account for contamination ($M=5$ resamples; Section~\ref{sec:contam_theory}) yields a total inference runtime under 50 minutes. For each resample, we run four parallel chains with 300 warmup steps and 300 production samples, yielding an average effective sample size (ESS) of 619 across global parameters. Convergence is confirmed across all parameters via Gelman-Rubin diagnostics ($\hat{R} < 1.01$; \citealt{gelman1992, vehtari2021}) and zero divergent transitions. For Savage-Dickey density ratio calculations (Eq.~\ref{eq:savdickey}), marginal posterior densities at $\theta_0$ are evaluated by fitting Gaussian kernel estimates directly to the posterior samples. 

The complete codebase for \textit{FlowSN2} is publicly available at: \url{https://github.com/bayesn/flowsn2}.

\section{Experiments}
\label{sec:toy}

In this section, we describe the experiments and results used to validate the \textit{FlowSN2} framework. The experiments use a simplified forward model similar to the one used in \cite{boyd2024} and later \cite{flowsn} that has an analytical likelihood solution. While elements of the simulations are intentionally simplistic, so we can compare with a tractable solution, it provides further confidence when moving to more complex regimes where there are no analytical solutions. This was demonstrated in \cite{flowsn}, where we showed the methodology originally developed on simplified models could naturally be extended to realistic state-of-the-art SNANA survey simulations \citep{kessler2009} of LSST \citep{lsst2019}.

\subsection{Dusty Photometric Redshift Analysis on Pure-Ia Samples}
\label{sec:ia_exp}
Here we describe the experiments used to validate our new framework can correctly account for photometric redshifts, as well as dust modelling, in a SN~Ia cosmology analysis that is under the influence of selection effects. We also take the opportunity to combine mock low redshift spectroscopic and high redshift photometric surveys for the first time using our framework. We introduce classification and contamination into the experiments in Section~\ref{sec:contam_exp}.

\subsubsection{Experimental Setup}
\label{sec:ia_model_forward}

In these experiments we follow the general forward model described in Section~\ref{sec:forward} whilst making additional assumptions to allow for a tractable analytical likelihood solution.

We draw peculiar velocities from a Gaussian distribution such that 
\begin{equation}
    v^s_{\text{pec}} \sim N( 0 , \sigma_{\text{pec}}^2),
\end{equation}
where $\sigma_{\text{pec}}=300\text{ km s}^{-1}$. For simplicity, we also set the observer's velocity relative to the CMB rest-frame projected along the line of sight to the SNe to zero $v_{\odot}^s=0$.

Observed photometric redshifts are modelled to have Gaussian uncertainties such that 
\begin{equation}
    \hat{z}_s^{\text{hel}} \sim N(z_s^{\text{hel}}, \sigma_{z,s}^2),
\end{equation}
where $\sigma_{z,s} = 0.03(1+ z_s^{\text{hel}})$. The methodology itself is flexible to allow for non-Gaussian and multimodal photometric redshift estimates, however, we leave demonstration of this for future work. We assume our spectroscopic redshift sample has negligible uncertainty of  $\sigma_{z,s} = (1\times 10^{-4})(1+ z_s^{\text{hel}})$. 

We sample light curve shapes $x_s$ from a Gaussian population distribution
\begin{equation}
    x_s \sim N(x_0,\sigma_x^2),
\end{equation}
where $x_0$ represents the population mean and population $\sigma_x^2$ variance as in \cite{mandel2017}.

The intrinsic SN colour is sampled from a Gaussian population distribution
\begin{equation}
    c^{\text{int}}_s \sim N(c_0^{\text{int}}+\alpha_c x_s,\sigma_{c,\text{int}}^2),
\end{equation}
where $\alpha_c$ allows for correlations with light curve shape, $c^{\text{int}}$ is the intrinsic colour population mean (when $\alpha_c=0$) and $\sigma_{c,\text{int}}^2$ is the intrinsic colour population variance. 

The extrinsic dust colour excess $E_s$ is drawn from the exponential distribution
\begin{equation}
    E_s \sim P(E|\,\tau) =   \begin{cases}
      \frac{1}{\tau} \exp\Big(-\frac{E_s}{\tau}\Big) & \text{if $E_s\geq $ 0}\\
      0 & \text{if $E_s < $ 0}\end{cases},
\end{equation}
where $\tau$ is the average colour excess.

Host galaxy masses are drawn randomly from
\begin{equation}
\log_{10}(\mathcal{M}_s/M_\odot) \sim U(8, 12).
\end{equation}
In this work, we assume no uncertainty on the observed masses and set the step threshold as $\mathcal{M}_{\text{step}}=10^{10} M_\odot$. Future work using the methodology may model uncertainties on galaxy masses by feeding these into the normalising flow.

The latent apparent magnitude is calculated using the Simple-BayeSN \citep{mandel2017} linear formula as defined in Eq.~\eqref{eq:simple_bayesn}. Instead of simulating the light curve data, we simulate summary statistics $\bm{\hat{d}}_s$ directly from latent variables $\bm{d}_s$ using a multivariate Gaussian distribution such that 
\begin{equation}
\label{eq:salt_like}
\bm{\hat{d}}_s \sim P(\bm{\hat{d}}|\ \bm{d}_s,\hat{z}_s^{\text{hel}}) = N(\hat{\bm{d}} |\, \bm{d}_s,\bm{\Sigma}_s). 
\end{equation}
We draw the summary statistic fit covariance matrices $\bm{\Sigma}_s$ using the relations in Appendix~\ref{app:cov_samp}, approximately based on those in the Pantheon+ dataset \citep{scolnic2022}.

We define two selection probabilities in terms of the observed summary parameters. The first selection probability that is a function of the observed SN light curve summary parameters is given by 
\begin{equation}
\label{eq:cdf_app1}
P(I_s = 1 |\, \bm{\hat{d}}_s) = \Phi \Bigg(\frac{m_{\text{cut}} -(\hat{m}_s + a_{\text{cut},m} \, \hat{x}_s + b_{\text{cut},m} \, \hat{c}_s)}{\sigma_{\text{cut},m}}\Bigg),
\end{equation} 
as used in the \citep{rubin2015} model and the second selection probability allows the observed redshifts to influence selection
\begin{equation}
\label{eq:cdf_app2}
P(I_s = 1 |\, \hat{z}_s) = \Phi \Bigg(\frac{z_{\text{cut}} -\hat{z}_s}{\sigma_{\text{cut},z}}\Bigg),
\end{equation}
where $\Phi(\cdot)$ is a Gaussian cumulative distribution function. In our simulations, the two surveys have different selection probability parameters.

For our high redshift photometric survey we simulate $5,000$ SNe~Ia between $0< z_s < 1.2$. The high redshift survey, chosen to have comparable settings to the Dark Energy Survey \citep{des5yr}, has $m_{\text{cut}}=24.5$ mag, $\sigma_{\text{cut},m}=0.5$  mag, $a_{\text{cut},m}=-0.1$, $b_{\text{cut},m}=1$, $z_{\text{cut}}=0.9$ and $\sigma_{\text{cut},z}=0.1$, yielding $N^{\text{obs}}\approx2,000$ SNe~Ia post-selection. Future SN cosmological analyses will still rely on low redshift spectroscopic samples from surveys, such as TiDES \citep{tides}, to anchor cosmological constraints. We simulate an additional $5,000$ low redshift spectroscopic SNe~Ia with negligible redshift uncertainties between $0< z_s < 0.3$. The low redshift survey has $m_{\text{cut}}=20.0$ mag, $\sigma_{\text{cut},m}=0.25$  mag, $a_{\text{cut},m}=-0.1$, $b_{\text{cut},m}=1$, $z_{\text{cut}}=0.1$ and $\sigma_{\text{cut},z}=0.25$, yielding $N^{\text{obs}}\approx300$ SNe~Ia post-selection.
The distribution of our selected SNe from the two simulated surveys is visualised in Figure~\ref{fig:surveys}.

The true SN~Ia population parameters $\bm{\Theta}_{\text{SN}}$ used in our simulations are in Table~\ref{tab:c1-c3_ia_only} and Table~\ref{tab:c4-c5_ia_only}. The priors on our global parameters used in inference are in Table~\ref{tab:priors}. We simulate from five different cosmologies $\bm{C}_1$ to $\bm{C}_5$, whilst also varying $\xi_{\text{rate}}$. For $\bm{C}_1$ to $\bm{C}_3$ we perform inference  with and without a CMB prior. For $\bm{C}_4$ and $\bm{C}_5$ we perform our $w_0-w_a$ inference with baryon acoustic oscillation (BAO) and CMB priors. Both the BAO and CMB priors used in our experiments are defined in Appendix~\ref{app:cmb_bao_prior}. The analytical likelihood solution for this forward model used for comparison is derived in Appendix~\ref{app:analytical}. 
\begin{figure}
\centering
\includegraphics[width=1.0\linewidth]{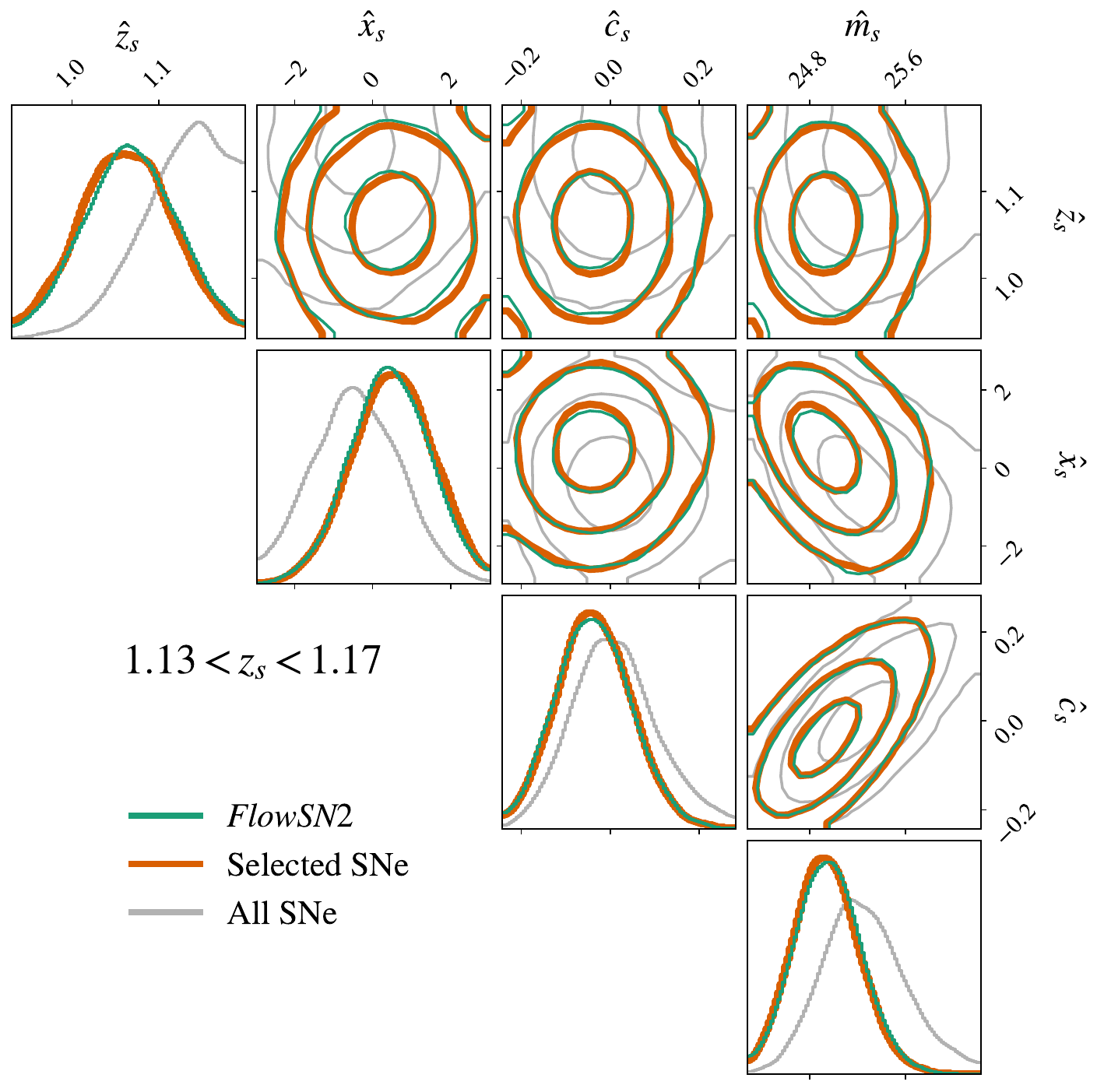}
    \vspace{-0.4cm}    \caption{Samples from our trained four-dimensional normalising flow $q_1^{\bm{\phi}_1}$ corresponding to the photometric redshift survey. We condition on the highest redshift SNe to highlight the maximum impact of selection effects. The distribution of unseen validation data demonstrates that the normalising flow has captured the influence of selection, relative to the full population shown in grey.}
    \label{fig:learnt_q1} 
\end{figure}
\subsubsection{Training}
\label{sec:exp_train}
To train our normalising flows, we use the same procedures described in Section~\ref{sec:training} and the implementations described in  Section~\ref{sec:implementation}. We train four normalising flows in total, including a $q_1^{\bm{\phi}_1}$ and $q_2^{\bm{\phi}_2}$ for each of the low redshift and high redshift surveys. This involves simulating $N^{\text{obs}}\approx 25$ million selected low redshift spectroscopic SNe~Ia and $N^{\text{obs}}\approx 25$ million selected high redshift photometric SNe~Ia using the forward model defined in Section~\ref{sec:ia_model_forward}. We do not train on contaminants. All the SNe were simulated with the same fiducial cosmology parameters $\bm{C}_1=(w_0=-1.0,w_a=0,\Omega_{m0}=0.315,\Omega_{\Lambda 0}=0.685,h=0.7)$, same fiducial rate $\xi^{\text{fid}}_{\text{rate}}=1.5$ and same fiducial absolute intrinsic magnitude zeropoint $M^{\text{int}}_{0,\text{fid}}=  -19.38$. We use a Gaussian distance smearing standard deviation of $\sigma_{\mu}=0.25$. The population parameter training distributions $\bm{\Theta}^t_{\text{SN}}\sim P_{\text{train}}(\bm{\Theta}_{\text{SN}})$ are shown in Table~\ref{tab:priors}.
\begin{figure*}
\centering
	\includegraphics[width=0.95\textwidth]{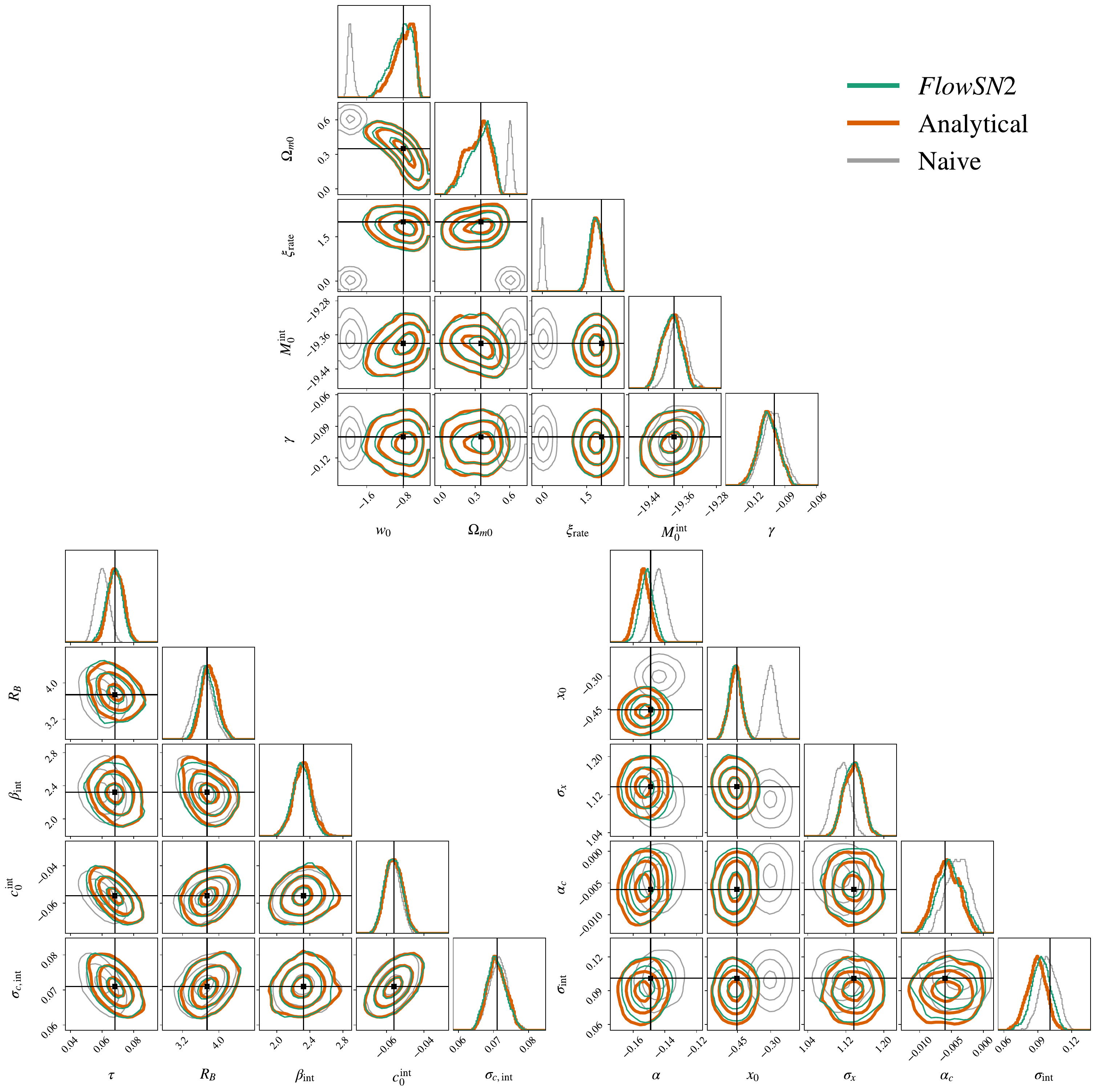}
    \vspace{0cm}    \caption{\textit{FlowSN2} posteriors on all our cosmological and SN population parameters. These inferences are using a pure-Ia sample with photometric redshifts under the influence of selection. We also show posteriors that use the analytical likelihood solution (defined in Appendix~\ref{app:analytical}) for the simplified model and the naive likelihood that does not take selection effects into account. This simulation is from cosmology $\bm{C}_2$ showing \textit{FlowSN2} is still effective when inferring cosmological parameters away from the fiducial $\Lambda$CDM cosmology $\bm{C}_{\text{fid}}=\bm{C}_1$ used in training the normalising flows. The true value for each parameter is indicated by the solid black line. } 
    \label{fig:ia_corners}
\end{figure*}
\subsubsection{SN~Ia Only Results}

Prior to inference, we assess the quality of our learnt $q_1^{\bm{\phi}_1}$ likelihood approximation by comparing its samples against held-out data distributions  as demonstrated in Figure~\ref{fig:learnt_q1}. We find that the four-dimensional normalising flow closely approximates the selected SN distribution, particularly at high redshifts where the selected and underlying population distributions diverge most strongly. Similarly, inspecting the $q_2^{\bm{\phi}_2}$ distributions (illustrated in Figure~\ref{fig:graphic}) reveals a decrease in selected SNe at high redshift on the latent Hubble diagram. Evaluating samples from the normalising flows against held-out data thus serves as a reliable validation step before proceeding to inference. The same four normalising flows (two for each survey) are utilised across both the pure-Ia and contamination experiments, as well as in further investigations into scaling and model selection.

We perform inference on the combined $N^{\text{obs}}\approx 2,300$ pure-Ia sample drawn
from the two surveys described in Section~\ref{sec:exp_train}. Our
\textit{FlowSN2} posteriors are compared against those obtained with the analytical likelihood
solution derived in Appendix~\ref{app:analytical}. Figure~\ref{fig:ia_corners} shows near-exact agreement between the \textit{FlowSN2} posterior and the analytical solution posterior across our cosmology parameters and 13 SN population parameters. The illustrated naive posteriors, that do not take selection effects into account, are strongly biased, particularly for the cosmology and rate parameters. We find the Simple-BayeSN \citep{mandel2017} intrinsic SN parameters and extrinsic dust parameters are well-constrained and posteriors are in agreement with the analytical solution.

The test simulations used in Figure~\ref{fig:ia_corners} are generated from a different set of cosmological parameters $\bm{C}_2$ and SN rates than the fiducial $\bm{C}_1$ cosmology and rates used in training. This demonstrates a key advantage of the \textit{FlowSN2} framework, as training requires only a single assumed cosmology and rate. By contrast, other simulation-based inference (SBI) approaches to this problem require a hypercube spanning the full parameter space \citep{karchev2023,karchev2025}.

To further investigate potential biases induced by our fiducial cosmology, we perform inference on simulations generated across five sets of cosmological parameters and rates. For each cosmology, we repeat the inference 25 times with different random seeds. The average posterior estimates and standard errors are reported in Table~\ref{tab:c1-c3_ia_only} and Table~\ref{tab:c4-c5_ia_only}. Cosmology $\bm{C}_1$ represents the fiducial $\Lambda$CDM cosmology used in training, whereas $\bm{C}_2$ and $\bm{C}_3$ are simulated under a $w$CDM cosmology, lying below and above the fiducial parameters on the Hubble diagram respectively. For these three cosmological parameter sets, we find statistically insignificant biases below $2\sigma$, falling within $0.2\sigma$ of the analytical solution. When combined with the CMB prior (defined in Appendix~\ref{app:cmb_prior}), our inferences remain robust, displaying negligible bias. By contrast, our previous analysis \citep{flowsn} showed that BBC inferences \citep{kessler2017} exhibited a $\sim 4\sigma$ bias toward the fiducial cosmology when combined with a CMB prior.

To demonstrate \textit{FlowSN2}'s ability to infer parameters for alternative cosmological models without retraining, we infer $w_0w_a$CDM parameters in Table~\ref{tab:c4-c5_ia_only}. Parameters $\bm{C}_4$ are equal to our fiducial parameters, while $\bm{C}_5$ are closer to the parameters inferred in the recent DESI results \citep{desi2025}. For additional constraining power we perform these inferences with a CMB and BAO prior as described in Appendix~\ref{app:cmb_bao_prior}. We find that again the \textit{FlowSN2} is able to infer the true cosmology under selection effects, with biases less than $0.1\sigma$ away from the analytical likelihood solution. 

As reported in Table~\ref{tab:c1-c3_ia_only} and Table~\ref{tab:c4-c5_ia_only}, we find that the SN population parameters, including SN rates, are well-constrained, with the majority exhibiting statistically insignificant biases. The \textit{FlowSN2} framework also successfully recovers host-galaxy intrinsic mass steps to within $0.1\sigma$ of the analytical solution. The parameters exhibiting the largest relative deviations from the analytical solution are $\sigma_{\text{int}}$ and $\tau$, indicating that noisy variance parameters are the most challenging to learn. Nevertheless, the absolute magnitudes of these biases remain well within acceptable tolerances for the scientific objectives of next-generation surveys. Furthermore, when additional typing uncertainty is introduced in the subsequent experiment, these biases become statistically insignificant. We assess the empirical frequentist coverage properties of our Bayesian credible regions for a pure-Ia sample, as shown in the right column of Figure~\ref{fig:calibration_grid}. These checks confirm that inferences for all cosmological parameters and the thirteen SN parameters fall within the 95\% Monte Carlo confidence bands.

\subsection{Inference with Contamination  }
\label{sec:contam_sec}
\begin{figure}
\centering
\includegraphics[width=1.0\linewidth]{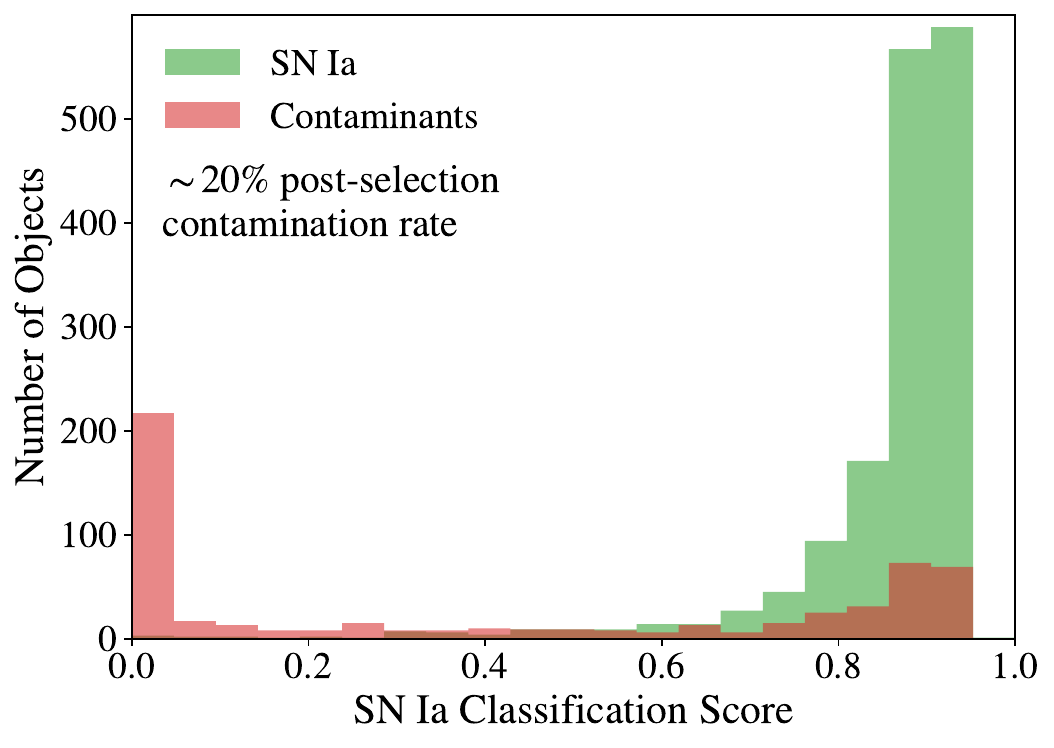}
    \vspace{-0.4cm}    \caption{Mock classification scores for SNe~Ia and contaminants in the high-redshift photometric sample used in our experiments. These are obtained using the known conditional class probabilities from the simulations. Further experimental details are in Section~\ref{sec:contam_exp}. }
    \label{fig:class_hist} 
\end{figure}

\subsubsection{Experimental Setup}
\label{sec:contam_exp}

In this experiment we introduce contamination from non-SN~Ia objects into our simulated
test sample. The SNe~Ia are simulated following exactly the procedure of
Section~\ref{sec:ia_model_forward}. Contaminants are introduced only into the high-redshift
photometric survey, since we model the low-redshift spectroscopic survey as
spectroscopically confirmed SNe~Ia.

We simulate the contaminants with the same generative forward model as the SNe~Ia with the same underlying cosmological parameters, but
give them different population parameters $\bm{\Theta}^{\text{non-Ia}}_{\text{SN}}$, chosen
to reflect the qualitative properties of core-collapse SNe \citep{vincenzi2021}.
Specifically, we assume the contaminants have no standardisation properties, with
$\alpha=0$, $\beta_{\text{int}}=0$ and $\alpha_c=0$, and a much larger intrinsic scatter
of $\sigma_{\text{int}}=1.2$ mag. We assume no correlation between host mass and object
brightness, $\gamma=0$. The contaminants are fainter than the SNe~Ia by $0.4$ mag, giving
$M^{\text{int}}_0=-18.98$ mag. It is this offset that particularly biases an analysis; without it, contamination would only inflate the scatter. Finally, we allow the contaminants to follow an alternative power-law rate,
$\xi^{\text{non-Ia}}_{\text{rate}}=3.0$, so that the two populations have different
redshift distributions.

Classification uses only the observed summary statistics $\hat{\bm{d}}_s$ and redshifts
$\hat{z}_s$. To demonstrate the method on a strongly contaminated sample, we deliberately give the
contaminants the same light curve shape $(x_0,\sigma_x)$, intrinsic colour
$(c^{\text{int}}_0,\sigma_{c,\text{int}})$ and extrinsic dust $(\tau,R_B)$ population
distributions as the SNe~Ia, so that the classes overlap in the observed summaries and
cannot be separated on colour or light curve shape alone. We note that in reality, we do not expect such a strong population overlap, as it is possible to separate contaminants using apparent colour-magnitude distributions \citep{vincenzi2021}.  

Measurement errors and the survey selection (Section~\ref{sec:ia_model_forward}) are
applied agnostically to the type of the object, so contaminants pass through exactly
the same procedure as the SNe~Ia. Of the $\approx 2500$ contaminants simulated in the
photometric survey, $\approx 500$ pass the selection criteria. This gives $\approx 2800$
selected objects in total, of which roughly $18\%$ are contaminants. Figure~\ref{fig:class_hist} shows the distribution of mock classification scores for the selected objects, separated by their true class.

As described in Section~\ref{sec:contam_theory}, a sufficient condition for reliable inference is that, among selected SNe with similar observed light curve summaries and redshifts, the average classification score matches the fraction that are actually SNe~Ia. In a real-data analysis, these scores may be obtained from a classifier trained on real data representative of the target population. In our simplified simulations, we instead compute the mock scores $\mathcal{C}^{\text{Ia}}_s$ using the exact conditional class probabilities from the known generative model, such that
\begin{multline}
  \mathcal{C}^{\text{Ia}}_s=P(\mathcal{Q}^{\text{Ia}}_s = 1 |\, \hat{\bm{d}}_s, \hat{z}_s, I_s=1, \bm{\Theta}_{\text{true}})  \\ = \frac{\bar{\pi}_{\text{Ia}}P\big(\bm{\hat{d}}_s, \hat{z}_s \big|\,I_s=1,\bm{\Theta}_{\text{true}}^{\text{Ia}} \big)}{\bar{\pi}_{\text{Ia}}P\big(\bm{\hat{d}}_s, \hat{z}_s \big|\,I_s=1,\bm{\Theta}_{\text{true}}^{\text{Ia}}\big) + (1-\bar{\pi}_{\text{Ia}})P\big(\bm{\hat{d}}_s, \hat{z}_s \big|\,I_s=1,\bm{\Theta}_{\text{true}}^{\text{non-Ia}}\big)} 
\end{multline}
where we condition likelihoods on the true SN~Ia global parameters $\bm{\Theta}^{\text{Ia}}_{\text{true}} $ and the true non-Ia global parameters $\bm{\Theta}^{\text{non-Ia}}_{\text{true}}$. We derive $P\big(\bm{\hat{d}}_s, \hat{z}_s \big|\,I_s=1,\bm{\Theta} \big)$ exactly for these simulations in Appendix~\ref{app:analytical}. The SN~Ia proportion post-selection $\bar{\pi}_{\text{Ia}}$ can be calculated empirically given we know the true labels in the simulation. In practice, the photometric classification scores will come from classifiers trained on spectroscopically confirmed SNe~Ia or simulations. The modular Bayesian resampling approach for handling contamination, along with alternative mixture model approaches \citep{beams}, will require classifier training data that is representative, or at least provide support, over the high redshift photometric target population. This remains an open challenge for the field, where the solution will need to be tested in conjunction with the \textit{FlowSN2} framework.  

Going forward in our experiments, we use these exact conditional classification probabilities as our mock classifier scores, following the methodology described in Section~\ref{sec:contam_theory}. We resample the data $M=5$ times based on these classification scores. Since the modular Bayesian approach of \cite{autenrieth2026} does not require an explicit likelihood model for the non-Ia population, we use the same normalising flows trained on the pure-Ia simulations from the pure-Ia experiment, described in Section~\ref{sec:exp_train}. We use the same SN~Ia population parameters $\bm{\Theta}^{\text{Ia}}_{\text{SN}}$ as in the previous experiment, detailed in Table~\ref{tab:c1-c3_contam} and Table~\ref{tab:c4-c5_contam}. Our priors on the population parameters in inference, detailed in Table~\ref{tab:priors}, are also the same. We repeat experiments again using five sets of cosmological parameters $\bm{C}_1$ to  $\bm{C}_5$.
\begin{table*}
\centering
\caption{Average of posterior median estimates over 25 repeats and the standard error on this average for the full photometric SN Ia cosmology experiments with non-Ia contamination. We include results using our default modular Bayesian approach of resampling of the data ($M=5$) to handle contamination, as well as results using the pure SN Ia sample with the true labels for comparison. Both use the \textit{FlowSN2} normalising flows in their inference. We also combine SNe with a CMB prior for additional constraining as detailed in Appendix~\ref{app:cmb_bao_prior}. Note uncertainties are the standard error on the average so are a factor of five smaller than those in individual experiments. The $\sigma$ values are calculated using unrounded averages and standard errors.}
\label{tab:c1-c3_contam}
\begin{tabular}{lccccc}
\hline
\multicolumn{6}{c}{Cosmology $\bm{C}_1$} \\
\hline
Test &  & \multicolumn{2}{c}{SN Only} & \multicolumn{2}{c}{SN + CMB Prior}\\
\hline
Global Parameter & Truth  & Bayesian Resampling & Pure SN Ia Sample & Bayesian Resampling & Pure SN Ia Sample \\
\hline
$w_0$ & -1 & -1.00 $\pm$ 0.04 (-0.1$\sigma$) & -1.01 $\pm$ 0.04 (-0.2$\sigma$) & -0.999 $\pm$ 0.003 (0.2$\sigma$) & -1.000 $\pm$ 0.003 (-0.1$\sigma$)\\
$\Omega_{m0}$ & 0.315 & 0.303 $\pm$ 0.011 (-1.1$\sigma$) & 0.297 $\pm$ 0.013 (-1.4$\sigma$) & 0.3148 $\pm$ 0.0006 (-0.3$\sigma$) & 0.3150 $\pm$ 0.0006 (0.0$\sigma$)\\
$M_0^{\mathrm{int}}$ & -19.38 & -19.381 $\pm$ 0.004 (-0.2$\sigma$) & -19.382 $\pm$ 0.004 (-0.6$\sigma$) & -19.377 $\pm$ 0.003 (1.1$\sigma$) & -19.376 $\pm$ 0.003 (1.2$\sigma$)\\
$\alpha$ & -0.149 & -0.1474 $\pm$ 0.0009 (1.7$\sigma$) & -0.1469 $\pm$ 0.0009 (2.4$\sigma$) & -0.1476 $\pm$ 0.0009 (1.6$\sigma$) & -0.1467 $\pm$ 0.0009 (2.4$\sigma$)\\
$\beta_\mathrm{int}$ & 2.322 & 2.32 $\pm$ 0.02 (-0.1$\sigma$) & 2.32 $\pm$ 0.02 (-0.3$\sigma$) & 2.32 $\pm$ 0.02 (0.1$\sigma$) & 2.32 $\pm$ 0.02 (-0.2$\sigma$)\\
$\alpha_c$ & -0.006 & -0.0058 $\pm$ 0.0004 (0.5$\sigma$) & -0.0059 $\pm$ 0.0004 (0.2$\sigma$) & -0.0058 $\pm$ 0.0004 (0.6$\sigma$) & -0.0060 $\pm$ 0.0004 (0.0$\sigma$)\\
$R_B$ & 3.74 & 3.70 $\pm$ 0.03 (-1.2$\sigma$) & 3.70 $\pm$ 0.03 (-1.3$\sigma$) & 3.70 $\pm$ 0.03 (-1.3$\sigma$) & 3.70 $\pm$ 0.03 (-1.4$\sigma$)\\
$c^{\mathrm{int}}_0$ & -0.056 & -0.0582 $\pm$ 0.0008 (-2.9$\sigma$) & -0.0586 $\pm$ 0.0008 (-3.3$\sigma$) & -0.0580 $\pm$ 0.0008 (-2.6$\sigma$) & -0.0586 $\pm$ 0.0008 (-3.3$\sigma$)\\
$\sigma_\mathrm{int}$ & 0.101 & 0.1046 $\pm$ 0.0014 (2.6$\sigma$) & 0.1043 $\pm$ 0.0015 (2.2$\sigma$) & 0.1050 $\pm$ 0.0014 (2.9$\sigma$) & 0.1046 $\pm$ 0.0015 (2.4$\sigma$)\\
$\sigma_{c,\mathrm{int}}$ & 0.071 & 0.0707 $\pm$ 0.0004 (-0.6$\sigma$) & 0.0704 $\pm$ 0.0004 (-1.5$\sigma$) & 0.0708 $\pm$ 0.0004 (-0.5$\sigma$) & 0.0704 $\pm$ 0.0004 (-1.5$\sigma$)\\
$\sigma_x$ & 1.137 & 1.139 $\pm$ 0.004 (0.5$\sigma$) & 1.140 $\pm$ 0.004 (0.6$\sigma$) & 1.139 $\pm$ 0.004 (0.5$\sigma$) & 1.140 $\pm$ 0.004 (0.6$\sigma$)\\
$\tau$ & 0.068 & 0.0688 $\pm$ 0.0009 (0.9$\sigma$) & 0.0692 $\pm$ 0.0009 (1.3$\sigma$) & 0.0686 $\pm$ 0.0009 (0.6$\sigma$) & 0.0691 $\pm$ 0.0009 (1.2$\sigma$)\\
$x_0$ & -0.452 & -0.450 $\pm$ 0.004 (0.4$\sigma$) & -0.448 $\pm$ 0.004 (0.9$\sigma$) & -0.450 $\pm$ 0.004 (0.5$\sigma$) & -0.447 $\pm$ 0.004 (1.1$\sigma$)\\
$\xi_\mathrm{rate}$ & 1.5 & 1.53 $\pm$ 0.04 (0.8$\sigma$) & 1.54 $\pm$ 0.04 (1.0$\sigma$) & 1.54 $\pm$ 0.04 (1.0$\sigma$) & 1.55 $\pm$ 0.04 (1.4$\sigma$)\\
$\gamma$ & -0.1 & -0.0979 $\pm$ 0.0017 (1.3$\sigma$) & -0.0991 $\pm$ 0.0019 (0.5$\sigma$) & -0.0974 $\pm$ 0.0017 (1.6$\sigma$) & -0.0989 $\pm$ 0.0019 (0.6$\sigma$)\\
\hline
\multicolumn{6}{c}{Cosmology $\bm{C}_2$} \\
\hline
Test &  & \multicolumn{2}{c}{SN Only} & \multicolumn{2}{c}{SN + CMB Prior}\\
\hline
Global Parameter & Truth  & Bayesian Resampling & Pure SN Ia Sample & Bayesian Resampling & Pure SN Ia Sample \\
\hline
$w_0$ & -0.8 & -0.81 $\pm$ 0.04 (-0.2$\sigma$) & -0.84 $\pm$ 0.05 (-0.8$\sigma$) & -0.804 $\pm$ 0.002 (-1.6$\sigma$) & -0.804 $\pm$ 0.002 (-1.8$\sigma$)\\
$\Omega_{m0}$ & 0.35 & 0.329 $\pm$ 0.016 (-1.3$\sigma$) & 0.332 $\pm$ 0.018 (-0.9$\sigma$) & 0.3509 $\pm$ 0.0007 (1.3$\sigma$) & 0.3511 $\pm$ 0.0007 (1.5$\sigma$)\\
$M_0^{\mathrm{int}}$ & -19.38 & -19.380 $\pm$ 0.003 (-0.1$\sigma$) & -19.384 $\pm$ 0.003 (-1.1$\sigma$) & -19.376 $\pm$ 0.003 (1.3$\sigma$) & -19.376 $\pm$ 0.003 (1.2$\sigma$)\\
$\alpha$ & -0.149 & -0.1472 $\pm$ 0.0009 (1.9$\sigma$) & -0.1468 $\pm$ 0.0008 (2.6$\sigma$) & -0.1472 $\pm$ 0.0009 (1.9$\sigma$) & -0.1467 $\pm$ 0.0009 (2.6$\sigma$)\\
$\beta_\mathrm{int}$ & 2.322 & 2.328 $\pm$ 0.020 (0.3$\sigma$) & 2.32 $\pm$ 0.02 (-0.3$\sigma$) & 2.33 $\pm$ 0.02 (0.5$\sigma$) & 2.32 $\pm$ 0.02 (-0.1$\sigma$)\\
$\alpha_c$ & -0.006 & -0.0057 $\pm$ 0.0004 (0.7$\sigma$) & -0.0058 $\pm$ 0.0004 (0.4$\sigma$) & -0.0058 $\pm$ 0.0004 (0.6$\sigma$) & -0.0059 $\pm$ 0.0004 (0.2$\sigma$)\\
$R_B$ & 3.74 & 3.69 $\pm$ 0.03 (-1.8$\sigma$) & 3.69 $\pm$ 0.03 (-1.7$\sigma$) & 3.69 $\pm$ 0.03 (-1.8$\sigma$) & 3.68 $\pm$ 0.03 (-1.8$\sigma$)\\
$c^{\mathrm{int}}_0$ & -0.056 & -0.0581 $\pm$ 0.0008 (-2.6$\sigma$) & -0.0587 $\pm$ 0.0008 (-3.3$\sigma$) & -0.0580 $\pm$ 0.0008 (-2.6$\sigma$) & -0.0588 $\pm$ 0.0008 (-3.4$\sigma$)\\
$\sigma_\mathrm{int}$ & 0.101 & 0.1058 $\pm$ 0.0014 (3.4$\sigma$) & 0.1050 $\pm$ 0.0015 (2.6$\sigma$) & 0.1059 $\pm$ 0.0014 (3.5$\sigma$) & 0.1051 $\pm$ 0.0015 (2.7$\sigma$)\\
$\sigma_{c,\mathrm{int}}$ & 0.071 & 0.0707 $\pm$ 0.0005 (-0.6$\sigma$) & 0.0703 $\pm$ 0.0005 (-1.6$\sigma$) & 0.0707 $\pm$ 0.0005 (-0.5$\sigma$) & 0.0703 $\pm$ 0.0005 (-1.6$\sigma$)\\
$\sigma_x$ & 1.137 & 1.141 $\pm$ 0.004 (0.9$\sigma$) & 1.140 $\pm$ 0.004 (0.8$\sigma$) & 1.141 $\pm$ 0.004 (0.9$\sigma$) & 1.140 $\pm$ 0.004 (0.8$\sigma$)\\
$\tau$ & 0.068 & 0.0687 $\pm$ 0.0009 (0.8$\sigma$) & 0.0694 $\pm$ 0.0009 (1.5$\sigma$) & 0.0687 $\pm$ 0.0009 (0.7$\sigma$) & 0.0694 $\pm$ 0.0009 (1.5$\sigma$)\\
$x_0$ & -0.452 & -0.450 $\pm$ 0.004 (0.4$\sigma$) & -0.447 $\pm$ 0.005 (1.0$\sigma$) & -0.450 $\pm$ 0.004 (0.5$\sigma$) & -0.447 $\pm$ 0.005 (1.1$\sigma$)\\
$\xi_\mathrm{rate}$ & 2 & 2.03 $\pm$ 0.04 (0.8$\sigma$) & 2.05 $\pm$ 0.04 (1.2$\sigma$) & 2.04 $\pm$ 0.04 (1.1$\sigma$) & 2.06 $\pm$ 0.04 (1.4$\sigma$)\\
$\gamma$ & -0.1 & -0.0987 $\pm$ 0.0016 (0.8$\sigma$) & -0.0997 $\pm$ 0.0019 (0.2$\sigma$) & -0.0987 $\pm$ 0.0015 (0.8$\sigma$) & -0.0996 $\pm$ 0.0018 (0.2$\sigma$)\\
\hline
\multicolumn{6}{c}{Cosmology $\bm{C}_3$} \\
\hline
Test &  & \multicolumn{2}{c}{SN Only} & \multicolumn{2}{c}{SN + CMB Prior}\\
\hline
Global Parameter & Truth  & Bayesian Resampling & Pure SN Ia Sample & Bayesian Resampling & Pure SN Ia Sample \\
\hline
$w_0$ & -1.2 & -1.19 $\pm$ 0.03 (0.3$\sigma$) & -1.20 $\pm$ 0.03 (0.1$\sigma$) & -1.205 $\pm$ 0.003 (-1.5$\sigma$) & -1.208 $\pm$ 0.003 (-2.3$\sigma$)\\
$\Omega_{m0}$ & 0.25 & 0.242 $\pm$ 0.006 (-1.2$\sigma$) & 0.238 $\pm$ 0.008 (-1.5$\sigma$) & 0.2504 $\pm$ 0.0004 (1.2$\sigma$) & 0.2507 $\pm$ 0.0004 (2.0$\sigma$)\\
$M_0^{\mathrm{int}}$ & -19.38 & -19.379 $\pm$ 0.004 (0.3$\sigma$) & -19.380 $\pm$ 0.004 (-0.1$\sigma$) & -19.378 $\pm$ 0.003 (0.6$\sigma$) & -19.377 $\pm$ 0.003 (0.9$\sigma$)\\
$\alpha$ & -0.149 & -0.1471 $\pm$ 0.0010 (2.0$\sigma$) & -0.1468 $\pm$ 0.0010 (2.3$\sigma$) & -0.1470 $\pm$ 0.0010 (2.0$\sigma$) & -0.1466 $\pm$ 0.0010 (2.4$\sigma$)\\
$\beta_\mathrm{int}$ & 2.322 & 2.32 $\pm$ 0.02 (-0.2$\sigma$) & 2.31 $\pm$ 0.02 (-0.5$\sigma$) & 2.32 $\pm$ 0.02 (-0.1$\sigma$) & 2.31 $\pm$ 0.02 (-0.4$\sigma$)\\
$\alpha_c$ & -0.006 & -0.0059 $\pm$ 0.0004 (0.3$\sigma$) & -0.0059 $\pm$ 0.0004 (0.1$\sigma$) & -0.0059 $\pm$ 0.0004 (0.3$\sigma$) & -0.0060 $\pm$ 0.0004 (-0.0$\sigma$)\\
$R_B$ & 3.74 & 3.71 $\pm$ 0.03 (-0.9$\sigma$) & 3.70 $\pm$ 0.04 (-1.2$\sigma$) & 3.71 $\pm$ 0.03 (-0.9$\sigma$) & 3.70 $\pm$ 0.04 (-1.2$\sigma$)\\
$c^{\mathrm{int}}_0$ & -0.056 & -0.0574 $\pm$ 0.0008 (-1.7$\sigma$) & -0.0579 $\pm$ 0.0008 (-2.3$\sigma$) & -0.0574 $\pm$ 0.0008 (-1.7$\sigma$) & -0.0580 $\pm$ 0.0008 (-2.3$\sigma$)\\
$\sigma_\mathrm{int}$ & 0.101 & 0.1047 $\pm$ 0.0014 (2.6$\sigma$) & 0.1047 $\pm$ 0.0017 (2.2$\sigma$) & 0.1048 $\pm$ 0.0014 (2.6$\sigma$) & 0.1048 $\pm$ 0.0016 (2.3$\sigma$)\\
$\sigma_{c,\mathrm{int}}$ & 0.071 & 0.0709 $\pm$ 0.0004 (-0.1$\sigma$) & 0.0706 $\pm$ 0.0004 (-0.8$\sigma$) & 0.0710 $\pm$ 0.0004 (-0.1$\sigma$) & 0.0706 $\pm$ 0.0004 (-0.9$\sigma$)\\
$\sigma_x$ & 1.137 & 1.141 $\pm$ 0.004 (1.1$\sigma$) & 1.142 $\pm$ 0.004 (1.2$\sigma$) & 1.142 $\pm$ 0.004 (1.1$\sigma$) & 1.142 $\pm$ 0.004 (1.2$\sigma$)\\
$\tau$ & 0.068 & 0.0682 $\pm$ 0.0009 (0.2$\sigma$) & 0.0686 $\pm$ 0.0010 (0.6$\sigma$) & 0.0682 $\pm$ 0.0009 (0.2$\sigma$) & 0.0686 $\pm$ 0.0010 (0.6$\sigma$)\\
$x_0$ & -0.452 & -0.450 $\pm$ 0.004 (0.5$\sigma$) & -0.447 $\pm$ 0.005 (1.1$\sigma$) & -0.450 $\pm$ 0.004 (0.5$\sigma$) & -0.446 $\pm$ 0.005 (1.2$\sigma$)\\
$\xi_\mathrm{rate}$ & 1 & 1.05 $\pm$ 0.04 (1.1$\sigma$) & 1.05 $\pm$ 0.05 (1.2$\sigma$) & 1.06 $\pm$ 0.04 (1.3$\sigma$) & 1.07 $\pm$ 0.05 (1.6$\sigma$)\\
$\gamma$ & -0.1 & -0.0976 $\pm$ 0.0018 (1.4$\sigma$) & -0.0983 $\pm$ 0.0019 (0.9$\sigma$) & -0.0976 $\pm$ 0.0017 (1.4$\sigma$) & -0.0982 $\pm$ 0.0019 (0.9$\sigma$)\\
\hline
\end{tabular}
\end{table*}

\begin{figure*}
\centering
	\includegraphics[width=0.95\textwidth]{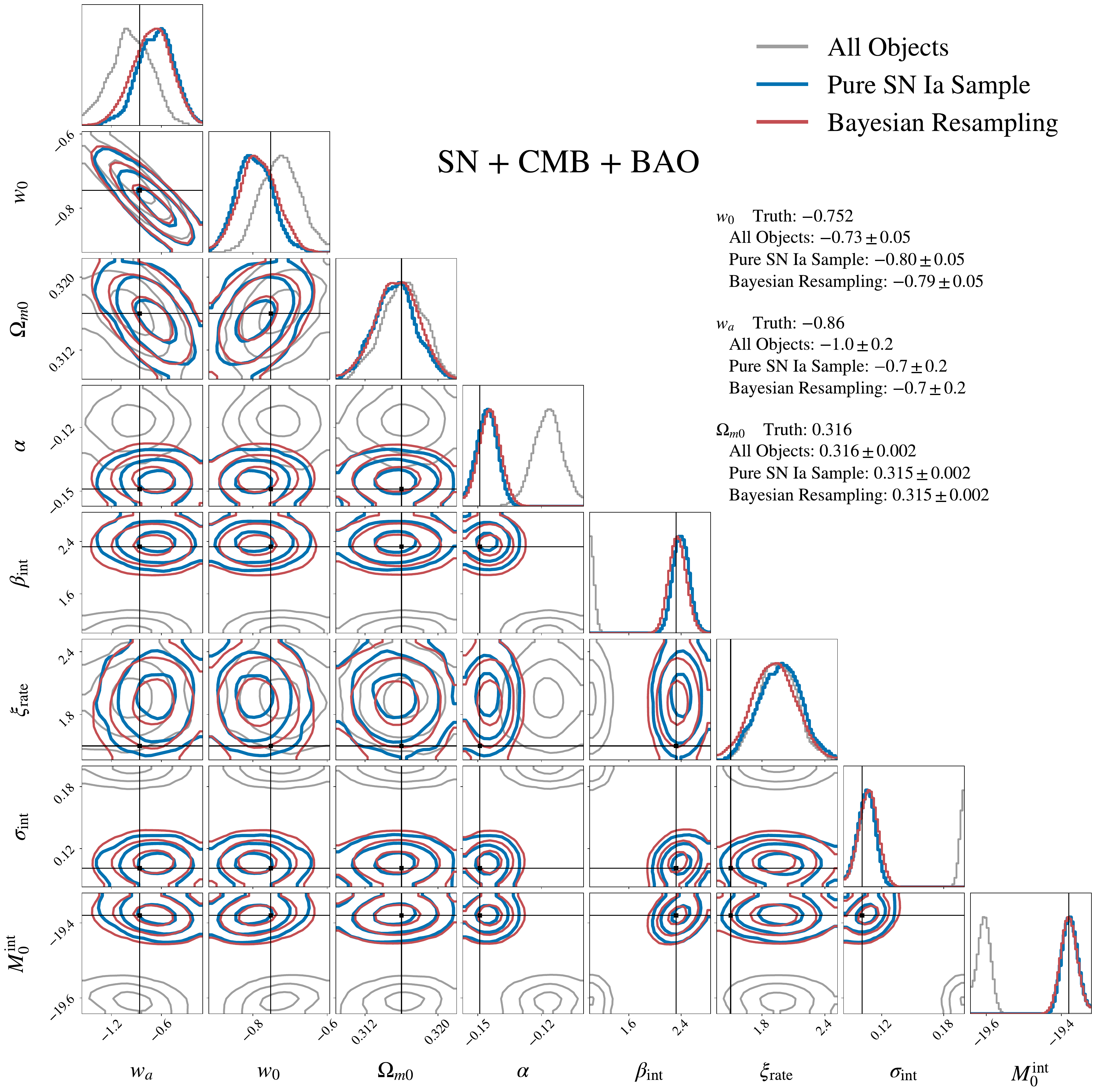}
    \vspace{0cm}    \caption{\textit{FlowSN2} $w_0 w_a$CDM posteriors under the influence of non-Ia contamination. We perform inference with all objects including contaminants to show how they can bias each parameter. We then include inferences using our default approach of modular Bayesian resampling ($M=5$) of the data to account for contamination as described in Section~\ref{sec:contam_theory}. For comparison we include inferences on the pure SN~Ia subsample using the true labels.
Our experiments include a $\sim$18\% post-selection contamination rate. Experimental details can be found in Section~\ref{sec:contam_exp}. } 
    \label{fig:contam_corner}
\end{figure*}

\subsubsection{Contamination Results}

\begin{table*}
\centering
\caption{Average of posterior median estimates over 25 repeats and the standard error on this average for the full photometric SN Ia cosmology experiments with non-Ia contamination. Here we inference results on $w_0w_a$CDM model parameters. We include results using our default modular Bayesian approach of resampling of the data ($M=5$) to handle contamination, as well as results using the pure SN Ia sample with the true labels for comparison. Both use the \textit{FlowSN2} normalising flows in their inference. In order to constrain the evolving dark energy model we include CMB and BAO priors as detailed in Appendix~\ref{app:cmb_bao_prior}. Note uncertainties are the standard error on the average so are a factor of five smaller than those in individual experiments. The $\sigma$ values are calculated using unrounded averages and standard errors.}
\label{tab:c4-c5_contam}
\begin{tabular}{lccc}
\hline
\multicolumn{4}{c}{Cosmology $\bm{C}_4$} \\
\hline
Test &  & \multicolumn{2}{c}{SN + CMB + BAO}\\
\hline
Global Parameter & Truth  & Bayesian Resampling & Pure SN Ia Sample \\
\hline
$w_0$ & -1 & -1.002 $\pm$ 0.010 (-0.2$\sigma$) & -1.006 $\pm$ 0.009 (-0.6$\sigma$)\\
$w_a$ & 0 & -0.01 $\pm$ 0.05 (-0.2$\sigma$) & 0.01 $\pm$ 0.05 (0.2$\sigma$)\\
$\Omega_{m0}$ & 0.315 & 0.3151 $\pm$ 0.0007 (0.2$\sigma$) & 0.3150 $\pm$ 0.0007 (-0.0$\sigma$)\\
$M_0^{\mathrm{int}}$ & -19.38 & -19.378 $\pm$ 0.004 (0.7$\sigma$) & -19.377 $\pm$ 0.003 (0.8$\sigma$)\\
$\alpha$ & -0.149 & -0.1476 $\pm$ 0.0009 (1.6$\sigma$) & -0.1467 $\pm$ 0.0009 (2.5$\sigma$)\\
$\beta_\mathrm{int}$ & 2.322 & 2.32 $\pm$ 0.02 (0.1$\sigma$) & 2.32 $\pm$ 0.02 (-0.1$\sigma$)\\
$\alpha_c$ & -0.006 & -0.0058 $\pm$ 0.0004 (0.6$\sigma$) & -0.0059 $\pm$ 0.0004 (0.2$\sigma$)\\
$R_B$ & 3.74 & 3.70 $\pm$ 0.03 (-1.2$\sigma$) & 3.69 $\pm$ 0.03 (-1.5$\sigma$)\\
$c^{\mathrm{int}}_0$ & -0.056 & -0.0580 $\pm$ 0.0008 (-2.6$\sigma$) & -0.0586 $\pm$ 0.0008 (-3.4$\sigma$)\\
$\sigma_\mathrm{int}$ & 0.101 & 0.1049 $\pm$ 0.0014 (2.8$\sigma$) & 0.1046 $\pm$ 0.0015 (2.4$\sigma$)\\
$\sigma_{c,\mathrm{int}}$ & 0.071 & 0.0708 $\pm$ 0.0004 (-0.5$\sigma$) & 0.0704 $\pm$ 0.0004 (-1.6$\sigma$)\\
$\sigma_x$ & 1.137 & 1.139 $\pm$ 0.004 (0.5$\sigma$) & 1.139 $\pm$ 0.004 (0.6$\sigma$)\\
$\tau$ & 0.068 & 0.0686 $\pm$ 0.0009 (0.7$\sigma$) & 0.0691 $\pm$ 0.0009 (1.2$\sigma$)\\
$x_0$ & -0.452 & -0.450 $\pm$ 0.004 (0.4$\sigma$) & -0.447 $\pm$ 0.005 (1.1$\sigma$)\\
$\xi_\mathrm{rate}$ & 1.5 & 1.54 $\pm$ 0.04 (0.9$\sigma$) & 1.56 $\pm$ 0.04 (1.4$\sigma$)\\
$\gamma$ & -0.1 & -0.0973 $\pm$ 0.0017 (1.6$\sigma$) & -0.0990 $\pm$ 0.0019 (0.5$\sigma$)\\
\hline
\multicolumn{4}{c}{Cosmology $\bm{C}_5$} \\
\hline
Test &  & \multicolumn{2}{c}{SN + CMB + BAO}\\
\hline
Global Parameter & Truth  & Bayesian Resampling & Pure SN Ia Sample \\
\hline
$w_0$ & -0.752 & -0.751 $\pm$ 0.010 (0.1$\sigma$) & -0.754 $\pm$ 0.011 (-0.2$\sigma$)\\
$w_a$ & -0.86 & -0.88 $\pm$ 0.05 (-0.4$\sigma$) & -0.86 $\pm$ 0.05 (-0.0$\sigma$)\\
$\Omega_{m0}$ & 0.316 & 0.3162 $\pm$ 0.0004 (0.5$\sigma$) & 0.3161 $\pm$ 0.0004 (0.3$\sigma$)\\
$M_0^{\mathrm{int}}$ & -19.38 & -19.376 $\pm$ 0.003 (1.2$\sigma$) & -19.376 $\pm$ 0.003 (1.3$\sigma$)\\
$\alpha$ & -0.149 & -0.1472 $\pm$ 0.0010 (1.9$\sigma$) & -0.1466 $\pm$ 0.0009 (2.5$\sigma$)\\
$\beta_\mathrm{int}$ & 2.322 & 2.330 $\pm$ 0.020 (0.4$\sigma$) & 2.32 $\pm$ 0.02 (0.0$\sigma$)\\
$\alpha_c$ & -0.006 & -0.0057 $\pm$ 0.0004 (0.8$\sigma$) & -0.0059 $\pm$ 0.0004 (0.3$\sigma$)\\
$R_B$ & 3.74 & 3.70 $\pm$ 0.03 (-1.4$\sigma$) & 3.69 $\pm$ 0.03 (-1.7$\sigma$)\\
$c^{\mathrm{int}}_0$ & -0.056 & -0.0584 $\pm$ 0.0007 (-3.3$\sigma$) & -0.0588 $\pm$ 0.0008 (-3.6$\sigma$)\\
$\sigma_\mathrm{int}$ & 0.101 & 0.1054 $\pm$ 0.0014 (3.1$\sigma$) & 0.1051 $\pm$ 0.0016 (2.6$\sigma$)\\
$\sigma_{c,\mathrm{int}}$ & 0.071 & 0.0708 $\pm$ 0.0005 (-0.5$\sigma$) & 0.0703 $\pm$ 0.0005 (-1.5$\sigma$)\\
$\sigma_x$ & 1.137 & 1.140 $\pm$ 0.004 (0.6$\sigma$) & 1.140 $\pm$ 0.004 (0.8$\sigma$)\\
$\tau$ & 0.068 & 0.0690 $\pm$ 0.0009 (1.1$\sigma$) & 0.0693 $\pm$ 0.0009 (1.4$\sigma$)\\
$x_0$ & -0.452 & -0.451 $\pm$ 0.004 (0.3$\sigma$) & -0.448 $\pm$ 0.004 (0.8$\sigma$)\\
$\xi_\mathrm{rate}$ & 1.5 & 1.53 $\pm$ 0.04 (0.9$\sigma$) & 1.55 $\pm$ 0.04 (1.2$\sigma$)\\
$\gamma$ & -0.1 & -0.0972 $\pm$ 0.0017 (1.7$\sigma$) & -0.0988 $\pm$ 0.0019 (0.6$\sigma$)\\
\hline
\end{tabular}
\end{table*}

In these contamination experiments, we perform inference with the \textit{FlowSN2} normalising flow likelihood for SNe~Ia exclusively. We compare results obtained across three regimes, using the biased full sample, the pure SN~Ia subsample and the modular Bayesian resampling approach \citep{autenrieth2026}.  Figure~\ref{fig:contam_corner} shows the resulting posteriors for each case.  When the full contaminated sample is analysed without accounting for contamination, the inferred SN population parameters, such as $\alpha$, $\beta_{\text{int}}$, $M_0^{\text{int}}$ and $\sigma_{\text{int}}$, are most strongly affected. As a secondary consequence, the contaminated sample introduces a $\sim 1\sigma$ deviation in the $w_0-w_a$ cosmological parameters despite guidance from the CMB and BAO priors, highlighting the critical need for robust non-Ia contamination mitigation in photometric SN cosmology.

Figure~\ref{fig:contam_corner} confirms that posteriors derived from the pure SN~Ia sample successfully recover the true parameters, consistent with our findings in Section~\ref{sec:ia_exp}. We find that the resampled posterior shows close agreement with the pure SN~Ia subsample posterior for both population and cosmological parameters. To further evaluate this performance, in Table~\ref{tab:c1-c3_contam} and Table~\ref{tab:c4-c5_contam} we repeated these tests across 25 realisations for each cosmological parameter set ($\bm{C}_1$ through $\bm{C}_5$), both with and without CMB and BAO priors. Cosmological parameter biases using this contaminated dataset remain within $\sim 2\sigma$ of the true values and fall within $0.8\sigma$ of the pure SN~Ia subsample mean posterior estimates. We use $M=5$ resamples in these experiments, noting that increasing $M$ would improve the Monte Carlo accuracy of the resampling approximation. Here, the posterior widths from resampling closely match those using the pure SN~Ia subsample. More generally, accounting for classification uncertainty through the resampling can broaden the posterior when fewer objects are retained in the resampled datasets \citep{autenrieth2026}. We again confirm the fiducial cosmology used during training exerts no influence on inferred cosmological biases and \textit{FlowSN2} reliably recovers SN~Ia rates even when contaminant rates differ.

We evaluate the calibration properties of the \textit{FlowSN2} posteriors in Figure~\ref{fig:calibration_grid}, showing the frequentist empirical coverage of our Bayesian credible regions. None of the parameter posteriors in any of our experiments exhibit significant overconfidence. Resampled posteriors that account for contamination appear slightly conservative, with the tails for $x_0$ and $\sigma_{c,\text{int}}$ being broader. The cosmological parameters $w_0$ and $\Omega_{m0}$ are likewise conservative at the tails when constrained solely by SNe~Ia and contaminants, but tighten once CMB and BAO priors are included. Further training is expected to refine this calibration, an avenue we reserve for future work on more realistic simulations.

\begin{figure*}
\centering
	\includegraphics[width=0.95\textwidth]{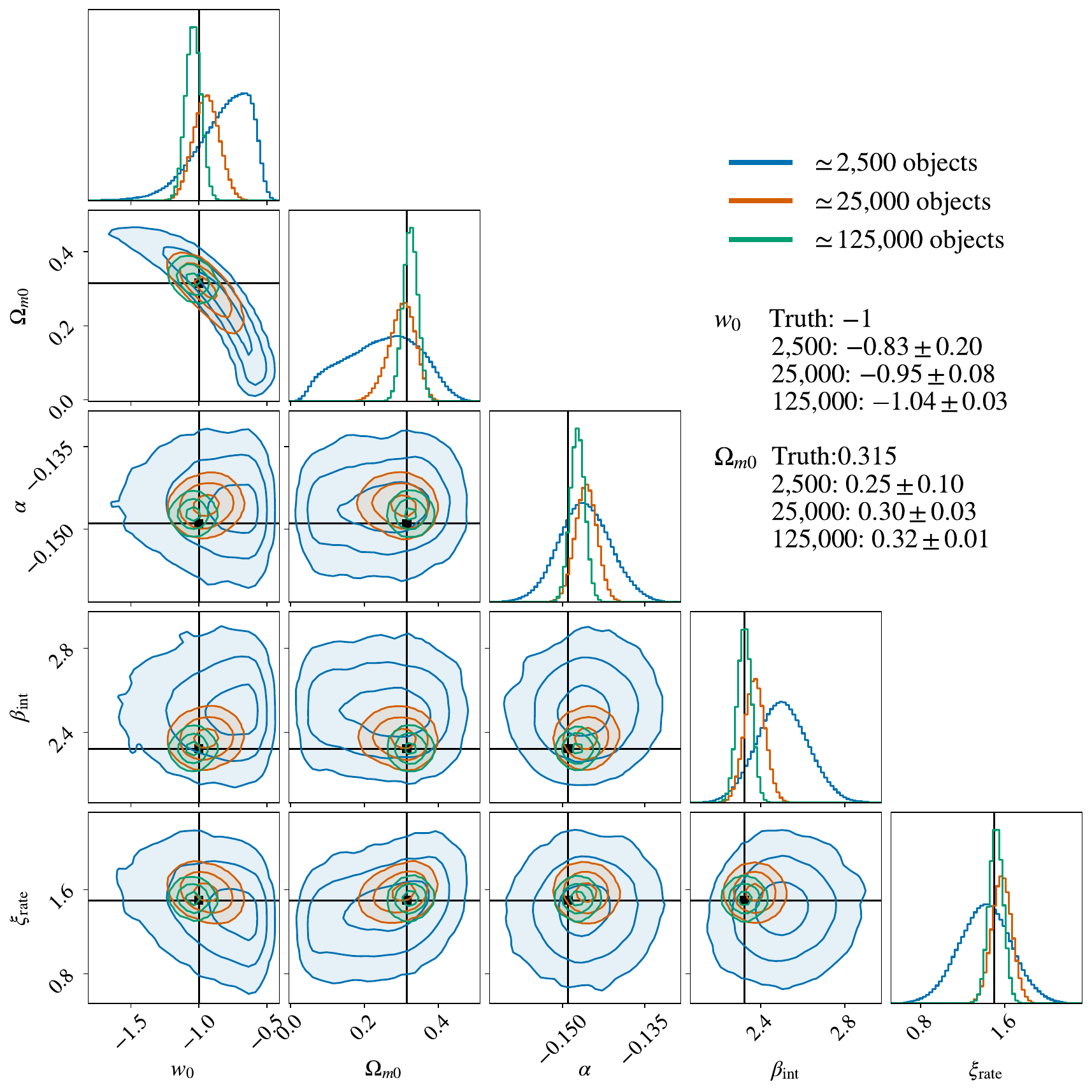}
    \vspace{0cm}    \caption{Plot demonstrating the scalability of the \textit{FlowSN2} framework inferring cosmological parameters and key SN~Ia parameters jointly on increasing sample sizes. These inferences were under the influence of selection effects, with 18\% non-Ia contamination rate and with photometric redshifts. The straight lines represent the true parameters in the simulations.}
    \label{fig:scaling}
\end{figure*}

\subsection{Scaling}

It is also important that our methodology scales to future sample sizes. For demonstration purposes,
we repeat the contamination experiments of Section~\ref{sec:contam_exp} at increasing
sample sizes, scaling the baseline of $\approx 2{,}000$ SNe~Ia and $\approx 500$
contaminants up by factors of ten and fifty, to $\approx 20{,}000$ SNe~Ia with
$\approx 5{,}000$ contaminants and $\approx 100{,}000$ SNe~Ia with $\approx 25{,}000$
contaminants. Both the low-redshift spectroscopic and high-redshift photometric samples
are scaled by the same factor, preserving their relative sizes and the contaminant
fraction. All experiments reuse the same four normalising flows trained for the pure-Ia
analysis (Section~\ref{sec:exp_train}), with no retraining.

The baseline sample of $\approx 2{,}500$ objects takes around ten minutes per resample,
totalling roughly 50 minutes for the $M=5$ resamples. Since the modelled likelihood marginalises over individual latent SN parameters, the sampled dimension is fixed by the
global parameters alone and does not grow with the sample size. Together with improved
GPU utilisation at larger batch sizes, this makes the runtime scale sub-linearly with
the number of objects. The
$\approx 25{,}000$ object inference completes in around five hours in total and the
$\approx 125{,}000$ object inference in around twenty hours, despite the tenfold and
fiftyfold increases in sample size.

Figure~\ref{fig:scaling} shows the three inferences for cosmology $\bm{C}_1$. The cosmological
constraints tighten as the statistical errors shrink with sample size and all three
recover the truth to within $2\sigma$, demonstrating that the trained normalising flows remain accurate at scale. We find all 15 global parameters are successfully inferred at each scale within $3\sigma$. Since likelihood approximation is re-used $N_{\text{obs}}$ times, one might expect compounding errors in the approximation to lead to biased inferences, but this result shows us that inferences are still reliable at this scale. We conclude the method is scalable and ready for the next generation of SN~Ia cosmology with large sample sizes.

\begin{table*}
    \centering
    \caption{Savage--Dickey Bayes factors for the nested model comparisons, each
    averaged over 25 simulated realisations with $M=5$ label resamples
    per realisation. Uncertainties are the standard error on the mean $\ln \text{BF}'_{1,0}$.
    Positive $\ln \text{BF}'_{1,0}$ favours the model in which the parameter is free. Strength of
    evidence follows the Jeffreys scale. Experimental details can be found in Section~\ref{sec:model_comparison}.}
    \label{tab:bayes_factors}
    \begin{tabular}{lcl}
        \hline
        Simulation & $\ln \text{BF}'_{1,0}$ & Interpretation \\
        \hline
        Varying dust only ($\gamma = 0$) & $-2.92 \pm 0.09$ & Strong evidence for $\gamma = 0$ \\
        Varying dust $+$ intrinsic step ($\gamma = -0.06$) & $+4.18 \pm 0.62$ & Strong evidence for $\gamma \neq 0$ \\
        Constant dark energy ($w_a = 0$) & $-1.93 \pm 0.15$ & Substantial evidence for $w_a = 0$ \\
        Evolving dark energy ($w_a = -0.86$) & $+3.64 \pm 0.47$ & Strong evidence for $w_a \neq 0$ \\
        \hline
    \end{tabular}
\end{table*}

\subsection{Model Comparison}
\label{sec:model_comparison}

Parameter estimation answers how well a given model fits the data, but many of the open
questions in SN~Ia cosmology are instead questions of \emph{which} model the data prefer.
A convenient feature of these questions is that they are often nested models. A simpler
null model $\mathscr{M}_0$ is recovered from a more complex model $\mathscr{M}_1$ by fixing
a single parameter to a reference value, $\theta = \theta_0$. Whether there is an intrinsic
host-galaxy mass step, whether dark energy evolves and whether a colour-luminosity slope
differs from its canonical value are all of this form.

The novelty of our approach is that we perform this model selection directly on a
photometric sample, propagating photometric-redshift uncertainties, non-Ia contamination
and survey selection effects consistently into the Bayes factor. Photometric redshifts are
treated as data with their own likelihood rather than as fixed inputs, selection is built into the trained normalising flows by construction and contamination is accounted for through the modular Bayesian
resampling approach of Section~\ref{sec:contam_theory}. Model selection then inherits all three
treatments at no extra cost, since the Bayes factor is evaluated on exactly the same
resampled posteriors used for parameter estimation. 

For nested models, we compare $\mathscr{M}_0$ and $\mathscr{M}_1$ using the Savage-Dickey
density ratio of Eq.~\eqref{eq:savdickey}, interpreting the result on the Jeffreys' scale
\citep{jeffreys1961theory}. The Bayes factor requires
only the marginal posterior density of $\theta$ under the more complex model, evaluated at
$\theta_0$, together with the prior density at the same point. We already produce that
posterior when we fit $\mathscr{M}_1$, so no additional sampling is needed. In the presence of contamination, we evaluate the ratio on the modular posterior as in Eq.~\eqref{eq:bf_sd_avg}, reusing the same $M=5$ resampled posteriors used for parameter estimation.

We validate this on simulations, where the true data-generating model is known. For each
test we construct two sets of simulations, one drawn from the null model and one from the
extended model, seeing whether the Bayes factor correctly favours the model that generated
the data. We repeat each experiment over 25 independent realisations and report the mean of
$\ln \text{BF}'_{1,0}$ together with its standard error. Our results for both tests are summarised in Table~\ref{tab:bayes_factors}. The models compared here are by no means the only ones worth testing, since there are many
variations of each. These experiments instead serve to demonstrate how such model
comparisons are carried out within the framework.

\subsubsection{Dust or Intrinsic Mass Step?}
\label{sec:dust_vs_intrinsic}

Whether the environmental mass-step \citep{kelly2010,sullivan2010} is intrinsic to the progenitor population
or an extrinsic consequence of host-galaxy dust remains an open and important question,
since an unmodelled or mis-attributed step propagates directly into inferred distances and
therefore into cosmology. \citet{brout2021} argued that the step is entirely a dust effect,
driven by host-mass-dependent dust properties, with more massive hosts being both dustier
and having a lower dust relation parameter $R_B$. Using a hierarchical Bayesian SED model, \citet{thorp2022} found, in contrast, that a residual intrinsic step of order 0.06~mag persists even once a mass-dependent dust distribution is taken into account. More recent work has found that the intrinsic step persists when applying the model to larger SN~Ia datasets \citep{grayling2024,grayling2025,ginolin2026}.

We cast the problem as a nested model comparison. The null model $\mathscr{M}_0$ allows the dust
distribution to vary with host mass, but has no intrinsic step, $\gamma = 0$. The extended
model $\mathscr{M}_1$ additionally allows a free intrinsic step $\gamma$. To make the dust
distribution host-mass dependent we split both the mean colour excess $\tau$ and the $R_B$ parameter at the mass threshold $\mathcal{M}_{\text{step}} = 10^{10}\,M_\odot$, following the
qualitative picture of \citet{brout2021}. Objects in high-mass hosts are simulated to have a larger
$\tau$ ($0.098$ against $0.068$) and a lower $R_B$ ($3.0$ against $3.7$). The high-mass $R_B$ is chosen to be within the training range, so is therefore more
conservative than the value inferred by \citet{brout2021}. Both flows are already amortised
over per-object $\tau$ and $R_B$, so the host-mass
dependence is introduced with no retraining. We then simulate two datasets, one purely from
varying dust ($\gamma = 0$) and one from varying dust together with an intrinsic step
($\gamma = -0.06$), the latter matching the residual step reported by \citet{thorp2022}.

The Bayes factors, given in Table~\ref{tab:bayes_factors}, recover the correct
data-generating model in both cases. For the varying dust only simulations, the Bayes factor
favours the null model, $\ln \text{BF}'_{1,0} = -2.92 \pm 0.09$, which on the Jeffreys' scale
is strong evidence for $\gamma = 0$. For the varying dust and intrinsic step simulations, it
favours the extended model, $\ln \text{BF}'_{1,0} = +4.18 \pm 0.62$, strong evidence for
$\gamma \neq 0$. In both cases the recovered model is the one that generated the data.

\subsubsection{Constant or Evolving Dark Energy?}
\label{sec:constant_vs_evolving}

Whether the dark energy equation of state is constant ($w_a = 0$) or evolves with time
($w_a \neq 0$) is one of the most actively debated questions in current cosmology. Recent
baryon acoustic oscillation (BAO) measurements from DESI, in combination with DES SNe~Ia
\citep{des5yr} and CMB data \citep{planck2020}, show a preference for an evolving equation of
state under the $w_0 w_a$CDM parameterisation at the $\sim 3\sigma$ level
\citep{desi2025,popovic2026}, a result that has attracted considerable attention as a
possible departure from $\Lambda$CDM. This makes the constant versus evolving comparison a
natural target for our framework.

Similar to \cite{popovic2026} and \cite{ormondroyd2026}, we use the Bayesian evidence to tell us which cosmological model the observations support. We treat this experiment as a nested test between the null model $\mathscr{M}_0$ with
$w_a = 0$ and the extended model $\mathscr{M}_1$ with $w_a$ free. SNe~Ia alone constrain $w_0w_a$CDM parameters poorly, so for this test we combine the SN likelihood with the CMB
and BAO priors described in Appendix~\ref{app:cmb_bao_posterior}. We simulate two datasets, one with constant
dark energy ($w_a = 0$) and one with an evolving equation of state at the DESI DR2 best-fit
value ($w_a = -0.86$).

The Bayes factors are given in Table~\ref{tab:bayes_factors} and recover the correct model in
both cases. For the constant dark energy simulations, the Bayes factor favours the null model,
$\ln \text{BF}'_{1,0} = -1.93 \pm 0.15$, substantial evidence for $w_a = 0$. For the
evolving dark energy simulations, it favours the extended model,
$\ln \text{BF}'_{1,0} = +3.64 \pm 0.47$, strong evidence for $w_a \neq 0$. This suggests that
\textit{FlowSN2} could provide a powerful means of determining which dark energy model real
data favour, whilst still accounting for selection effects, photometric redshifts and
contamination.

\section{Conclusions and Future Work}
\label{sec:conc}

This work has shown that \textit{FlowSN2} is a powerful framework for inferring cosmological and population parameters under the influence of selection effects, photometric redshift uncertainties and contamination. \textit{FlowSN2} reframes SN~Ia photometric cosmology from a regression problem to a joint inference problem where redshifts, alongside the light curves themselves, are treated as data. We validate that the posteriors are statistically motivated and equivalent to the analytical solution in tractable regimes. Furthermore, the methodology used for handling non-Ia contamination \citep{autenrieth2026} yields results that are in strong agreement with the pure SN~Ia posteriors. Although this initial demonstration was conducted under simplified conditions, \cite{flowsn} showed that this line of work performs equally well in more complex regimes where no analytical solution exists.

With this new methodology, we are capable of reanalysing previous photometric studies, such as the full DES 5YR sample \citep{des5yr}. This will involve training the normalising flows on realistic simulations \citep{kessler2009,skysurvey,dai2026,jaxsnana} before performing inference on the real data. Within this framework, survey-specific systematics can be modelled explicitly as covariates, or alternatively, if the systematics are poorly understood, we may average over posteriors from an ensemble of normalising flows trained on simulations where the systematic parameter varies. Our joint modelling of low and high redshift samples with varying data quality and selection effects demonstrates that this framework is a natural fit for analysing SN compilations from various surveys, such as Pantheon+ \citep{scolnic2022,brout2022}.

The \textit{FlowSN2} framework will be particularly valuable for the era of LSST \citep{lsst2019} and Roman \citep{roman2018}, as we have shown that it can scale up to over a hundred thousand SNe without retraining. The ongoing surveys will bring about challenging new systematics, such as varying survey selection. If each observing year involves discrete changes in strategy, we may train a set of normalising flows per year and trivially combine them at inference. If changes are more gradual and can be simulated faithfully, we can condition the likelihood approximation on the time of observation. Further work will be required to test the robustness of our framework against model misspecification, relying on a host of standardised tests developed by the SBI community \citep{montel2025,ocallaghan2025}.

We demonstrated that \textit{FlowSN2} is able to model more complex SN~Ia standardisations beyond the \cite{tripp1998} relationship using the Simple-BayeSN model \citep{mandel2017}, separating intrinsic parameters from extrinsic dust parameters. It is straightforward to go one step further and use the full BayeSN \citep{mandel2022} SED parameter fits as summary statistics in our framework, mapping $(\hat{m}_s,\hat{c}_s,\hat{x}_s) \rightarrow (\mu_s,\theta_s,A^s_V)$. To enable further enhancements in SN standardisation, we may wish to include residual SED parameters $\epsilon_s$ from \cite{mandel2017} or learnt hybrid summary statistics \citep{makinen2024} in our modelling.

We demonstrated that at no extra cost, we can use the \textit{FlowSN2} posterior for Bayes factor calculations when comparing nested models. Current debates persist regarding whether the apparent SN host-galaxy mass step \citep{kelly2010,sullivan2010,lampeitl2010,childress2013} stems from intrinsic progenitor differences or extrinsic dust differences in host-galaxy environments \citep{brout2021,thorp2022,murakami2025,ginolin2025a}. Most previous analyses were forced to evaluate this using volume-limited spectroscopic samples. We demonstrate through our ``dust or intrinsic step'' example that \textit{FlowSN2} is capable of identifying the correct generative forward model using photometric data while accounting for selection effects. In the future, we aim to incorporate additional complexity in host-galaxy modelling, as demonstrated in \cite{karchev2025cigars}. Our modular SBI framework naturally allows upstream parameter dependencies to be changed without retraining the normalising flow that learns the downstream influence of selection on observed data.

The \textit{FlowSN2} framework provides unique advantages over other inference methodologies in the context of evolving dark energy debates \citep{desi2025,efstathiou2024,vincenzi2025,dhawan2025,popovic2026}. The framework is amortised over the Hubble diagram, enabling us to infer parameters for any parameterisation of distance as a function of redshift. This allows us to naturally compare posterior constraints across different cosmological models under the influence of selection effects, contamination and photometric redshift uncertainties without retraining. As shown in \cite{flowsn}, alternative approaches to handling selection effects can converge to different cosmological posteriors than those derived using this probabilistic prescription. Furthermore, we benefit from performing Bayesian model comparison in these settings at minimal additional computational cost. This scalable yet interpretable approach to SBI for SN cosmology will be a valuable tool in uncovering the nature of our Universe using data from the next generation of surveys.

\section*{Acknowledgements}
We thank Konstantin Karchev and Richard Stiskalek for useful discussions related to this work.

BMB is supported by the Cambridge Centre for Doctoral Training in Data-Intensive Science funded by the UK Science and Technology Facilities Council (STFC) and two G-Research Early Career Researcher grants used for equipment and travel. BMB also acknowledges travel support provided by Christ's College Cambridge's Clayton Fund, LSST DESC supported by DOE funds administered by SLAC and STFC for UK participation in LSST through grant ST/S006206/1. Supernova and astrostatistics research at Cambridge University is supported in part by the European Union’s Horizon 2020 research and innovation programme under European Research Council Grant
Agreement No. 101002652 (BayeSN; PI K. Mandel). ST has been supported by the European Research Council (ERC) under the European Union’s Horizon 2020 research and innovation programme (grant agreement no. 101018897 CosmicExplorer). LK acknowledges support for an Early Career Fellowship from the Leverhulme Trust through grant ECF-2024-054 and the Isaac Newton Trust through grant 24.08(w).

\section*{Data Availability}

The implementation of \textit{FlowSN2} and the experiments from this paper are publicly available at: \url{https://github.com/bayesn/flowsn2}.



\bibliographystyle{mnras}
\bibliography{references} 



\newpage
\onecolumn
\appendix

\section{Analytical Derivation}
\label{app:analytical}

Here we derive the analytical likelihood solution to a dusty Simple-BayeSN model \citep{mandel2017} under the influence of Gaussian cumulative distribution function (CDF) selection effects in both the magnitude and redshift axis. The generative forward model used in experiments is described in Section~\ref{sec:ia_model_forward}. The likelihood solution is used in Section~\ref{sec:ia_exp} as an analytical baseline and also in Section~\ref{sec:contam_sec} to provide perfect classification scores.

We begin by deriving the likelihood (conditioned on latent parameters) without selection effects
\begin{multline}
\label{eq:lik_joint}
P(\bm{\hat{d}}_s,\hat{z}_s |\, \mu_s, z_s,E_s,\bm{\Theta}_{\text{SN}})\\ \hspace{4cm}= \int^{\infty}_{-\infty}\int^{\infty}_{-\infty} P(\hat{\bm{d}}_s , \hat{z}_s |\, \bm{d}_s,z_s,v^s_{\text{pec}}) P(\bm{d}_s |\, \mu_s, v^s_{\text{pec}},E_s, \bm{\Theta}_{\text{SN}}) \text{ d}\bm{d}_s \text{ d}v^s_{\text{pec}} \\ = 
N\Big((\bm{\hat{d}}_s,\hat{z}_s)^T\Big|\,\bm{\rho}^s_{\bm{d}|E}, W^s_{\bm{d}|E}\Big)
\end{multline}
\[\hspace{2cm}
\bm{\rho}^s_{\bm{d}} =
\begin{pmatrix}
\mu_s+M^{\text{int}}_0 + \alpha x_0 + \beta_{\text{int}} (c^{\text{int}}_0 +\alpha_c x_0)\\
c^{\text{int}}_0 +\alpha_c x_0\\
x_0\\
z_s
\end{pmatrix},
\qquad
\bm{\rho}^s_{\bm{d}|E} =
\begin{pmatrix}
\mathbb{E}[\hat{m}_s|\,E_s]\\
\mathbb{E}[\hat{c}_s|\,E_s]\\
\mathbb{E}[\hat{x}_s]\\
\mathbb{E}[\hat{z}_s]
\end{pmatrix}
= \bm{\rho}^s_{\bm{d}} + E_s \bm{\rho}_E,
\qquad
\bm{\rho}_E = \begin{pmatrix} R_B\\ 1\\ 0\\ 0\end{pmatrix}
\]
\[\hspace{4cm}
 \bm{W}^s_{\bm{d}|E} = \begin{pmatrix}
\text{Var}[\hat{m}_s|\, E_s] & \text{Cov}[\hat{m}_s,\hat{c}_s|\, E_s] &\text{Cov}[\hat{m}_s,\hat{x}_s]  &
\text{Cov}[\hat{m}_s,\hat{z}_s]

\\
 \text{Cov}[\hat{m}_s,\hat{c}_s|\, E_s] & \text{Var}[\hat{c}_s|\, E_s]  &  \text{Cov}[\hat{c}_s,\hat{x}_s] 
 & 0
 \\
\text{Cov}[\hat{m}_s,\hat{x}_s]  & \text{Cov}[\hat{c}_s,\hat{x}_s]  & \text{Var}[\hat{x}_s]  & 0
\\ 
\text{Cov}[\hat{m}_s,\hat{z}_s] & 0 & 0 & \text{Var}[\hat{z}_s]
\end{pmatrix}
\]
\begin{alignat*}{2}
\hspace{2cm}
  \text{Var}[\hat{m}_s| \, E_s] &= \frac{100 \sigma_{\text{pec}}^2}{\ln(10)^2c^2}+\sigma_{\text{int}}^2+\beta_{\text{int}}^2\sigma_{c,\text{int}}^2
  &\qquad \text{Cov}[\hat{m}_s,\hat{c}_s| \, E_s] &= \alpha\alpha_c\sigma_x^2 +\beta_{\text{int}}(\sigma_{c,\text{int}}^2+\alpha_c^2\sigma_x^2)+\Sigma_s^{mc},\\
  &\quad \hspace{1cm}+(\alpha+\beta_{\text{int}}\alpha_c)^2 \sigma_x^2+\Sigma^{mm}_s,
  &\qquad \text{Cov}[\hat{m}_s,\hat{x}_s] &= \alpha\sigma_x^2 + \beta_{\text{int}} \alpha_c \sigma_x^2+\Sigma_s^{mx},\\
  \text{Var}[\hat{c}_s| \, E_s] &= \sigma_{c,\text{int}}^2 + \alpha_c^2\sigma_x^2+\Sigma^{cc}_s,
  &\qquad \text{Cov}[\hat{c}_s,\hat{x}_s] &= \alpha_c \sigma_x^2 +\Sigma_s^{cx},\\
  \text{Var}[\hat{x}_s] &=\sigma_x^2+\Sigma^{xx}_s,
  &\qquad \text{Cov}[\hat{m}_s,\hat{z}_s] &= \frac{10\sigma_{\text{pec}}^2}{\ln(10)c^2}.\\
  \text{Var}[\hat{z}_s] &= \frac{\sigma_{\text{pec}}^2}{c^2} + \sigma^2_{z,s},
\end{alignat*}

We can now define two selection probabilities in terms of the observed summary parameters and observed redshifts
\begin{equation}
\label{eq:cdf_app}
\hspace{2cm}
P(I_s = 1 |\, \bm{\hat{d}}_s) = \Phi \Bigg(\frac{m_{\text{cut}} -(\hat{m}_s + a_{\text{cut},m} \, \hat{x}_s + b_{\text{cut},m} \, \hat{c}_s)}{\sigma_{\text{cut},m}}\Bigg),
\qquad
P(I_s = 1 |\, \hat{z}_s) = \Phi \Bigg(\frac{z_{\text{cut}} -\hat{z}_s}{\sigma_{\text{cut},z}}\Bigg),
\end{equation}
where $\Phi(\cdot)$ is a Gaussian cumulative distribution function.
The colour excess $E_s$ has an exponential prior parametrised by average $\tau$
\begin{equation}
\hspace{6cm}
    P(E_s| \, \tau) = \frac{1}{\tau}\exp(-E_s/\tau).
\end{equation}
The analytical selected likelihood marginalised over dust is
\begin{multline}
\label{eq:sel_lik}
    P(\hat{\bm{d}}_s,\hat{z}_s |\, I_s=1,\mu_s,z_s,\bm{\Theta}_{\text{SN}}) \\ \hspace{3cm}=\frac{P(I_s = 1 |\, \bm{\hat{d}}_s)P(I_s = 1 |\, \hat{z}_s) \int^{\infty}_0P(\bm{\hat{d}}_s,\hat{z}_s |\, \mu_s, z_s,E_s,\bm{\Theta}_{\text{SN}}) P(E_s|\, \tau)\text{ d}E_s}{\int^{\infty}_{0} \int^{\infty}_0 \int^{\infty}_{-\infty} P(I_s = 1 |\, \bm{\hat{d}}_s)P(I_s = 1 |\, \hat{z}_s) P(\bm{\hat{d}}_s,\hat{z}_s |\, \mu_s, z_s,E_s,\bm{\Theta}_{\text{SN}}) P(E_s|\, \tau) \text{ d}\hat{\bm{d}}_s \text{ d}\hat{z}_s \text{ d}E_s}  \\ 
= \frac{\mathcal{A}}{\mathcal{B}}.
\end{multline}
The  numerator $\mathcal{A}$ is
\begin{equation}
\hspace{2cm}
       \mathcal{A} = P(I_s = 1 |\, \bm{\hat{d}}_s)P(I_s = 1 |\, \hat{z}_s) 
    \frac{|2\pi s_{E}^2|^{1/2}}{\tau} N\Big((\bm{\hat{d}}_s,\hat{z}_s)^T \Big| \, \bm{\rho}^s_{\bm{d}} + \bm{\rho_{E}}\hat{E}_s, \bm{W}^s_
{\bm{d}|E} \Big) \text{exp}\Bigg[\frac{1}{2}\Bigg(\frac{s_E}{\tau}\Bigg)^2  - \frac{\hat{E}_s}{\tau}\Bigg] \Phi\Bigg(\frac{\hat{E}_s}{s_E}-\frac{s_E}{\tau}\Bigg),
\end{equation}
 \begin{align*}
 \hspace{4cm}
    s_E^{-2} &\equiv \bm{\rho}_E^T \bm{W}^{s\text{ }-1}_{\bm{d}|E} \bm{\rho}_E,
    &\qquad
    \hat{E}_s &\equiv s_E^2  \bm{\rho}^T_E \bm{W}^{s\text{ }-1}_{\bm{d}|E}\bigg((\bm{\hat{d}}_s,\hat{z}_s)^T - \bm{\rho}^s_{\bm{d}}\bigg).
\end{align*}
\newpage
To solve the denominator $\mathcal{B}$ we first rewrite Eq.~\eqref{eq:lik_joint} using the conditional properties of multivariate Gaussians
\begin{multline}
P(\bm{\hat{d}}_s,\hat{z}_s |\, \mu_s, z_s,E_s,\bm{\Theta}_{\text{SN}})=N\Big((\bm{\hat{d}}_s,\hat{z}_s)^T\Big|\,\bm{\rho}^s_{\bm{d}|E}, W^s_{\bm{d}|E}\Big) \\= N\big(\hat{m}_s \big|\, \mathbb{E}[\hat{m}_s|\,\hat{c}_s,\hat{x}_s,\hat{z}_s,E_s],\text{Var} [\hat{m}_s|\,\hat{c}_s,\hat{x}_s,\hat{z}_s,E_s]\big)  N\big(\hat{c}_s \big|\, \mathbb{E}[\hat{c}_s|\,\hat{x}_s,E_s],\text{Var} [\hat{c}_s|\,\hat{x}_s, E_s]\big)  N\big(\hat{x}_s \big|\, \mathbb{E}[\hat{x}_s],\text{Var} [\hat{x}_s]\big) N\big(\hat{z}_s \big|\, \mathbb{E}[\hat{z}_s],\text{Var} [\hat{z}_s]\big),
\end{multline}

\begin{align*}
\hspace{5cm}
    \mathbb{E}[\hat{c}_s|\,\hat{x}_s,E_s] &= \mathbb{E}[\hat{c}_s|\,E_s]
      + \frac{\text{Cov}[\hat{c}_s,\hat{x}_s]}{\text{Var}[\hat{x}_s]}\big(\hat{x}_s - \mathbb{E}[\hat{x}_s]\big),\\
    \text{Var}[\hat{c}_s|\,\hat{x}_s,E_s] &= \text{Var}[\hat{c}_s|\,E_s]
      - \frac{\text{Cov}[\hat{c}_s,\hat{x}_s]^2}{\text{Var}[\hat{x}_s]},\\[6pt]
    \mathbb{E}[\hat{m}_s|\,\hat{x}_s,E_s] &= \mathbb{E}[\hat{m}_s|\,E_s]
      + \frac{\text{Cov}[\hat{m}_s,\hat{x}_s]}{\text{Var}[\hat{x}_s]}\big(\hat{x}_s - \mathbb{E}[\hat{x}_s]\big),\\
    \text{Var}[\hat{m}_s|\,\hat{x}_s,E_s] &= \text{Var}[\hat{m}_s|\,E_s]
      - \frac{\text{Cov}[\hat{m}_s,\hat{x}_s]^2}{\text{Var}[\hat{x}_s]},\\
    \text{Cov}[\hat{m}_s,\hat{c}_s|\,\hat{x}_s,E_s] &= \text{Cov}[\hat{m}_s,\hat{c}_s|\,E_s]
      - \frac{\text{Cov}[\hat{m}_s,\hat{x}_s]\,\text{Cov}[\hat{c}_s,\hat{x}_s]}{\text{Var}[\hat{x}_s]},\\[6pt]
    \mathbb{E}[\hat{m}_s|\,\hat{c}_s,\hat{x}_s,\hat{z}_s,E_s] &= \mathbb{E}[\hat{m}_s|\,\hat{x}_s,E_s]
      + \frac{\text{Cov}[\hat{m}_s,\hat{c}_s|\,\hat{x}_s,E_s]}{\text{Var}[\hat{c}_s|\,\hat{x}_s,E_s]}
        \big(\hat{c}_s - \mathbb{E}[\hat{c}_s|\,\hat{x}_s,E_s]\big)\\
      &\quad \hspace{4cm}+ \frac{\text{Cov}[\hat{m}_s,\hat{z}_s]}{\text{Var}[\hat{z}_s]}\big(\hat{z}_s - \mathbb{E}[\hat{z}_s]\big),\\
    \text{Var}[\hat{m}_s|\,\hat{c}_s,\hat{x}_s,\hat{z}_s,E_s] &= \text{Var}[\hat{m}_s|\,\hat{x}_s,E_s]
      - \frac{\text{Cov}[\hat{m}_s,\hat{c}_s|\,\hat{x}_s,E_s]^2}{\text{Var}[\hat{c}_s|\,\hat{x}_s,E_s]}
      - \frac{\text{Cov}[\hat{m}_s,\hat{z}_s]^2}{\text{Var}[\hat{z}_s]}.
\end{align*}
The three dimensions can be marginalised over using the identity
\begin{equation}
\label{eq:indentity}
\hspace{4cm}
\int^{\infty}_{-\infty} \Phi((\mu_1-x)/\sigma_1)N(x|\,\mu_2,\sigma_2^2)\,\text{d}x
= \Phi\Bigg(\frac{\mu_1 - \mu_2}{\sigma_1} \times \Big(1+\frac{\sigma_2^2}{\sigma_{1}^2}\Big)^{-\frac{1}{2}}\Bigg).
\end{equation}
The first dimension $\hat{m}_s$ is integrated to get 
\begin{multline}
    \int^{\infty}_{-\infty} \Phi \Bigg(\frac{m_{\text{cut}} -(\hat{m}_s + a_{\text{cut},m} \hat{x}_s + b_{\text{cut},m} \hat{c}_s)}{\sigma_{\text{cut},m}}\Bigg) 
   N\big(\hat{m}_s \big|\, \mathbb{E}[\hat{m}_s|\,\hat{c}_s,\hat{x}_s,\hat{z}_s,E_s],\text{Var} [\hat{m}_s|\,\hat{c}_s,\hat{x}_s,\hat{z}_s,E_s]\big)\,\text{d}\hat{m}_s 
\\= \Phi \Bigg(\frac{m_{\text{cut}} -( \mathbb{E}[\hat{m}_s|\,\hat{c}_s,\hat{x}_s,\hat{z}_s,E_s] + a_{\text{cut},m} \hat{x}_s + b_{\text{cut},m} \hat{c}_s)}{\sigma'_{\text{cut},m}}\Bigg),
\end{multline}
with
\begin{equation*}
\hspace{6cm}
    \sigma'_{\text{cut},m}= \sqrt{\sigma^2_{\text{cut},m}+ \text{Var}[\hat{m}_s|\,\hat{c}_s,\hat{x}_s,\hat{z}_s,E_s]}.
\end{equation*}
The second dimension $\hat{c}_s$
is then integrated out to get
\begin{multline}
    \int^{\infty}_{-\infty} \Phi \Bigg(\frac{m_{\text{cut}} -(\mathbb{E}[\hat{m}_s|\,\hat{c}_s,\hat{x}_s,\hat{z}_s,E_s] +a_{\text{cut},m} \hat{x}_s + b_{\text{cut},m} \hat{c}_s)}{\sigma'_{\text{cut},m}}\Bigg) 
   N\big(\hat{c}_s \big|\, \mathbb{E}[\hat{c}_s|\,\hat{x}_s,E_s],\text{Var} [\hat{c}_s|\,\hat{x}_s,E_s]\big)\,\text{d}\hat{c}_s \\   = \int^{\infty}_{-\infty} \Phi \Bigg(\frac{m_{\text{cut}} -(\mathbb{E}[\hat{m}_s|\,\hat{x}_s,E_s]+\gamma_c (\hat{c}_s - \mathbb{E}[\hat{c}_s|\,\hat{x}_s,E_s])+\gamma_z(\hat{z}_s-\mathbb{E}[\hat{z}_s]) +a_{\text{cut},m} \hat{x}_s + b_{\text{cut},m} \hat{c}_s)}{\sigma'_{\text{cut},m}}\Bigg)
   N\big(\hat{c}_s \big|\, \mathbb{E}[\hat{c}_s|\,\hat{x}_s,E_s],\text{Var} [\hat{c}_s|\,\hat{x}_s,E_s]\big)\,\text{d}\hat{c}_s \\
\hspace{3cm}= \Phi \Bigg(\frac{m_{\text{cut}} -\big( \mathbb{E}[\hat{m}_s|\,\hat{x}_s,E_s] -\gamma_c \mathbb{E}[\hat{c}_s|\,\hat{x}_s,E_s] +a_{\text{cut},m} \hat{x}_s + (b_{\text{cut},m} + \gamma_c )\mathbb{E}[\hat{c}_s|\,\hat{x}_s,E_s]+\gamma_z(\hat{z}_s-\mathbb{E}[\hat{z}_s])\big)}{\sigma''_{\text{cut},m}}\Bigg) \\
= \Phi \Bigg(\frac{m_{\text{cut}} -\big( \mathbb{E}[\hat{m}_s|\,\hat{x}_s,E_s] +a_{\text{cut},m} \hat{x}_s + b_{\text{cut},m}\mathbb{E}[\hat{c}_s|\,\hat{x}_s,E_s]+\gamma_z(\hat{z}_s-\mathbb{E}[\hat{z}_s])\big)}{\sigma''_{\text{cut},m}}\Bigg) ,
\end{multline}
where 
\begin{align*}
\hspace{1.5cm}
    \sigma''_{\text{cut},m} &= \sqrt{\sigma'^2_{\text{cut},m}+ (b_{\text{cut},m} + \gamma_c)^2\text{Var}[\hat{c}_s|\,\hat{x}_s,E_s]},
    &\qquad \gamma_c &= \frac{\text{Cov}[\hat{m}_s,\hat{c}_s|\,\hat{x}_s,E_s]}{\text{Var}[\hat{c}_s|\,\hat{x}_s,E_s]},
    &\qquad \gamma_z &= \frac{\text{Cov}[\hat{m}_s,\hat{z}_s]}{\text{Var}[\hat{z}_s]}.
\end{align*}
\newpage
Then the third dimension $\hat{x}_s$ is integrated to get
\begin{multline}
\int^{\infty}_{-\infty} \Phi \Bigg(\frac{m_{\text{cut}} -\big( \mathbb{E}[\hat{m}_s|\,\hat{x}_s,E_s] +a_{\text{cut},m} \hat{x}_s + b_{\text{cut},m}\mathbb{E}[\hat{c}_s|\,\hat{x}_s,E_s]+\gamma_z(\hat{z}_s-\mathbb{E}[\hat{z}_s])\big)}{\sigma''_{\text{cut},m}}\Bigg)  
   N\big(\hat{x}_s \big|\, \mathbb{E}[\hat{x}_s],\text{Var} [\hat{x}_s]\big)\,\text{d}\hat{x}_s\\=
       \int^{\infty}_{-\infty} \Phi \Bigg(\frac{m_{\text{cut}} -\big( \mathbb{E}[\hat{m}_s|\,E_s]+\gamma_x(\hat{x}_s -\mathbb{E}[\hat{x}_s]) +a_{\text{cut},m} \hat{x}_s + b_{\text{cut},m}(\mathbb{E}[\hat{c}_s|\, E_s]+\delta_x(\hat{x}_s -\mathbb{E}[\hat{x}_s]))+\gamma_z(\hat{z}_s-\mathbb{E}[\hat{z}_s])\big)}{\sigma''_{\text{cut},m}}\Bigg)  N\big(\hat{x}_s \big|\, \mathbb{E}[\hat{x}_s],\text{Var} [\hat{x}_s]\big)\,\text{d}\hat{x}_s\\
\hspace{2cm}= \Phi \Bigg(\frac{m_{\text{cut}} -\big( \mathbb{E}[\hat{m}_s|\,E_s] - \gamma_x \mathbb{E}[\hat{x}_s] +(a_{\text{cut},m}+\gamma_x + \delta_x b_{\text{cut},m} )\mathbb{E}[\hat{x}_s]+ b_{\text{cut},m} (\mathbb{E}[\hat{c}_s|\,E_s]-\delta_x \mathbb{E}[\hat{x}_s])+\gamma_z(\hat{z}_s-\mathbb{E}[\hat{z}_s])\big)}{\sigma'''_{\text{cut},m}}\Bigg) \\
    \hspace{4cm}=\Phi \Bigg(\frac{m_{\text{cut}} -\big( \mathbb{E}[\hat{m}_s|\,E_s] +a_{\text{cut},m} \mathbb{E}[\hat{x}_s]+ b_{\text{cut},m} \mathbb{E}[\hat{c}_s|\,E_s]+\gamma_z(\hat{z}_s-\mathbb{E}[\hat{z}_s])\big)}{\sigma'''_{\text{cut},m}}\Bigg) ,
\end{multline}
where 
\begin{align*}
    \sigma'''_{\text{cut},m} &= \sqrt{\sigma''^2_{\text{cut},m}+ (a_{\text{cut},m}+\gamma_x + \delta_x b_{\text{cut},m})^2\text{Var}[\hat{x}_s]},
    &\qquad \gamma_x &= \frac{\text{Cov}[\hat{m}_s,\hat{x}_s]}{\text{Var}[\hat{x}_s]},
    &\qquad \delta_x &= \frac{\text{Cov}[\hat{c}_s,\hat{x}_s]}{\text{Var}[\hat{x}_s]}.
\end{align*}
Next we integrate out $\hat{z}_s$ to get
\begin{multline}
   \int^{\infty}_0 \Phi \Bigg(\frac{m_{\text{cut}} -\big( \mathbb{E}[\hat{m}_s|\,E_s] +a_{\text{cut},m} \mathbb{E}[\hat{x}_s]+ b_{\text{cut},m} \mathbb{E}[\hat{c}_s|\,E_s]+\gamma_z(\hat{z}_s-\mathbb{E}[\hat{z}_s])\big)}{\sigma'''_{\text{cut},m}}\Bigg) \Phi\Bigg(\frac{z_{\text{cut}} -\hat{z}_s}{\sigma_{\text{cut},z}}\Bigg) N\big(\hat{z}_s \big|\, \mathbb{E}[\hat{z}_s],\text{Var}[\hat{z}_s]\big) \text{ d}\hat{z}_s\\ 
=\text{BVN}\Bigg(h=\frac{m_{\text{cut}} -\big( \mathbb{E}[\hat{m}_s|\,E_s] +a_{\text{cut},m} \mathbb{E}[\hat{x}_s]+ b_{\text{cut},m} \mathbb{E}[\hat{c}_s|\,E_s]\big)}{\sqrt{\sigma'''^2_{\text{cut},m}+\frac{\text{Cov}[\hat{m}_s,\hat{z}_s]^2}{\text{Var}[\hat{z}_s]}}}\text{ },\text{ }k=\frac{z_{\text{cut}}-\mathbb{E}[\hat{z}_s]}{\sqrt{\sigma^2_{\text{cut},z}+\text{Var}[\hat{z}_s]}} \text{ ,  } \rho = \frac{\text{Cov}[\hat{m}_s,\hat{z}_s]}{\sqrt{\sigma'''^2_{\text{cut},m}+\frac{\text{Cov}[\hat{m}_s,\hat{z}_s]^2}{\text{Var}[\hat{z}_s]}}\sqrt{\sigma^2_{\text{cut},z}+\text{Var}[\hat{z}_s]}}\Bigg),
\end{multline}
where BVN is a cumulative bivariate normal distribution
\begin{equation}
\hspace{4cm}
    \text{BVN}(h, k; \rho) = \int_{-\infty}^{h} \int_{-\infty}^{k} \frac{1}{2\pi\sqrt{1-\rho^2}} \exp\left( -\frac{x^2 - 2\rho xy + y^2}{2(1-\rho^2)} \right) \text{ d}y  \text{ d}x.
\end{equation}
Then finally we can solve for $\mathcal{B}$ by integrating over $E_s$ using integration by parts 
\begin{multline}
\mathcal{B}=\int^{\infty}_0 P(I_s = 1|\, \mu_s,z_s,E_s, \bm{\Theta}_{\text{SN}}) P(E_s|\, \tau) \text{ d}E_s  = \int^{\infty}_0 \text{BVN}(h_a + h_bE_s,k; \rho) \frac{1}{\tau} \exp{(-E_s/\tau)} \text{ d}E_s \\ \hspace{8cm}
= \text{BVN}(h_a, k; \rho) - \exp\left( \frac{1}{2h_b^2\tau^2}+\frac{h_a}{h_b\tau} \right) \text{BVN}\Big(h_a +\frac{1}{h_b\tau}, k + \frac{\rho}{h_b\tau}; \rho \Big) \\
\end{multline}
where
\begin{align*}
\hspace{2cm}
    h_a &= \frac{m_{\text{cut}} -\big( \mathbb{E}[\hat{m}_s|\,E_s=0] +a_{\text{cut},m} \mathbb{E}[\hat{x}_s]+ b_{\text{cut},m} \mathbb{E}[\hat{c}_s|\,E_s=0]\big)}{\sqrt{\sigma'''^2_{\text{cut},m}+\frac{\text{Cov}[\hat{m}_s,\hat{z}_s]^2}{\text{Var}[\hat{z}_s]}}},
    &\qquad h_b &= -\frac{R_B+b_{\text{cut},m}}{\sqrt{\sigma'''^2_{\text{cut},m}+\frac{\text{Cov}[\hat{m}_s,\hat{z}_s]^2}{\text{Var}[\hat{z}_s]}}},\\
    \rho &= \frac{\text{Cov}[\hat{m}_s,\hat{z}_s]}{\sqrt{\sigma'''^2_{\text{cut},m}+\frac{\text{Cov}[\hat{m}_s,\hat{z}_s]^2}{\text{Var}[\hat{z}_s]}}\sqrt{\sigma^2_{\text{cut},z}+\text{Var}[\hat{z}_s]}},
    &\qquad k &= \frac{z_{\text{cut}}-\mathbb{E}[\hat{z}_s]}{\sqrt{\sigma^2_{\text{cut},z}+\text{Var}[\hat{z}_s]}},
\end{align*}
thus fully solving Eq.~\eqref{eq:sel_lik} for $P(\hat{\bm{d}}_s,\hat{z}_s |\, I_s=1,\mu_s,z_s,\bm{\Theta}_{\text{SN}}) $ analytically.

The solution to Eq.~\eqref{eq:sel_lik}  is then used to calculate $P(\bm{\hat{d}}_s, \hat{z}_s |\, I_s = 1,\bm{\Theta})$ where we numerically marginalise latent over $z_s$. The  marginalised likelihood $P(\bm{\hat{d}}_s, \hat{z}_s |\, I_s = 1,\bm{\Theta})$ is used as an analytical  baseline comparison in Section~\ref{sec:ia_exp} and also to give us perfect classification scores in Section~\ref{sec:contam_sec}.

\newpage

\section{Simplified Forward Model Summary Fit Covariances}
\label{app:cov_samp}
In our experiments, we draw realistic $3 \times 3$ light curve summary statistic fit covariance matrices $\bm{\Sigma}_s \sim P(\bm{\Sigma})$ for each simulated SN. We follow the empirical relations introduced by \citet{flowsn}, which are motivated by observational covariance distributions in the Pantheon+ sample \citep{scolnic2022}. The diagonal and off-diagonal elements are sampled according to
\begin{align*}
\hspace{5cm}
\log \Sigma_s^{mm} &\sim N\big(0.2(m_s - 56),\,1.2^2\big) \\
\log \Sigma_s^{cc} &\sim N(-7,\,0.6^2) \\
\log \Sigma_s^{xx} &\sim N(-3,\,1^2) \\
\Sigma_s^{mc} &\sim N\Big(
5.2\times10^{-4} + 0.289\, \Sigma_s^{mm},\,
(0.126\, \Sigma_s^{mm})^2
\Big) \\
\Sigma_s^{mx} &\sim N\Big(
8.1\times10^{-4} + 0.0384\, \Sigma_s^{xx},\,
(0.0203\, \Sigma_s^{xx})^2
\Big) \\
\Sigma_s^{cx} &\sim N\Big(
1.69\times10^{-4} + 0.0136\, \Sigma_s^{xx},\,
(0.0118\, \Sigma_s^{xx})^2
\Big)
\end{align*}
where we enforce a lower bound on the eigenvalues ($\lambda_{\text{min}} > 10^{-5}$) to guarantee that $\bm{\Sigma}_s$ is strictly positive-definite. We apply this identical covariance distribution across both the training and test samples.

\section{Global Parameter Priors and Training Ranges}
\begin{table}
\centering
\caption{Priors on the global parameters of used in our experiments. Both the \textit{FlowSN2} and the analytical posteriors use the same priors. We denote a normal distribution truncated between $[a,b]$ as $N(\mu,\sigma^2)_{[a,b]}$. Some parameters are not varied in training and are kept at their fiducial values.  }
\label{tab:priors}
\begin{tabular}{lcc}
\hline
Global Parameter & Prior & Training Range \\
\hline
$\Omega_{m0}$ & $U(0,1)$ & 0.315 \\
$w_0$  & $U(-2,0)$ & -1 \\
$w_a$ ( for $w_0w_a$CDM) & $U(-3,3)$ &0 \\
$M_0^{\mathrm{int}}$ & $U(-20.5,-18.5)$ & -19.38 \\
$\xi_\mathrm{rate}$ & $U(0,5)$ & 1.5 \\
$\gamma$ & $U(-0.5,0.5)$ & 0 \\
$\alpha$ & $U(-0.2,0)$ & $U(-0.2,0)$ \\
$\beta_\mathrm{int}$ & $U(1,3)$ & $U(1,3)$ \\
$\alpha_c$ & $U(-0.02,0)$ & $U(-0.02,0)$ \\
$R_B$ & $U(2.2,5)$ & $U(2.2,5)$ \\
$\tau$ & $U(0,0.15)$ & $U(0,0.15)$ \\
$c^{\mathrm{int}}_0$ & $U(-0.1,0.1)$ & $U(-0.1,0.1)$ \\
$x_0$ & $U(-0.6,-0.2)$ & $U(-0.6,-0.2)$ \\
$\sigma_\mathrm{int}$ & $N(0,1^2)_{[0,0.2]}$ & $U(0,0.2)$ \\
$\sigma_{c,\mathrm{int}}$ & $N(0,1^2)_{[0,0.1]}$ & $U(0,0.1)$ \\
$\sigma_x$ & $N(0,2^2)_{[0,1.5]}$ & $U(0,1.5)$ \\
\hline
\end{tabular}
\end{table}

In Table~\ref{tab:priors}, we include the priors on our cosmological and SN population parameters used in inference. We also include the training distributions sampled per event when simulating the training set. Since our normalising flows are only conditioned (for $q_1^{\bm{\phi}_1}$) and evaluated  (for $q_2^{\bm{\phi}_2}$) on latent redshift $z_s$ and distance modulus $\mu_s$, there is no need to vary some of the top-level parameters during training. We instead add a Gaussian scatter of $\sigma_{\mu}=0.25$ to latent distances pre-selection to provide support for different cosmological models on the Hubble diagram.

\newpage
\section{CMB and BAO Priors}
\label{app:cmb_bao_prior}

For the experiments $\bm{C}_1$ to $\bm{C}_3$ that infer the $w$CDM model parameters we add a CMB prior for additional constraining power, whilst for experiments $\bm{C}_4$ and $\bm{C}_5$ we also include a BAO prior for constraining $w_0 w_a $CDM.

\subsection{CMB}
\label{app:cmb_prior}

To represent the constraints from the CMB we use a standard two-parameter compression
\citep{komatsu2009}. We define the ratio (or shift) parameter $R$ and the acoustic scale
$\ell_A$
\begin{equation}
\label{eq:cmb_obs_def}
\hspace{5cm}
    R(\bm{C}) = \frac{\sqrt{\Omega_{m0}}\,H_0\,d_c(z_*;\bm{C})}{c},
    \qquad
    \ell_A(\bm{C}) = \frac{\pi\,d_c(z_*;\bm{C})}{r_s(z_*;\bm{C})},
\end{equation}
where $d_c$ is the transverse co-moving distance function in Mpc and $r_s$ is the co-moving
sound horizon, both functions of $\bm{C}$. We evaluate the functions at the redshift of
recombination $z_* $ which we obtain for each set of cosmological parameters using the \citet{hu1996} approximation. In our calculations we assume
constant physical baryon density $\omega_b = 0.02237$, Hubble constant
$H_0 = 70 \text{ km s}^{-1} \text{ Mpc}^{-1}$ and speed of light $c = 299792.458$ km s$^{-1}$.

For each realisation we draw one correlated observation
\begin{equation}
\label{eq:cmb_draw}
\hspace{4.5cm}
    \begin{pmatrix}\hat{R}\\ \hat{\ell}_A\end{pmatrix} \sim
    N\!\left(\begin{pmatrix}R(\bm{C}_\mathrm{true})\\ \ell_A(\bm{C}_\mathrm{true})\end{pmatrix},
    \bm{\Sigma}_\mathrm{CMB}\right),
    \qquad
    \bm{\Sigma}_\mathrm{CMB} = \begin{pmatrix}
    \sigma_R^2 & \rho\,\sigma_R\sigma_{\ell_A}\\
    \rho\,\sigma_R\sigma_{\ell_A} & \sigma_{\ell_A}^2\end{pmatrix},
\end{equation}
with the Planck 2018 distance prior precisions $\sigma_R=0.0046$, $\sigma_{\ell_A}=0.09$ and
$\rho=0.46$ \citep{planck2020}.

\subsection{BAO}
\label{app:bao_prior}

We condense BAO observations into a 13 dimensional vector, using seven effective redshifts
$z_1$ to $z_7$ between $0.295$ and $2.33$, such that
\begin{equation}
\hspace{4cm}
\label{eq:bao_vector}
    \bm{B}(\bm{C}) = \left(\frac{d_V(z_1; \, \bm{C})}{r_d(\bm{C})},\;
    \frac{d_c(z_2; \, \bm{C})}{r_d(\bm{C})},\;  \frac{d_H(z_2;\,\bm{C})}{r_d(\bm{C}) },\;
    \ldots,\;
    \frac{d_c(z_7; \, \bm{C})}{r_d(\bm{C})},\; \frac{d_H(z_7; \, \bm{C})}{r_d(\bm{C})}\right)^T,
\end{equation}
where $d_H$ is the radial distance, $d_V$ is the volume averaged distance
and $r_d$ is the co-moving sound horizon at the baryon drag epoch, evaluated with an approximation  from \citet{aubourg2015}.

The observational covariance $\bm{\Sigma}_\mathrm{BAO}$ is defined to match DESI DR2
constraining power \citep{desi2025}. We draw each BAO observation from a multivariate
Gaussian centred at the true BAO observations
\begin{equation}
\label{eq:bao_draw}
\hspace{7.5cm}
    \bm{\hat{B}} \sim N\!\left(\bm{B}(\bm{C}_\mathrm{true}),\bm{\Sigma}_\mathrm{BAO}\right).
\end{equation}

\subsection{Combined Posterior}
\label{app:cmb_bao_posterior}
With both CMB and BAO priors the posterior of Eq.~\eqref{eq:joint_post} becomes
\begin{multline}
\label{eq:hmc_cmb_bao_post}
P\big(\bm{\Theta} | \, \bm{\hat{D}}_\text{obs}, \bm{\hat{Z}}_\text{obs},
\bm{I}_{\text{obs}}, \hat{R}, \hat{\ell}_A, \bm{\hat{B}}\big) \propto
P\big(\bm{\hat{D}}_\text{obs}, \bm{\hat{Z}}_\text{obs} \big|\,
\bm{I}_{\text{obs}},\bm{\Theta} \big)
P\big(\hat{R},\hat{\ell}_A|\,\bm{C}\big)P\big(\bm{\hat{B}}|\,\bm{C}\big)P(\bm{\Theta}|\,\bm{I}_{\text{obs}})\\
\text{where}\quad
P\big(\hat{R},\hat{\ell}_A|\,\bm{C}\big) =
N\bigg(\big(\hat{R},\hat{\ell}_A\big)^T\bigg|\big(R(\bm{C}),\ell_A(\bm{C})\big)^T,\bm{\Sigma}_\mathrm{CMB}\bigg) \text{ and }
P\big(\bm{\hat{B}}|\,\bm{C}\big) = N\!\left(\bm{\hat{B}}\big|\,\bm{B}(\bm{C}),\bm{\Sigma}_\mathrm{BAO}\right).
\end{multline}

\section{Further Investigation into Posterior Biases and Calibration}
\label{app:further}
\begin{figure*}
\centering
	\includegraphics[width=0.95\textwidth]{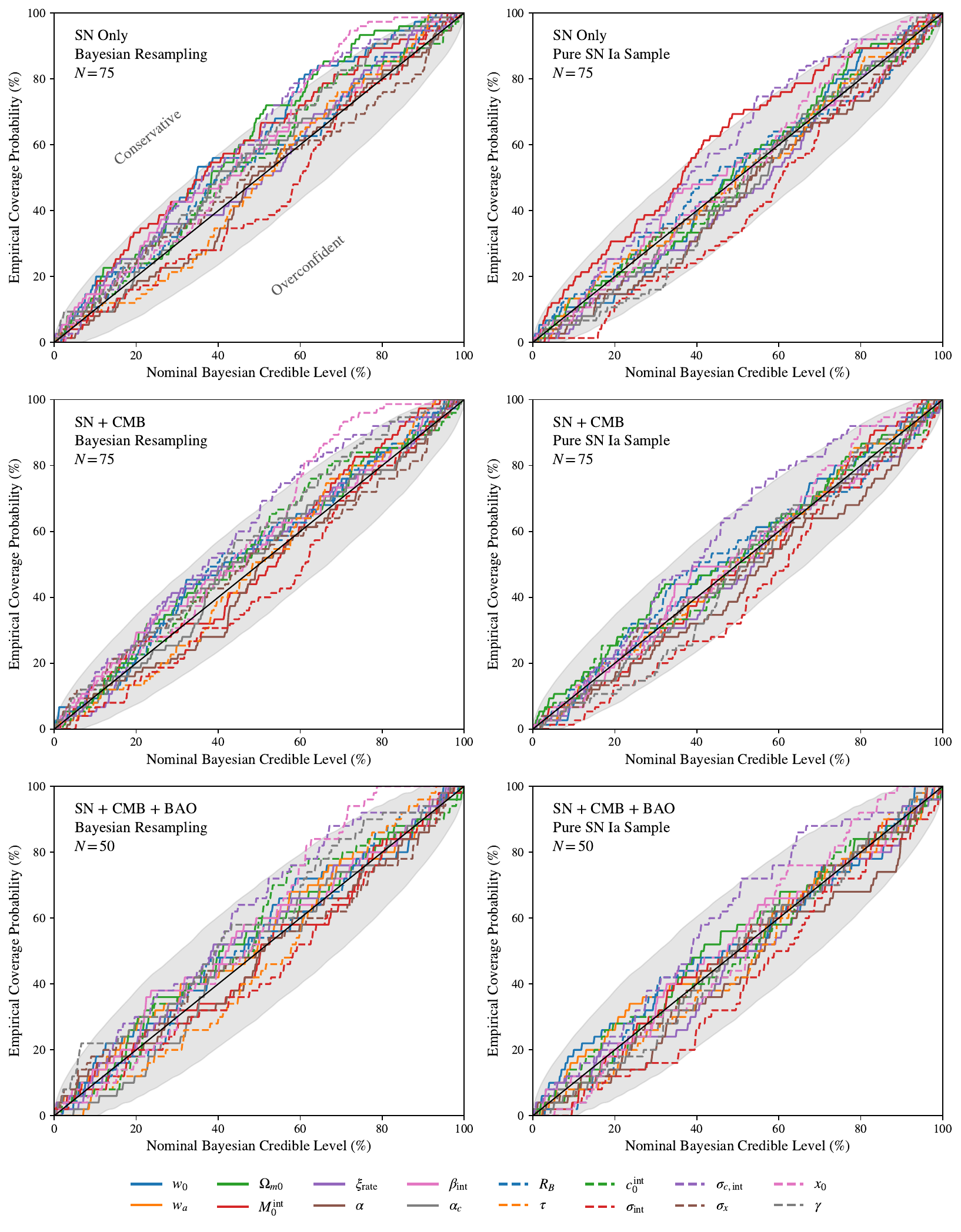}
    \vspace{0cm}   
    
    \vspace{-0.3cm}\caption{Coverage of the \textit{FlowSN2} posteriors under photometric contamination,
photometric redshift uncertainty and selection effects. Each panel shows the
empirical coverage of every inferred parameter, pooled across the simulated
cosmologies $\bm{C}_1$ to $\bm{C}_5$, using $N$ independent realisations. The left column resamples the objects ($M=5$) with the modular Bayesian resampling treatment of the photometric
classification probabilities, while the right column shows coverage using a pure SN~Ia sample with the true labels. The middle row adds a CMB prior, while both CMB and BAO priors are added in the bottom
row. Only the bottom row
infers $w_a$, using the BAO prior, while the upper rows hold it fixed at $w_a=0$. The grey shaded region is the 95\% Monte Carlo confidence
band for the empirical coverage under perfect calibration.}
    \label{fig:calibration_grid}
\end{figure*}

Here, we present additional experimental results. For the pure-Ia experiments described in Section~\ref{sec:ia_exp}, posterior averages over 25 realisations for cosmologies $\bm{C}_1$--$\bm{C}_3$ are reported in Table~\ref{tab:c1-c3_ia_only}, comparing \textit{FlowSN2} against the analytical baseline. Table~\ref{tab:c4-c5_ia_only} presents the corresponding results for $\bm{C}_4$--$\bm{C}_5$, which incorporate CMB and BAO priors to constrain the $w_0 w_a$CDM model parameters. In Figure~\ref{fig:calibration_grid}, we demonstrate the empirical calibration of our \textit{FlowSN2} posteriors for both pure-Ia samples (right column) and contaminated samples (left column).

\begin{table*}
\centering
\caption{Average of posterior median estimates over 25 repeats and the standard error on this average for the photometric SN Ia cosmology experiments without contamination. We include \textit{FlowSN2} inferences, as well as result that use the analytical likelihood solution for comparison. We also combine SNe with a CMB prior for additional constraining as detailed in Appendix~\ref{app:cmb_bao_prior}. Note uncertainties are the standard error on the average so are a factor of five smaller than those in individual experiments. The $\sigma$ values are calculated using unrounded averages and standard errors.}
\label{tab:c1-c3_ia_only}
\begin{tabular}{lccccc}
\hline
\multicolumn{6}{c}{Cosmology $\bm{C}_1$} \\
\hline
Test &  & \multicolumn{2}{c}{SN Only} & \multicolumn{2}{c}{SN + CMB Prior}\\
\hline
Global Parameter & Truth  & \textit{FlowSN2} & Analytical & \textit{FlowSN2} & Analytical \\
\hline
$w_0$ & -1 & -1.07 $\pm$ 0.04 (-1.8$\sigma$) & -1.06 $\pm$ 0.04 (-1.6$\sigma$) & -1.000 $\pm$ 0.003 (-0.1$\sigma$) & -1.000 $\pm$ 0.003 (-0.1$\sigma$)\\
$\Omega_{m0}$ & 0.315 & 0.320 $\pm$ 0.013 (0.4$\sigma$) & 0.317 $\pm$ 0.013 (0.1$\sigma$) & 0.3150 $\pm$ 0.0006 (-0.0$\sigma$) & 0.3150 $\pm$ 0.0006 (0.1$\sigma$)\\
$M_0^{\mathrm{int}}$ & -19.38 & -19.374 $\pm$ 0.004 (1.5$\sigma$) & -19.375 $\pm$ 0.004 (1.2$\sigma$) & -19.366 $\pm$ 0.004 (3.9$\sigma$) & -19.367 $\pm$ 0.004 (3.5$\sigma$)\\
$\alpha$ & -0.149 & -0.1469 $\pm$ 0.0008 (2.5$\sigma$) & -0.1493 $\pm$ 0.0008 (-0.4$\sigma$) & -0.1469 $\pm$ 0.0008 (2.5$\sigma$) & -0.1492 $\pm$ 0.0008 (-0.3$\sigma$)\\
$\beta_\mathrm{int}$ & 2.322 & 2.379 $\pm$ 0.019 (3.0$\sigma$) & 2.39 $\pm$ 0.02 (3.3$\sigma$) & 2.379 $\pm$ 0.019 (2.9$\sigma$) & 2.39 $\pm$ 0.02 (3.4$\sigma$)\\
$\alpha_c$ & -0.006 & -0.0055 $\pm$ 0.0004 (1.1$\sigma$) & -0.0060 $\pm$ 0.0004 (-0.0$\sigma$) & -0.0056 $\pm$ 0.0004 (1.1$\sigma$) & -0.0060 $\pm$ 0.0004 (0.0$\sigma$)\\
$R_B$ & 3.74 & 3.68 $\pm$ 0.04 (-1.7$\sigma$) & 3.69 $\pm$ 0.04 (-1.3$\sigma$) & 3.68 $\pm$ 0.04 (-1.8$\sigma$) & 3.69 $\pm$ 0.03 (-1.4$\sigma$)\\
$c^{\mathrm{int}}_0$ & -0.056 & -0.0563 $\pm$ 0.0009 (-0.4$\sigma$) & -0.0560 $\pm$ 0.0008 (-0.0$\sigma$) & -0.0564 $\pm$ 0.0008 (-0.5$\sigma$) & -0.0560 $\pm$ 0.0008 (0.1$\sigma$)\\
$\sigma_\mathrm{int}$ & 0.101 & 0.1077 $\pm$ 0.0016 (4.3$\sigma$) & 0.1050 $\pm$ 0.0017 (2.4$\sigma$) & 0.1078 $\pm$ 0.0015 (4.5$\sigma$) & 0.1051 $\pm$ 0.0017 (2.5$\sigma$)\\
$\sigma_{c,\mathrm{int}}$ & 0.071 & 0.0706 $\pm$ 0.0004 (-0.9$\sigma$) & 0.0708 $\pm$ 0.0005 (-0.4$\sigma$) & 0.0706 $\pm$ 0.0004 (-0.8$\sigma$) & 0.0709 $\pm$ 0.0005 (-0.3$\sigma$)\\
$\sigma_x$ & 1.137 & 1.133 $\pm$ 0.005 (-0.8$\sigma$) & 1.131 $\pm$ 0.005 (-1.1$\sigma$) & 1.133 $\pm$ 0.005 (-0.8$\sigma$) & 1.132 $\pm$ 0.005 (-1.1$\sigma$)\\
$\tau$ & 0.068 & 0.0673 $\pm$ 0.0009 (-0.8$\sigma$) & 0.0682 $\pm$ 0.0009 (0.2$\sigma$) & 0.0674 $\pm$ 0.0008 (-0.8$\sigma$) & 0.0681 $\pm$ 0.0008 (0.1$\sigma$)\\
$x_0$ & -0.452 & -0.451 $\pm$ 0.006 (0.2$\sigma$) & -0.454 $\pm$ 0.006 (-0.3$\sigma$) & -0.450 $\pm$ 0.006 (0.3$\sigma$) & -0.453 $\pm$ 0.006 (-0.2$\sigma$)\\
$\xi_\mathrm{rate}$ & 1.5 & 1.55 $\pm$ 0.05 (1.0$\sigma$) & 1.55 $\pm$ 0.05 (1.1$\sigma$) & 1.54 $\pm$ 0.04 (0.9$\sigma$) & 1.55 $\pm$ 0.04 (1.1$\sigma$)\\
$\gamma$ & -0.1 & -0.097 $\pm$ 0.002 (1.3$\sigma$) & -0.097 $\pm$ 0.002 (1.3$\sigma$) & -0.097 $\pm$ 0.002 (1.2$\sigma$) & -0.097 $\pm$ 0.002 (1.2$\sigma$)\\
\hline
\multicolumn{6}{c}{Cosmology $\bm{C}_2$} \\
\hline
Test &  & \multicolumn{2}{c}{SN Only} & \multicolumn{2}{c}{SN + CMB Prior}\\
\hline
Global Parameter & Truth  & \textit{FlowSN2} & Analytical & \textit{FlowSN2} & Analytical \\
\hline
$w_0$ & -0.8 & -0.87 $\pm$ 0.04 (-1.6$\sigma$) & -0.86 $\pm$ 0.04 (-1.5$\sigma$) & -0.804 $\pm$ 0.002 (-1.9$\sigma$) & -0.804 $\pm$ 0.002 (-1.9$\sigma$)\\
$\Omega_{m0}$ & 0.35 & 0.350 $\pm$ 0.016 (0.0$\sigma$) & 0.347 $\pm$ 0.016 (-0.2$\sigma$) & 0.3511 $\pm$ 0.0007 (1.6$\sigma$) & 0.3511 $\pm$ 0.0007 (1.6$\sigma$)\\
$M_0^{\mathrm{int}}$ & -19.38 & -19.386 $\pm$ 0.004 (-1.5$\sigma$) & -19.389 $\pm$ 0.004 (-2.2$\sigma$) & -19.377 $\pm$ 0.003 (1.1$\sigma$) & -19.380 $\pm$ 0.003 (0.1$\sigma$)\\
$\alpha$ & -0.149 & -0.1467 $\pm$ 0.0008 (2.7$\sigma$) & -0.1494 $\pm$ 0.0008 (-0.4$\sigma$) & -0.1467 $\pm$ 0.0008 (2.8$\sigma$) & -0.1493 $\pm$ 0.0008 (-0.3$\sigma$)\\
$\beta_\mathrm{int}$ & 2.322 & 2.302 $\pm$ 0.015 (-1.3$\sigma$) & 2.312 $\pm$ 0.017 (-0.6$\sigma$) & 2.304 $\pm$ 0.015 (-1.2$\sigma$) & 2.314 $\pm$ 0.017 (-0.5$\sigma$)\\
$\alpha_c$ & -0.006 & -0.0054 $\pm$ 0.0004 (1.5$\sigma$) & -0.0057 $\pm$ 0.0004 (0.6$\sigma$) & -0.0055 $\pm$ 0.0004 (1.4$\sigma$) & -0.0057 $\pm$ 0.0004 (0.7$\sigma$)\\
$R_B$ & 3.74 & 3.75 $\pm$ 0.03 (0.3$\sigma$) & 3.78 $\pm$ 0.03 (1.4$\sigma$) & 3.75 $\pm$ 0.03 (0.2$\sigma$) & 3.77 $\pm$ 0.03 (1.1$\sigma$)\\
$c^{\mathrm{int}}_0$ & -0.056 & -0.0565 $\pm$ 0.0009 (-0.6$\sigma$) & -0.0560 $\pm$ 0.0009 (-0.0$\sigma$) & -0.0566 $\pm$ 0.0009 (-0.7$\sigma$) & -0.0561 $\pm$ 0.0009 (-0.1$\sigma$)\\
$\sigma_\mathrm{int}$ & 0.101 & 0.1034 $\pm$ 0.0015 (1.6$\sigma$) & 0.0999 $\pm$ 0.0016 (-0.7$\sigma$) & 0.1037 $\pm$ 0.0015 (1.8$\sigma$) & 0.1002 $\pm$ 0.0016 (-0.5$\sigma$)\\
$\sigma_{c,\mathrm{int}}$ & 0.071 & 0.0712 $\pm$ 0.0004 (0.5$\sigma$) & 0.0714 $\pm$ 0.0004 (1.0$\sigma$) & 0.0712 $\pm$ 0.0004 (0.5$\sigma$) & 0.0714 $\pm$ 0.0005 (1.0$\sigma$)\\
$\sigma_x$ & 1.137 & 1.138 $\pm$ 0.003 (0.4$\sigma$) & 1.137 $\pm$ 0.003 (-0.0$\sigma$) & 1.138 $\pm$ 0.003 (0.4$\sigma$) & 1.137 $\pm$ 0.003 (-0.1$\sigma$)\\
$\tau$ & 0.068 & 0.0669 $\pm$ 0.0008 (-1.4$\sigma$) & 0.0677 $\pm$ 0.0008 (-0.4$\sigma$) & 0.0669 $\pm$ 0.0008 (-1.3$\sigma$) & 0.0677 $\pm$ 0.0008 (-0.4$\sigma$)\\
$x_0$ & -0.452 & -0.453 $\pm$ 0.005 (-0.2$\sigma$) & -0.455 $\pm$ 0.005 (-0.5$\sigma$) & -0.453 $\pm$ 0.006 (-0.2$\sigma$) & -0.455 $\pm$ 0.006 (-0.5$\sigma$)\\
$\xi_\mathrm{rate}$ & 2 & 1.92 $\pm$ 0.04 (-1.8$\sigma$) & 1.96 $\pm$ 0.04 (-0.9$\sigma$) & 1.92 $\pm$ 0.03 (-2.2$\sigma$) & 1.96 $\pm$ 0.03 (-1.1$\sigma$)\\
$\gamma$ & -0.1 & -0.0998 $\pm$ 0.0020 (0.1$\sigma$) & -0.1003 $\pm$ 0.0019 (-0.1$\sigma$) & -0.0999 $\pm$ 0.0020 (0.0$\sigma$) & -0.1004 $\pm$ 0.0020 (-0.2$\sigma$)\\
\hline
\multicolumn{6}{c}{Cosmology $\bm{C}_3$} \\
\hline
Test &  & \multicolumn{2}{c}{SN Only} & \multicolumn{2}{c}{SN + CMB Prior}\\
\hline
Global Parameter & Truth  & \textit{FlowSN2} & Analytical & \textit{FlowSN2} & Analytical \\
\hline
$w_0$ & -1.2 & -1.26 $\pm$ 0.04 (-1.5$\sigma$) & -1.27 $\pm$ 0.04 (-1.7$\sigma$) & -1.207 $\pm$ 0.003 (-2.6$\sigma$) & -1.207 $\pm$ 0.003 (-2.5$\sigma$)\\
$\Omega_{m0}$ & 0.25 & 0.251 $\pm$ 0.011 (0.1$\sigma$) & 0.254 $\pm$ 0.011 (0.4$\sigma$) & 0.2507 $\pm$ 0.0003 (2.2$\sigma$) & 0.2507 $\pm$ 0.0003 (2.1$\sigma$)\\
$M_0^{\mathrm{int}}$ & -19.38 & -19.381 $\pm$ 0.005 (-0.3$\sigma$) & -19.383 $\pm$ 0.005 (-0.7$\sigma$) & -19.375 $\pm$ 0.004 (1.4$\sigma$) & -19.377 $\pm$ 0.004 (0.8$\sigma$)\\
$\alpha$ & -0.149 & -0.1486 $\pm$ 0.0010 (0.4$\sigma$) & -0.1513 $\pm$ 0.0010 (-2.3$\sigma$) & -0.1485 $\pm$ 0.0009 (0.6$\sigma$) & -0.1512 $\pm$ 0.0009 (-2.3$\sigma$)\\
$\beta_\mathrm{int}$ & 2.322 & 2.29 $\pm$ 0.02 (-1.2$\sigma$) & 2.31 $\pm$ 0.03 (-0.6$\sigma$) & 2.29 $\pm$ 0.02 (-1.3$\sigma$) & 2.31 $\pm$ 0.02 (-0.5$\sigma$)\\
$\alpha_c$ & -0.006 & -0.0057 $\pm$ 0.0004 (0.7$\sigma$) & -0.0059 $\pm$ 0.0004 (0.2$\sigma$) & -0.0057 $\pm$ 0.0004 (0.8$\sigma$) & -0.0060 $\pm$ 0.0004 (0.1$\sigma$)\\
$R_B$ & 3.74 & 3.71 $\pm$ 0.04 (-1.0$\sigma$) & 3.72 $\pm$ 0.03 (-0.5$\sigma$) & 3.71 $\pm$ 0.04 (-0.9$\sigma$) & 3.72 $\pm$ 0.03 (-0.5$\sigma$)\\
$c^{\mathrm{int}}_0$ & -0.056 & -0.0570 $\pm$ 0.0010 (-1.0$\sigma$) & -0.0566 $\pm$ 0.0010 (-0.6$\sigma$) & -0.0570 $\pm$ 0.0010 (-0.9$\sigma$) & -0.0566 $\pm$ 0.0010 (-0.6$\sigma$)\\
$\sigma_\mathrm{int}$ & 0.101 & 0.1055 $\pm$ 0.0014 (3.2$\sigma$) & 0.1018 $\pm$ 0.0015 (0.5$\sigma$) & 0.1054 $\pm$ 0.0013 (3.2$\sigma$) & 0.1017 $\pm$ 0.0014 (0.5$\sigma$)\\
$\sigma_{c,\mathrm{int}}$ & 0.071 & 0.0704 $\pm$ 0.0006 (-1.0$\sigma$) & 0.0706 $\pm$ 0.0006 (-0.6$\sigma$) & 0.0705 $\pm$ 0.0006 (-0.9$\sigma$) & 0.0706 $\pm$ 0.0006 (-0.7$\sigma$)\\
$\sigma_x$ & 1.137 & 1.142 $\pm$ 0.005 (1.1$\sigma$) & 1.140 $\pm$ 0.005 (0.7$\sigma$) & 1.142 $\pm$ 0.004 (1.1$\sigma$) & 1.140 $\pm$ 0.004 (0.7$\sigma$)\\
$\tau$ & 0.068 & 0.0677 $\pm$ 0.0012 (-0.3$\sigma$) & 0.0685 $\pm$ 0.0011 (0.4$\sigma$) & 0.0676 $\pm$ 0.0012 (-0.4$\sigma$) & 0.0685 $\pm$ 0.0011 (0.5$\sigma$)\\
$x_0$ & -0.452 & -0.455 $\pm$ 0.008 (-0.4$\sigma$) & -0.461 $\pm$ 0.008 (-1.2$\sigma$) & -0.456 $\pm$ 0.008 (-0.5$\sigma$) & -0.462 $\pm$ 0.008 (-1.2$\sigma$)\\
$\xi_\mathrm{rate}$ & 1 & 1.03 $\pm$ 0.05 (0.7$\sigma$) & 1.03 $\pm$ 0.05 (0.6$\sigma$) & 1.02 $\pm$ 0.04 (0.6$\sigma$) & 1.01 $\pm$ 0.04 (0.4$\sigma$)\\
$\gamma$ & -0.1 & -0.096 $\pm$ 0.003 (1.4$\sigma$) & -0.097 $\pm$ 0.003 (1.3$\sigma$) & -0.096 $\pm$ 0.003 (1.3$\sigma$) & -0.097 $\pm$ 0.003 (1.2$\sigma$)\\
\hline
\end{tabular}
\end{table*}

\begin{table*}
\centering
\caption{Average of posterior median estimates over 25 repeats and the standard error on this average for the photometric SN Ia cosmology experiments without contamination. Here we inference results on $w_0w_a$CDM model parameters. We include \textit{FlowSN2} inferences, as well as result that use the analytical likelihood solution for comparison. In order to constrain the evolving dark energy model we include CMB and BAO priors as detailed in Appendix~\ref{app:cmb_bao_prior}. Note uncertainties are the standard error on the average so are a factor of five smaller than those in individual experiments. The $\sigma$ values are calculated using unrounded averages and standard errors.}
\label{tab:c4-c5_ia_only}
\begin{tabular}{lccc}
\hline
\multicolumn{4}{c}{Cosmology $\bm{C}_4$} \\
\hline
Test &  & \multicolumn{2}{c}{SN + CMB + BAO}\\
\hline
Global Parameter & Truth  & \textit{FlowSN2} & Analytical \\
\hline
$w_0$ & -1 & -1.012 $\pm$ 0.009 (-1.2$\sigma$) & -1.011 $\pm$ 0.009 (-1.2$\sigma$)\\
$w_a$ & 0 & 0.03 $\pm$ 0.04 (0.7$\sigma$) & 0.03 $\pm$ 0.05 (0.7$\sigma$)\\
$\Omega_{m0}$ & 0.315 & 0.3149 $\pm$ 0.0007 (-0.2$\sigma$) & 0.3149 $\pm$ 0.0007 (-0.2$\sigma$)\\
$M_0^{\mathrm{int}}$ & -19.38 & -19.378 $\pm$ 0.004 (0.6$\sigma$) & -19.380 $\pm$ 0.004 (-0.1$\sigma$)\\
$\alpha$ & -0.149 & -0.1466 $\pm$ 0.0011 (2.3$\sigma$) & -0.1489 $\pm$ 0.0011 (0.1$\sigma$)\\
$\beta_\mathrm{int}$ & 2.322 & 2.31 $\pm$ 0.02 (-0.6$\sigma$) & 2.32 $\pm$ 0.02 (-0.0$\sigma$)\\
$\alpha_c$ & -0.006 & -0.0059 $\pm$ 0.0004 (0.2$\sigma$) & -0.0061 $\pm$ 0.0004 (-0.2$\sigma$)\\
$R_B$ & 3.74 & 3.68 $\pm$ 0.04 (-1.6$\sigma$) & 3.73 $\pm$ 0.04 (-0.3$\sigma$)\\
$c^{\mathrm{int}}_0$ & -0.056 & -0.0581 $\pm$ 0.0009 (-2.4$\sigma$) & -0.0573 $\pm$ 0.0009 (-1.4$\sigma$)\\
$\sigma_\mathrm{int}$ & 0.101 & 0.1055 $\pm$ 0.0015 (2.9$\sigma$) & 0.1017 $\pm$ 0.0015 (0.5$\sigma$)\\
$\sigma_{c,\mathrm{int}}$ & 0.071 & 0.0697 $\pm$ 0.0005 (-2.4$\sigma$) & 0.0701 $\pm$ 0.0006 (-1.5$\sigma$)\\
$\sigma_x$ & 1.137 & 1.143 $\pm$ 0.004 (1.5$\sigma$) & 1.141 $\pm$ 0.004 (1.1$\sigma$)\\
$\tau$ & 0.068 & 0.0684 $\pm$ 0.0009 (0.5$\sigma$) & 0.0688 $\pm$ 0.0009 (0.8$\sigma$)\\
$x_0$ & -0.452 & -0.450 $\pm$ 0.006 (0.3$\sigma$) & -0.452 $\pm$ 0.006 (-0.0$\sigma$)\\
$\xi_\mathrm{rate}$ & 1.5 & 1.44 $\pm$ 0.04 (-1.5$\sigma$) & 1.45 $\pm$ 0.04 (-1.3$\sigma$)\\
$\gamma$ & -0.1 & -0.1013 $\pm$ 0.0018 (-0.7$\sigma$) & -0.1013 $\pm$ 0.0019 (-0.7$\sigma$)\\
\hline
\multicolumn{4}{c}{Cosmology $\bm{C}_5$} \\
\hline
Test &  & \multicolumn{2}{c}{SN + CMB + BAO}\\
\hline
Global Parameter & Truth  & \textit{FlowSN2} & Analytical \\
\hline
$w_0$ & -0.752 & -0.745 $\pm$ 0.012 (0.6$\sigma$) & -0.746 $\pm$ 0.012 (0.5$\sigma$)\\
$w_a$ & -0.86 & -0.90 $\pm$ 0.06 (-0.8$\sigma$) & -0.90 $\pm$ 0.06 (-0.8$\sigma$)\\
$\Omega_{m0}$ & 0.316 & 0.3163 $\pm$ 0.0004 (0.8$\sigma$) & 0.3163 $\pm$ 0.0004 (0.8$\sigma$)\\
$M_0^{\mathrm{int}}$ & -19.38 & -19.374 $\pm$ 0.003 (1.9$\sigma$) & -19.376 $\pm$ 0.004 (1.2$\sigma$)\\
$\alpha$ & -0.149 & -0.1478 $\pm$ 0.0008 (1.4$\sigma$) & -0.1502 $\pm$ 0.0009 (-1.4$\sigma$)\\
$\beta_\mathrm{int}$ & 2.322 & 2.343 $\pm$ 0.019 (1.1$\sigma$) & 2.356 $\pm$ 0.019 (1.7$\sigma$)\\
$\alpha_c$ & -0.006 & -0.0059 $\pm$ 0.0003 (0.5$\sigma$) & -0.0062 $\pm$ 0.0003 (-0.7$\sigma$)\\
$R_B$ & 3.74 & 3.71 $\pm$ 0.05 (-0.7$\sigma$) & 3.74 $\pm$ 0.05 (-0.0$\sigma$)\\
$c^{\mathrm{int}}_0$ & -0.056 & -0.0562 $\pm$ 0.0006 (-0.4$\sigma$) & -0.0557 $\pm$ 0.0006 (0.4$\sigma$)\\
$\sigma_\mathrm{int}$ & 0.101 & 0.1030 $\pm$ 0.0014 (1.5$\sigma$) & 0.0997 $\pm$ 0.0014 (-0.9$\sigma$)\\
$\sigma_{c,\mathrm{int}}$ & 0.071 & 0.0714 $\pm$ 0.0005 (0.9$\sigma$) & 0.0716 $\pm$ 0.0005 (1.4$\sigma$)\\
$\sigma_x$ & 1.137 & 1.136 $\pm$ 0.004 (-0.3$\sigma$) & 1.135 $\pm$ 0.004 (-0.6$\sigma$)\\
$\tau$ & 0.068 & 0.0667 $\pm$ 0.0006 (-2.3$\sigma$) & 0.0674 $\pm$ 0.0006 (-1.0$\sigma$)\\
$x_0$ & -0.452 & -0.443 $\pm$ 0.004 (2.4$\sigma$) & -0.445 $\pm$ 0.004 (1.7$\sigma$)\\
$\xi_\mathrm{rate}$ & 1.5 & 1.52 $\pm$ 0.03 (0.6$\sigma$) & 1.54 $\pm$ 0.03 (1.2$\sigma$)\\
$\gamma$ & -0.1 & -0.102 $\pm$ 0.002 (-1.0$\sigma$) & -0.102 $\pm$ 0.002 (-1.0$\sigma$)\\
\hline
\end{tabular}
\end{table*}

\bsp	
\label{lastpage}
\end{document}